\documentclass[aps,prx,onecolumn,superscriptaddress,amsmath,amssymb,noeprint,preprint]{revtex4-2}

\usepackage{hyperref}
\usepackage{graphicx}
\usepackage{float}
\usepackage{amsmath}
\usepackage{mathrsfs}
\usepackage{soul}
\usepackage{amssymb}
\usepackage{booktabs}
\usepackage{siunitx}
\usepackage{braket}
\usepackage{upgreek}
\usepackage{bm}
\usepackage{xcolor}
\usepackage[normalem]{ulem}
\usepackage{setspace}
\usepackage{amsfonts}
\usepackage{stackengine}
\usepackage[most]{tcolorbox}

\newtcolorbox{algobox}{
  colback=blue!5,
  colframe=black,
  arc=3mm,
  boxrule=0.6pt,
  left=0.5mm,
  right=0.5mm,
  top=1mm,
  bottom=1mm
}

\newcommand{\papertitle}{%
Crossing over universal scaling laws in two-dimensional driven dissipative condensates
}

\newcommand{\supplementarytocentry}[2]{%
  \noindent
  \hyperref[#1]{#2}%
  \leaders\hbox{.\kern0.5pt}\hfill
  \pageref{#1}\par
}

\DeclareGraphicsExtensions{.png,.jpg,.eps}

\begin{document}

\title{Crossing over universal scaling laws in two-dimensional driven dissipative condensates}

\author{Q. Fontaine}
\affiliation{Universit\'{e} Paris-Saclay, CNRS, Centre de Nanosciences et de Nanotechnologies (C2N), 91120, Palaiseau, France}

\author{F. Helluin}
\affiliation{Universit\'{e} Paris-Saclay, CNRS, Centre de Nanosciences et de Nanotechnologies (C2N), 91120, Palaiseau, France}

\author{M. Escalera}
\affiliation{Universit\'{e} Paris-Saclay, CNRS, Centre de Nanosciences et de Nanotechnologies (C2N), 91120, Palaiseau, France}

\author{D. Pinto Dias}
\affiliation{Universit\'{e} Paris-Saclay, CNRS, Centre de Nanosciences et de Nanotechnologies (C2N), 91120, Palaiseau, France}

\author{A. Lema\^{i}tre}
\affiliation{Universit\'{e} Paris-Saclay, CNRS, Centre de Nanosciences et de Nanotechnologies (C2N), 91120, Palaiseau, France}

\author{M.~Morassi}
\affiliation{Universit\'{e} Paris-Saclay, CNRS, Centre de Nanosciences et de Nanotechnologies (C2N), 91120, Palaiseau, France}

\author{M.~Wouters}
\affiliation{TQC, Universiteit Antwerpen, Universiteitsplein 1, B-2610 Antwerpen, Belgium.}

\author{A. Minguzzi}
\affiliation{Univ. Grenoble Alpes and CNRS, Laboratoire de Physique et Mod\'elisation des Milieux Condens\'es, 38000 Grenoble, France.}

\author{L. Canet}
\affiliation{Univ. Grenoble Alpes and CNRS, Laboratoire de Physique et Mod\'elisation des Milieux Condens\'es, 38000 Grenoble, France.}

\author{S. Ravets}
\affiliation{Universit\'{e} Paris-Saclay, CNRS, Centre de Nanosciences et de Nanotechnologies (C2N), 91120, Palaiseau, France}

\author{J.~Bloch}
\affiliation{Universit\'{e} Paris-Saclay, CNRS, Centre de Nanosciences et de Nanotechnologies (C2N), 91120, Palaiseau, France}

\date{\today}

\begin{abstract}

\bigskip

\textbf{In low dimensional systems, fluctuations are enhanced and prevent the spontaneous breaking of continuous symmetries. As a result, spatial and temporal correlation functions decay at large distances and long times. A well established example is given by two-dimensional bosonic condensates at equilibrium, which do not display long-range order of the coherence but algebraic decay belonging to the Berezinski--Kosterlitz--Thouless universality class. In contrast, the universal behaviors of non-equilibrium bosonic condensates are more diverse and many open questions remain. Here, we explore the spatio-temporal coherence properties of two-dimensional driven-dissipative polariton condensates in semiconductor optical microcavities. By tuning microscopic parameters, we observe a cross-over between two scaling laws that we attribute to the Edwards--Wilkinson (EW) and the Kardar--Parisi--Zhang (KPZ) universality classes. We demonstrate the collapse of the measured first-order correlations onto the EW and KPZ universal scaling functions and obtain critical exponents, well matching the values predicted theoretically. Our results highlight the intrinsic non-equilibrium nature of polariton condensates and establish them  as a platform of choice for controlled exploration of the two-dimensional KPZ universality class.}
\end{abstract}

\maketitle

\newpage

Bose Einstein condensation (BEC) is a fascinating illustration of quantum physics, where an ensemble of bosons massively occupy a single quantum state below a critical temperature and share a single macroscopic coherent wavefunction~\cite{Bose1924}. BECs have been extensively studied using cold atom gases~\cite{Anderson1995, Davis1995}, that are well isolated from their environment and effectively behave as closed systems. Reducing the dimensionality from three-dimensional to two-dimensional (2D), leads to the formation of quasi-condensates, where the effect of quantum and thermal
fluctuations is enhanced, prohibiting the establishment of true long-range order at any nonzero temperature. Below a critical temperature $T_{\rm BKT}$, phase fluctuations caused by long wavelength phonons lead to an algebraic spatio-temporal coherence decay characteristic of the Berezinskii--Kosterlitz--Thouless (BKT)~\cite{Berezinsky_1970} universality class. Above $T_{\rm BKT}$, vortex unbinding and proliferation occurs, leading to a disordered phase with exponential coherence decay.

Subsequent observations of 2D quasi-condensates were obtained in the solid state using various quasi-particles such as indirect excitons~\cite{Butov2002, Snoke2002, Dang2020}, exciton and plasmon polaritons~\cite{Kasprzak2006, Hakala2018} as well as magnons~\cite{Demokritov2006}. While 2D indirect exciton quasi-condensates effectively form a closed system and display BKT physics~\cite{Dang2020}, the rest of the aforementioned platforms are driven-dissipative in nature. A constant drive is required to compensate particle losses and reach a steady state. Therefore, these platforms offer a unique playground to explore driven-dissipative condensates and reveal new physics specific to open systems~\cite{bloch2022}.

In recent years, multiple facets of 2D exciton-polariton quasi-condensates (polariton BEC) have been theoretically explored, suggesting a particularly rich phase diagram~\cite{Altman2015, Zamora2017, Dagvadorj2021, Helluin2025}. For instance, algebraic coherence decays have been predicted, with exponents departing from equilibrium BKT physics~\cite{Szymanska2006, Dagvadorj2015, Comaron2021, Helluin2025}. This diffusive regime has recently been connected to the Edwards--Wilkinson (EW) universality class~\cite{Helluin2025}. Strikingly, theoretical predictions have highlighted the existence of a super-diffusive regime~\cite{Altman2015, Zamora2017, Mei2021, Ferrier2022, deligiannis2022, Helluin2025} belonging to the Kardar--Parisi--Zhang (KPZ) universality class~\cite{Kardar1986}. In this case, stretched exponential decays of the coherence are expected, with universal exponents corresponding to the KPZ universality class. Interestingly, by tuning the system microscopic parameters, one can tune across these universality classes \cite{Zamora2017, Helluin2025} as well as explore other phases such as solitons, spiral vortices or disordered phases.

The ability to observe KPZ scaling in polariton BECs was first proposed in 1D~\cite{He2015, Ji2015, He2017}. In 2D, the observability of the KPZ universality class with polaritons has stirred intense debate. On the one hand, it was suggested that the nucleation and proliferation of vortices would prevent the establishment of KPZ correlations at length scales that are relevant for an experiment~\cite{Altman2015}. On the other hand, other works have observed signatures of KPZ scaling laws in numerical simulations at realistic distances~\cite{Mei2021}. In parallel, various techniques have been proposed to favor the observability of KPZ correlations, namely the implementation of an anisotropic Optical Parametric Oscillation pumping scheme~\cite{Zamora2017}, or the use of discrete lattice systems~\cite{deligiannis2022,Helluin2025} to impose a cut-off on the vortex size.

Experimentally, algebraic decays of the spatio-temporal coherence in 2D polariton condensates have been reported in several works~\cite{Roumpos2012, Caputo2018, Comaron2025}. However convincing demonstration of EW scaling is still lacking. Regarding the KPZ universality class, the first demonstration was achieved in 1D discrete polariton lattices~\cite{Fontaine2022}, and recent attempts to extend this work to 2D~\cite{Widmann2026} have remained inconclusive~\cite{bloch2026}. Overall, important fundamental questions remain open regarding the ability to observe quasi-ordered phases (EW or KPZ) in a continuous 2D polariton system, where no cutoff prevents proliferation of vortices.

In this work, we experimentally establish 2D polariton BEC as a versatile platform to investigate the rich physics of open BECs and their associated universality classes. Using a planar microcavity, we probe the spatio-temporal first-order coherence of non-resonantly driven polariton BECs. We clearly identify a KPZ super-diffusive regime, where the coherence exhibits stretched exponential decays with critical exponents characteristic of the 2D KPZ universality class. Importantly, by tuning the system parameters, we demonstrate a crossover to an EW diffusive phase showing algebraic decay of the first-order coherence with the characteristic ratio equal to two between spatial and temporal exponents. Using numerical simulations, we analyze the role of topological defects. We evidence that vortices and anti-vortices are mostly paired, indicating that KPZ and EW regimes are the out-of-equilibrium analog of the ordered BKT phase. 

Cavity polaritons (polaritons) are hybrid exciton-photon quasi-particles emerging from the strong coupling of excitons confined in a quantum well (QW) and presenting a sharp resonance to photons confined in a cavity~\cite{carusotto2013}. These quasi-particles are intrinsically dissipative and escape the system on timescales comparable to the photon lifetime in the cavity. In order to reach a steady state, the losses must be compensated using a continuous laser as a driving field. The use of a non-resonant laser far blue-detuned from the polariton modes leads to the formation of an exciton reservoir around the exciton energy and to an incoherent population of the polariton bands. Increasing the pump fluence, the exciton reservoir populates the polariton modes via stimulated scattering, and plays the role of a gain medium. For large enough values of the gain, a single polariton state becomes macroscopically occupied. This defines a polariton BEC with condensate wavefunction $\psi(\boldsymbol{r},t)$, whose dynamics is well described by two coupled equations (see Methods). The first equation (Eq.~\ref{eq:GPE} of the Methods) describes the dynamics of the polariton wavefunction and takes the form of a stochastic driven-dissipative Gross-Pitaevskii equation. It includes a complex Gaussian white noise that acts as a stochastic drive on the polariton field. The second equation (Eq.~\ref{eq:reservoir} of the Methods) is a rate equation describing the dynamics of the reservoir population $n_R(t)$, and includes a stimulated scattering term into the condensate.

Analogously to closed systems, the spontaneous breaking of the $U(1)$ phase symmetry above condensation threshold in the mean field description is accompanied by a phase-like Goldstone mode. As a result, the long-time, large-distance condensate fluctuations are mainly governed by the dynamics of its phase component $\theta(\boldsymbol{r}, t)$. As a hallmark of nonequilibrium condensation, the associated Goldstone mode is however diffusive~\cite{Wouters2007, Sieberer2016}. In this regime, remarkably, the condensate phase dynamics is governed by the Kardar--Parisi--Zhang (KPZ) equation~\cite{Gladilin_Wouters_PRA2014, Gladilin_2015_EW_to_KPZ, He_Diehl_roughness, squizzato2018KPZsubclasses, deligiannis2021KPZsubclasses, Diehlspacetimevortex_PRL2017, Vercesi_PRR2023_1d_phase_diag}:
\begin{equation}
    \partial_{t} \theta(\boldsymbol{r}, t) = \nu \nabla^{2} \theta (\boldsymbol{r}, t) + \frac{\lambda}{2} \left[ \boldsymbol{\nabla} \theta (\boldsymbol{r}, t) \right]^{2} + \sqrt{D} \, \eta(\boldsymbol{r}, t) \, ,\label{eq:main_KPZ_mapping}
\end{equation}
where $\nu$, $\lambda$ and $D$ are coefficients that depend on the system microscopic parameters, and $\eta(\boldsymbol{r}, t)$ corresponds to a real-valued Gaussian white noise, see Supplementary Information (SI). For a finite-size system, the condensate phase dynamics (Eq.~\ref{eq:main_KPZ_mapping}) can yield an EW diffusive regime when the KPZ effective nonlinearity $g_{\rm KPZ} = \lambda^2 D/\nu^3$ is weak, or a KPZ super-diffusive regime otherwise. As a result, the condensate phase is predicted to develop spatio-temporal correlations characterized by universal KPZ or EW scalings. Notably, one can show that, in these regimes, the first-order coherence $g^{(1)}(\Delta \boldsymbol{ r}, \Delta t)$ is directly linked to the two-point phase correlation function $C_{\theta \theta} (\Delta\boldsymbol{ r}, \Delta t)$ by the following expression  (see SI):
\begin{equation}
  g^{(1)}( \boldsymbol{ \Delta r}, \Delta t) \, \propto \, \exp(- C_{\theta \theta} (\Delta \boldsymbol{ r}, \Delta t)) \, .
\end{equation}
This shows that scaling laws of the phase correlations can be accessed by measuring the first-order coherence via interferometry experiments. 

\begin{figure*}[t]
    \centering
    \includegraphics[width=\linewidth]{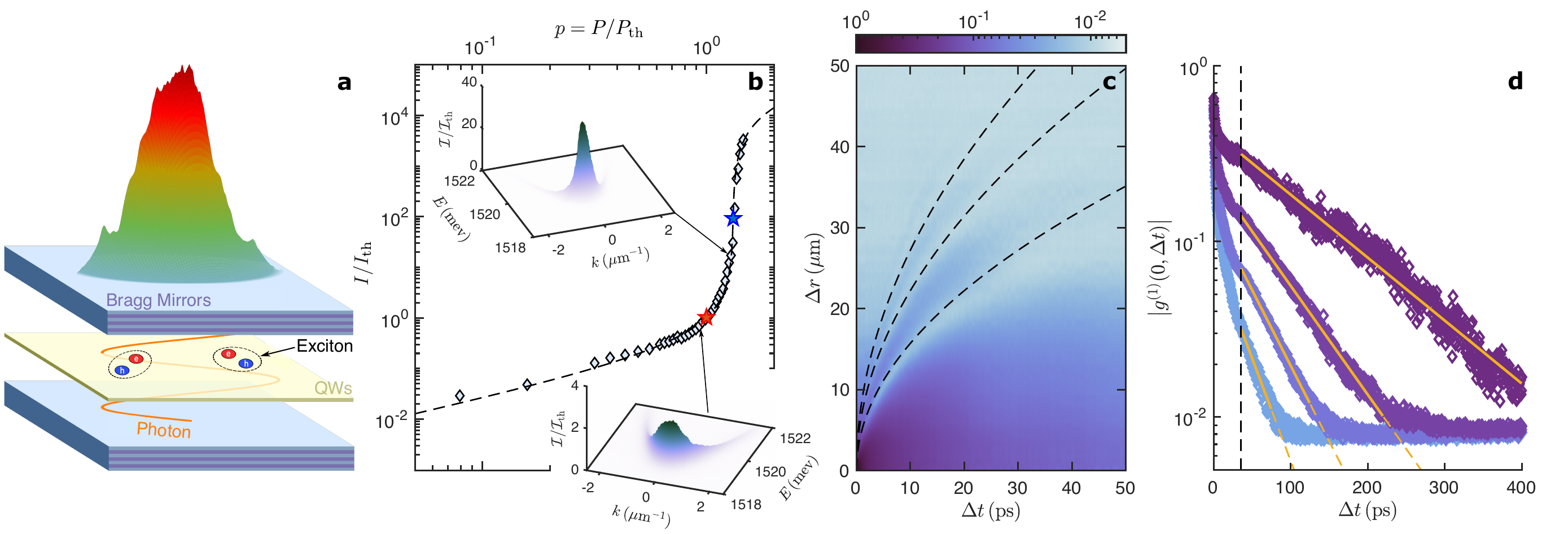}
    \caption{
    \textbf{a} Sketch of the microcavity sample operating in the strong exciton-photon regime, where an extended polariton condensate is optically generated.
    \textbf{b} Normalized emission intensity $I/I_{th}$ as a function of the normalized excitation power $p=P/P_{th}$, where $P_{th}$ ($I_{th}$) is the excitation power (measured intensity) at condensation threshold. The black dashed line shows a fit using a single-mode lasing model  from which $P_{th}$ (red star) is determined as well as $P_{trp}$ (blue star) the transparency threshold (see Methods and SI). Bottom (top) inset: measured emission intensity as a function of energy and momentum represented in 3D for $p \simeq 0.89 $ (resp. $p \simeq 1.22$). 
    \textbf{c} Measured $|g^{(1)}(\Delta r, \Delta t)|$ as a function of $\Delta r$ (averaged over the angular coordinate) and $\Delta t$ for $p \simeq 1.21$. 
    The black dashed lines correspond to a fit to the three observed dispersive branches (see Methods). 
    \textbf{d} Temporal coherence $|g^{(1)}(0, \Delta t)|$ measured as a function of $\Delta t$ for $p \simeq 1.16, 1.21 ,1.25, 1.27$. 
    The orange lines indicate exponential decays.
    For panels \textbf{b}-\textbf{d}, $\hbar \delta = -5.2 \, \mathrm{meV}$.}
   \label{fig:condensation} 
\end{figure*}

The sample we use is schematically represented in Fig.~\ref{fig:condensation}a. It consists of a planar cavity surrounded by two Bragg mirrors and embedding QWs (see Methods). The cavity and mirror layer thicknesses present an intentional spatial gradient enabling changing the energy detuning $\hbar \delta = \hbar \omega_C(\boldsymbol{k}=0) - \hbar \omega_X$ between the $\boldsymbol{k} = 0$ cavity mode energy $\hbar \omega_{C}(\boldsymbol{k}=0)$ and the QW exciton resonance $\hbar \omega_X$. Polaritons are excited using a CW laser highly blue detuned with respect to the polariton resonances and focused on the sample surface within a $30~{\rm \mu m}$ radius spot with flat top intensity profile.

We first analyze the polariton emission spectrum in momentum space. The $\boldsymbol{k}=0$ emission intensity is reported with diamond symbols in Fig.~\ref{fig:condensation}b as a function of excitation power $P$. Above a threshold power $P_{\rm th}$ marking the condensation threshold (red star), we observe a pronounced non-linear increase of the emission intensity (see Methods and Extended Data Fig.~\ref{fig:AF:comp_lin_mod} for a description of the determination of $P_{th}$ as well as the transparency threshold $P_{\rm trp}$ marked by a blue star). As shown in the figure insets, the emission becomes highly peaked around $\boldsymbol{k}=0$, which signals a macroscopic occupation of this polariton state and the onset of polariton condensation. Notably, due to the finite size of the pump, the condensate remains stable in vicinity of $P_{\rm th}$. Nevertheless, beyond a certain power value $p=P/P_{\rm th} \gtrsim 1.40$, a modulational instability develops~\cite{Bobrovska2014}, leading to the fragmentation of the condensate into spatially separated regions that are not mutually coherent, preventing investigation of higher $p$ values~\cite{baboux2018}.

Using a Michelson interferometer we overlap the real space emission image with its $\pi$-rotated image around the condensate center. We induce a controlled time-delay $\Delta t$ between both images in the interferometer (see Methods and Extended Data Fig.~\ref{fig:AF:setup}). From the fringe visibility, we extract the first-order coherence (see Methods) defined as:
\begin{equation}
   g^{(1)}(\Delta \boldsymbol{r}, \Delta t) = \frac{\langle \, \psi^{*} (\boldsymbol{r}/2, t_{0} ) \, \psi (-\boldsymbol{r}/2, t_{0} + \Delta t) \, \rangle}{\sqrt{\langle | \psi (\boldsymbol{r}/2, t_{0} ) |^{2} \rangle} \, \sqrt{\langle | \psi (-\boldsymbol{r}/2, t_{0} + \Delta t) |^{2} \rangle}}, 
   \label{eq:first_order_coherence}
\end{equation}
where angle brackets $\langle\cdot\rangle$ denote ensemble averaging over noise realizations during the acquisition time. Typical interference patterns and corresponding extracted $g^{(1)}$ are displayed in Extended Data Fig.~\ref{fig:AF:interfero}. Gathering such measurements for many values of $\Delta t$, we obtain the $g^{(1)}$ heat map shown in Fig.~\ref{fig:condensation}. Apart from the observed spatio-temporal decay of the coherence, we notice the presence of dispersive branches, which originate from an incoherent population of uncondensed polaritons distributed over the polariton band. These modes are well fitted by Fourier transforming the polariton parabolic band as shown with the dotted lines and explained in the Methods.

Focusing on the coherence temporal decay, we display, in Fig.~\ref{fig:condensation}d the  values of $g^1(\Delta \boldsymbol{r} = 0, \Delta t)$ measured for four excitation powers above condensation threshold. The curves shown in semi-logarithmic scale, exhibits an exponential decay at time delays $\Delta t > 40~{\rm ps}$  that we attribute to finite size effects and corresponds to the Schawlow-Townes limit found in laser physics~\cite{Schawlow1958,Keeling2010, Helluin2025}. In the rest of the paper, we thus search for the EW or KPZ universal regimes at time delays shorter than $40~{\rm ps}$.

\begin{figure*}[t]
    \centering
    \includegraphics[width=\linewidth]{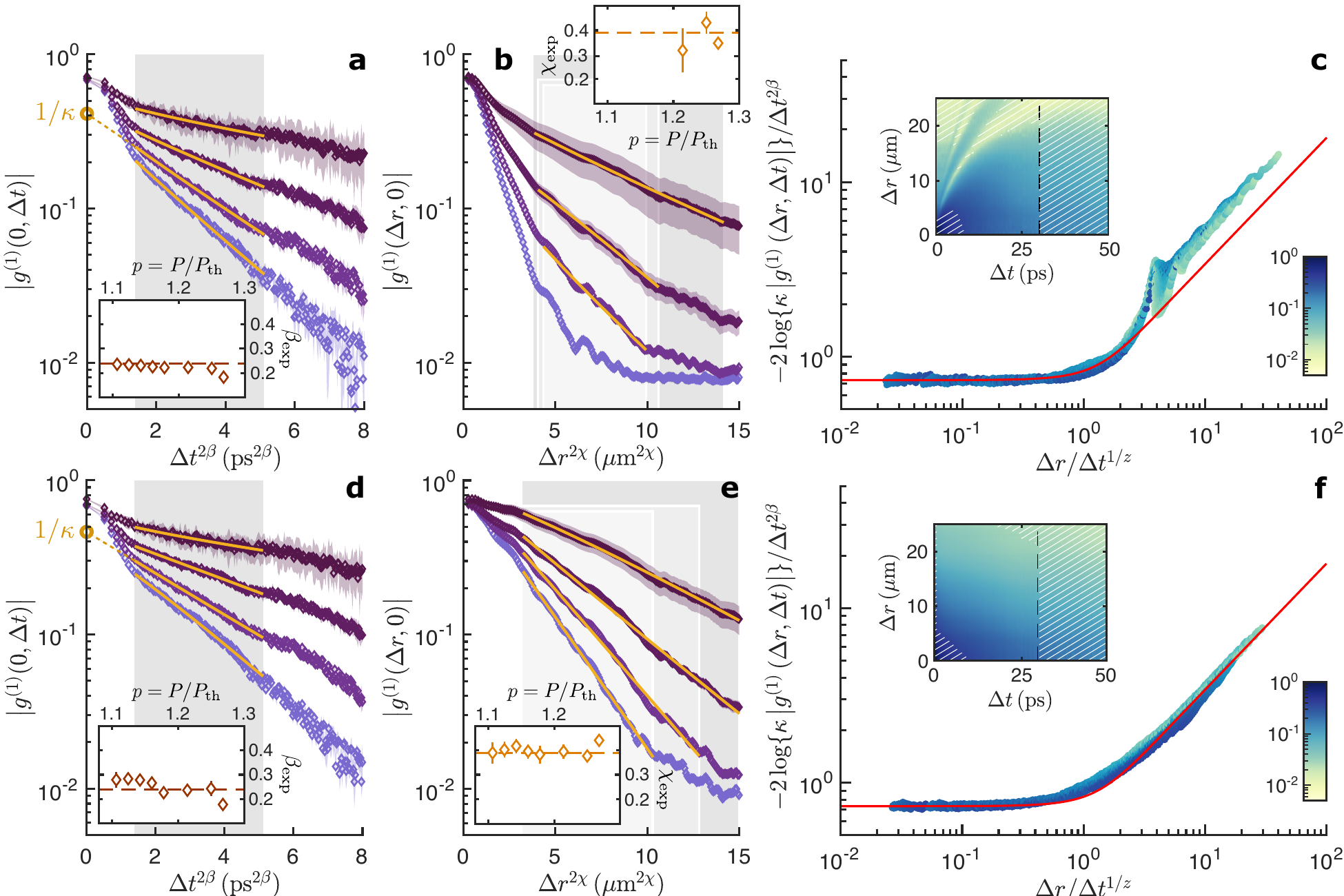}
    \caption{\textbf{Observation of KPZ universal scaling in 2D polariton condensates.}
    \textbf{a-b.} Semi-logarithmic plots of the measured values of $|g^{(1)}(0, \Delta t)|$ versus $\Delta t^{2 \beta}$ (panel \textbf{a}), and $|g^{(1)}(\Delta r, 0)|$) versus $\Delta r^{2 \chi}$ (panel \textbf{b}) for $p = 1.14,\, 1.21,\, 1.25, \, 1.27$. The experimental data points are represented by symbols forming thick lines in some parts of the graph. The orange lines are stretched exponential fits over the KPZ windows indicated by darker areas. The errorbars around each curve are obtained from multiple interferograms analysis. Insets: the diamond symbols plot the experimental values of $\beta_{\mathrm{exp}}$ (panel \textbf{a}) and $\chi_{\mathrm{exp}}$ (panel \textbf{b}) extracted from stretched exponential fits to the data, as a function of $p$. Error bars represent the one-standard-deviation ($\pm 1 \sigma$) uncertainties obtained from the fits. Horizontal dashed lines indicate the numerically calculated KPZ exponents.
    \textbf{c.} Plot of $-2 \, \mathrm{log}(\kappa |g^{(1)}(\Delta r, \Delta t)|)/\Delta t^{2\beta}$ measured for $p=1.21$ as a function of $\Delta r/\Delta t^{1/z}$ for values of $|g^{(1)}(\Delta r, \Delta t) |$ lying within the identified KPZ window (non-hatched region of the inset, showing the heat map of $|g^{(1)}(\Delta r, \Delta t) |$). The red solid line corresponds to the KPZ scaling function  $F = C_{0} F_{\mathrm{KPZ}} (y_0 \Delta r/\Delta t^{1/z})$, adjusted to the experimental data by tuning the values of $C_{0}$ and $y_{0}$.
    \textbf{d-e.} Same as \textbf{a}-\textbf{b}
    after filtering out the dispersive branches. The normalization constants $\kappa$ are determined as shown in panels \textbf{a} ($\kappa = 2.4$) and \textbf{d} ($\kappa = 2.1$).
    \textbf{f.} Same as \textbf{c} after filtering out the dispersive branches.
    All data points shown in this figure are measured for $\hbar \delta = - 5.2~\mathrm{meV}$.
    }
    \label{fig:scalingKPZ}
\end{figure*}

We now focus on the coherence decay of polariton condensates occurring before the onset of the Schawlow-Townes regime. We first consider a small negative detuning $\hbar \delta = \SI{-5.2}{\milli \eV}$ corresponding to polaritons with $25 \%$ excitonic fraction. Fig.~\ref{fig:scalingKPZ}a shows the temporal coherence decay at $\Delta r = 0$ for excitation powers ranging from $p=1.14$ to $p=1.27$. In order to highlight KPZ scaling in the temporal coherence decay, we plot the data in semi-logarithmic scale using rescaled coordinates according to $\Delta t ^{2 \beta}$, with $\beta=0.24$ the numerical estimate of the 2D KPZ growth exponent~\cite{Pagnani2015}. We identify a time window (grey-shaded area) where the coherence decays appear as straight lines, thus evidencing a stretched exponential scaling. In Extended Data Fig.~\ref{fig:AF:KPZ_EW_all_scales}, we plot the same data in logarithmic and semi-logarithmic scales. There, we observe that the data form curved lines, illustrating that the scalings are neither of power law nor of exponential nature. Departure from the stretched exponential scaling at short times $\Delta t \leq 2~{\rm ps}$ is attributed to the fast coherence decay of uncondensed polaritons (signal seen close to $\Delta x = \Delta y = 0$ in Extended Data Fig.~\ref{fig:AF:interfero}).

We perform the same analysis for the spatial coherence decay. In Fig.~\ref{fig:scalingKPZ}b, we report the measured $g^{(1)}(\Delta r, 0)$ in semi-logarithmic scale as a function of $r^{2 \chi}$ with $\chi = 0.39 $ the numerical estimate of the 2D KPZ roughness exponent~\cite{Pagnani2015}. For $p \geq 1.21$, we find spatial windows where the decay curves appear as straight lines, revealing stretched exponential spatial decays. Here again, photoluminescence emission of uncondensed polaritons dominates at short distances $\Delta r \leq 6~\rm{\mu m}$, while departure from stretched exponential scaling at large distances is attributed to the finite size of the condensate created under a $60~\rm{\mu m}$ diameter excitation spot.

We now explore the full spatio-temporal behavior of the coherence decay for $p=1.21$. In the KPZ and exponential rescaled units shown in Fig.~\ref{fig:scalingKPZ}c, we expect all measured values of $-2 \log(\kappa g^{(1)}(\Delta r, \Delta t))/\Delta t^{2\beta}$ belonging to the KPZ regime to collapse onto the KPZ universal scaling function. We point out that the renormalization coefficient $\kappa$ is crucial when using these rescaled axes~\cite{Fontaine2022,bloch2026} and is determined by normalizing to $1$ at $\Delta t = 0$ the stretched exponential fits in Fig.~\ref{fig:scalingKPZ}a (see Methods). We plot, in Fig.~\ref{fig:scalingKPZ}c, all data points belonging to the identified KPZ window, while the rest of the data points are filtered out (hatched regions). Remarkably, the data points collapse onto a single curve, with asymptotes that align parallel to those of the theoretical universal KPZ scaling function~\cite{Canet2012_Scaling_fct_amplitude_ratios_1d_2d_3d} (red line in Fig.~\ref{fig:scalingKPZ}c).

Departure from the theoretical curve at intermediate values of $\Delta r/{(\Delta t^{1/z})}$ is attributed to the dispersive modes. Using Fourier analysis, we filter out these modes (see SI), leading to the clear collapse onto the theoretical KPZ curve of all selected experimental datapoints (see Fig.~\ref{fig:scalingKPZ}f). For $p=1.21$, the effect of the filtering procedure has been to shift the oblique asymptote and align it with the KPZ universal function. At lower powers, where the dispersive modes are more pronounced, we stress that the filter may enable revealing otherwise masked KPZ stretched exponential decays. This is actually what we observe Fig.~\ref{fig:scalingKPZ}d-e, where we plot the filtered spatial and temporal decays. While the filtering has a minimal effect on the temporal scaling (less influenced by the dispersive modes), we recover clear spatial stretched exponential decays for a broader range of powers extending down to $p=1.14$. 

We quantitatively estimate the measured critical exponents $\beta_{\rm exp}$ and $\chi_{\rm exp}$ by performing stretched exponential fits to these data. The fitted values of these exponents are displayed in insets of Fig.~\ref{fig:scalingKPZ}a-b-d-e. All fitted exponents show good agreement with the 2D KPZ exponents. This analysis of the coherence decay thus unambiguously demonstrate the existence of a regime of parameters where polariton condensates fall into the 2D KPZ universality class. Additional data analysis provided in Fig.~\ref{fig:sup_map_AD_collapse_KPZ} and Fig.~\ref{fig:sup_map_AD_KPZ2} of the SI for different values of $p$ and $\delta$ further validate this conclusion.

Numerical simulations of the generalized Gross-Pitaevskii equation in presence of a reservoir with experimentally realistic parameters provide results in overall agreement with our experimental observations. Most importantly, our simulations show that all the assumptions needed to obtain the KPZ equation for the phase dynamics are satisfied, namely density fluctuations are negligible in the KPZ window so that the coherence decay is due to phase fluctuations (see Fig.~\ref{fig:supmat_theory_EW_KPZ}a-b of the SI).

\begin{figure}[t!]
    \centering
    \includegraphics[scale=0.45]{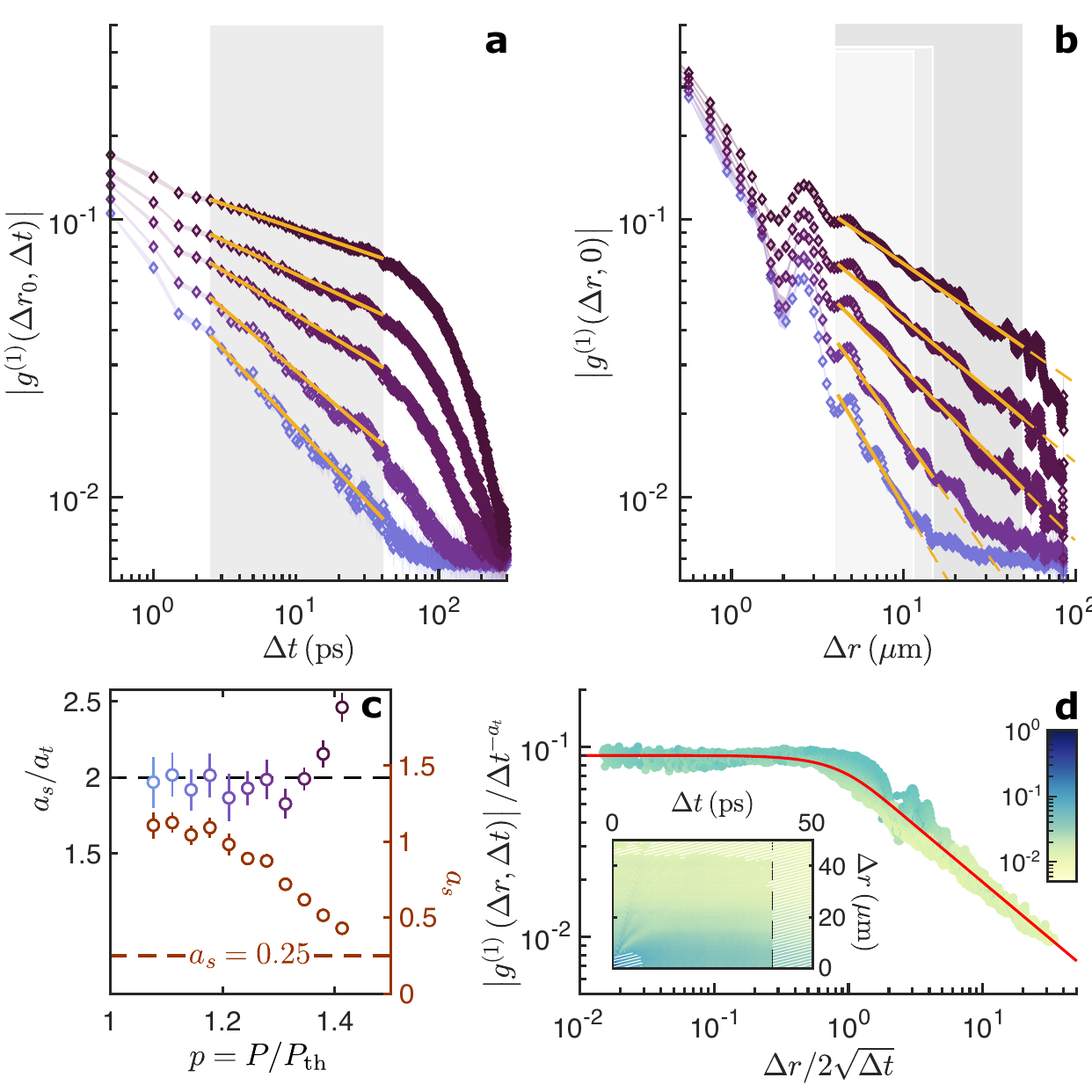}
    \caption{
    \textbf{Observation of EW universal scaling in 2D polariton condensates.}  
    \textbf{a} (\textbf{b}): Measured values of $|g^{(1)}(\Delta r_0, \Delta t) |$ ($|g^{(1)}(\Delta r, 0) |$) as a function of $\Delta t$  ($\Delta r$) in logarithmic scale for $p = 1.14,\, 1.28,\, 1.35,\, 1.38, \, 1.41$. In \textbf{a} $\Delta r_0 = 1 \, \mathrm{\mu m}$. 
   Straight orange lines: power-law fits over the EW windows indicated with dark areas. 
    Errorbars are calculated by performing a repeatability analysis on multiple interferograms.
    \textbf{c.} Measured values of the spatial exponent $a_{s}$ and of $a_{s}/a_{t}$ ($a_{t}$ is the temporal exponent) as a function of $p$.
    Errorbars are determined from the $\pm \sigma$ uncertainties on the fitted exponents in \textbf{a} and \textbf{b}.
     \textbf{d}: Plot of $|g^{(1)}(\Delta r, \Delta t)|)/\Delta t^{-a_t}$ measured for $p=1.35$ as a function of $\Delta r/2\sqrt{\Delta t}$ in logarithmic scale, for values of $|g^{(1)}(\Delta r, \Delta t)|$ lying in the non-hatched region of the inset. 
     The red line shows the universal EW scaling function~\cite{Nattermann_PRA1992_solutionEW}. Inset of  \textbf{d}: $|g^{(1)}(\Delta r, \Delta t)|$ in the $(\Delta r, \Delta t)$ plane.  All data points shown in this figure are measured for $\hbar \delta = - 12.6~\mathrm{meV}$.}
\label{fig:scalingEW}
\end{figure}

We now move to a different set of parameters by adjusting the cavity-exciton detuning to $\hbar \delta = \SI{-12.6}{\milli \eV}$ (more photonic polaritons with $10 \% $ exciton fraction). We measure $g^{(1)}(\Delta \mathbf{r}, \Delta t)$ for power values ranging from $p=1.14$ to $p=1.41$, and follow a similar approach as in Fig.~\ref{fig:scalingKPZ}. We show, in Fig.~\ref{fig:scalingEW}a-b, the temporal (spatial) coherence decay for $\Delta t=0$ ($\Delta r \simeq 1~\mathrm{\mu m}$) plotted in logarithmic scale. We find spatial and temporal windows (gray areas) where the coherence decay curves appear as straight lines, thus evidencing power law decays. Plotting the same curves using different axes coordinates, we confirm in Extended Data Fig.~\ref{fig:AF:KPZ_EW_all_scales} that the coherence decays neither follow stretched exponential nor exponential scalings.

Fitting the coherence decays by power laws, we obtain the spatial and temporal power-law exponents $a_s$ and $a_t$ plotted in Fig.~\ref{fig:scalingEW}c. Strikingly, the ratio $a_s/a_t$ approaches $2$ for a broad range of powers, which is a characteristic feature of the EW regime that emerges when the KPZ effective nonlinearity  is weak. Moreover, we find $a_s > 0.25$, which contrasts with the  BKT ordered phase, where $0.25$ represents the upper limit to the spatial exponent. Finally, in Fig.~\ref{fig:scalingEW}d, we show the collapse of the data points measured for $p=1.35$ on a properly chosen set of coordinates, thus confirming our interpretation in terms of EW scaling. For this value of the detuning $\delta$, the impact of the dispersive modes on $g^{(1)}(\Delta r, \Delta t)$ has been found to be much weaker than in the KPZ case, as they do not induce major modifications of the spatial and temporal coherence decays, and mostly show as weakly contrasted oscillations in the spatial decay at short distances $\Delta r \leq 4~{\rm \mu m}$ or in the data collapse. As a consequence, all data analysis in the EW case has been realized without performing any filtering of the dispersive modes. Finally, emergence of EW regime is also found in numerical simulations with experimentally realistic microscopic parameters (see Fig.~\ref{fig:supmat_theory_EW_KPZ}c-d in SI).

\begin{figure}[t!]
    \centering
    \includegraphics[scale=0.5]{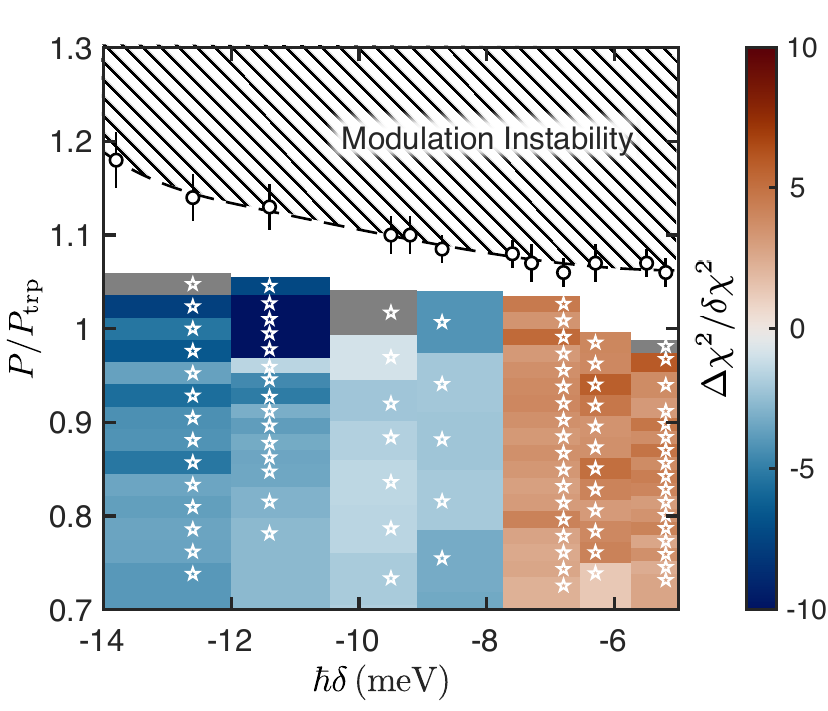}
    \caption{
    \textbf{Phase diagram of the 2D driven-dissipative polariton condensate} 
    White stars indicate experimental measurements as function of $\delta$ and normalized pump power $P/P_{trp}$. The colors around these points are determined from a comparative likelihood analysis between the KPZ and EW  fits of the first-order coherence temporal decay based on the value of $\Delta\chi ^2 / \delta\chi^2$  (see Methods, and SI Fig.~\ref{fig:sup_mat_phase_diagram}). Red (blue) colored areas indicate that the KPZ (EW) stretched exponential (power-law) decay is the best fit to experimental data, while for gray colors, both scaling laws cannot be discriminated. The black dashed line is a guide to the eye linking experimental points (open circles) marking the power limits above which modulational instability occurs.
    }
    \label{fig:Phase_Diagram}
\end{figure}

So far, our results have highlighted the change of universality class when going from a more excitonic to a more photonic polariton condensate. In the following, we explore the evolution of the coherence decay when progressively changing $\delta$. We perform a statistical analysis (see methods and SI) to determine which of stretched exponential or power law behavior best describes the measured coherence decay versus detuning and power. The result of this analysis is plotted in Fig.~\ref{fig:Phase_Diagram}, where blue colors (red colors) encode coherence decays that are statistically proven to behave as power law (stretched exponential). Gray colors indicate undecidable cases according to our statistical analysis. The black dashed line delimits the power range above which modulational instabilities appears. Overall, we evidence a clear crossover between the EW and KPZ universality classes as we tune $\delta$ from the photonic to the excitonic regime.

The observed crossover between universality classes is possible due to the finite size of the experimental system. Indeed for 2D systems of infinite size, KPZ is always the stable fixed point. Nevertheless when the system is finite, reaching this fixed point requires large enough spatiotemporal scales for fluctuations to develop. Therefore if the KPZ effective nonlinearity is weak enough, the system may remain in the transient EW regime.

Finally, we analyze the noise-generated vortices found in the simulations both for KPZ or EW regimes (see SI for a detailed discussion). As for 2D Bose gases at thermal equilibrium \cite{Foster_Davis_PRA2010_Vortex_pairing_2dBose_gas} we find that vortices disappear upon coarse-graining (see Fig.~\ref{fig:sup_map_Gaussian_smoothening} in SI). Moreover, our simulations show that vortices are correlated to antivortices at short distances and form vortex-antivortex pairs, while same-sign vortices exhibit short-range anti-correlations, reflecting their repulsive interactions (see Fig.~\ref{fig:sup_map_defect_correlation} in SI). These features evidence that EW and KPZ regimes found in our experiments are the non-equilibrium analog of the ordered BKT phase in closed systems at equilibrium.
 
To conclude, we experimentally provide an answer to the active debate concerning the possible existence of a KPZ ordered phase in a continuous 2D driven dissipative system. Indeed, exploring the spatiotemporal coherence of polariton condensates generated in a planar cavity, we evidence the tuning across KPZ and EW universality classes when changing some microscopic parameters. This work opens the way to further exploration of the rich phase diagram~\cite{Helluin2025} of polariton condensates, with for instance transitions to solitonic, spiral vortices~\cite{Chate_PhysicaA1996_phase_diagram_2DCGLE} or disordered phases. Further investigations are needed to explore the impact of this physics on other systems, such as photon and plasmonic condensates or vertical cavity lasers. We emphasize that semiconductor microcavities are so far quite unique, as experimental platforms enabling the study of 2D KPZ physics via classical interfaces are currently scarce~\cite{almeida2014, Almeida2017}. It could for example offer the ability to dynamically tune 2D KPZ physics using a resonant probe \cite{Stazzu2026}, or to explore generalized KPZ phases in presence of engineered noise \cite{Squizzato2019}.

\section*{Methods}

\subsection*{Modeling polariton BECs using two-coupled equations}

The dynamics of the condensate wavefuntion and the exciton reservoir density is modeled by two coupled equations:
\begin{align}
    i \hbar \partial_{t} \psi &= \left[ \epsilon(\boldsymbol{\hat{k}}) + g|\psi|^{2} + 2 g_{\mathrm{R}} n_{\mathrm{R}} +\frac{i \hbar}{2} \left( R n_{\mathrm{R}} - \gamma(\boldsymbol{\hat{k}}) \right)\right] 
    \psi(\boldsymbol{r}, t) + \xi(\mathbf{r}, t) \, ,    \label{eq:GPE}\\
    \partial_{t} n_{\mathrm{R}} &= P(\mathbf{r})-(\gamma_{\mathrm{R}}+R |\psi|^{2}) n_{\mathrm{R}}(\mathbf{r}, t) \, ,
    \label{eq:reservoir}
\end{align}
\noindent where $n_{\mathrm{R}}$ is the excitonic reservoir density driven by a nonresonant pump at rate $P(\boldsymbol{r})$, $\boldsymbol{\hat{k}} = -i \boldsymbol{\nabla}$ is the momentum operator, $\epsilon(\boldsymbol{\hat{k}})$ is the polariton energy dispersion, $\gamma(\boldsymbol{\hat{k}})$ is the momentum-dependent decay rate, and $g$ is the polariton-polariton interaction strength (set to zero in all simulations). Excitons either scatter into the polariton condensate through stimulated processes at a rate $R|\psi|^2$, or decay via alternative channels at a rate $\gamma_{\mathrm{R}}$. The term $2 g_{\mathrm{R}} n_{\mathrm{R}}$ describes repulsive interactions between polaritons and reservoir excitons. It dominates the polariton blueshift $g|\psi|^{2}$ close to threshold, and induces dephasing through inhomogeneous spectral broadening. Finally, $\xi(\boldsymbol{r}, t)$ is a complex Gaussian white noise, acting as a stochastic drive on the polariton field. In the following, we consider a quadratic momentum dependence of the polariton$\,$linewidth, $\gamma(k) = \gamma_{0}+\gamma_{2}k^{2}$, which, in the context of photon condensation, originates from the energy dependence of$\,$gain$\,$and$\,$loss, as dictated by the Kennard–Stepanov relation~\cite{Wouters2020}.

\subsection*{Sample structure}

The cavity sample was grown by molecular beam epitaxy. It consists of a $5\lambda/2$ ${\rm Al}_{0.95}{\rm Ga}_{0.05}{\rm As}$ optical cavity containing five pairs of $20~{\rm nm}$ GaAs quantum wells (QW) separated by $5~{\rm nm}$ ${\rm Al}_{0.95}{\rm Ga}_{0.05}{\rm As}$ barriers positioned at the five anti-nodes of the cavity mode electromagnetic field. The cavity is surrounded by two ${\rm Al}_{0.95}{\rm Ga}_{0.05}{\rm As}/{\rm Al}_{0.2}{\rm Ga}_{0.8}{\rm As}$ Bragg mirrors with 38 (26) pairs in the bottom (top) mirrors. A spatial wedge in the cavity and mirror thicknesses is intentionally introduced to enable spectral tuning of the cavity mode with respect to the QW exciton resonance energy around $1.52~\rm{ eV}$ by probing different sample regions (see SI for a detailed description of the sample parameters).

\subsection*{Linearized model and determination of the condensation threshold}
 
To precisely determine the condensation threshold and the emergence of extended coherence, we plot on Fig.~\ref{fig:AF:comp_lin_mod}a the measured values of $|g^{(1)} (\Delta r = r_0, \Delta t)|$ as a function of $\Delta t$, for the pump powers corresponding to the two spectra shown with insets in Fig.~\ref{fig:condensation}b corresponding to $p \simeq 0.89$ and $p \simeq 1.22$. The same graph also displays the predictions of a linearized model (solid red lines), derived from Eqs.~\ref{eq:GPE} and~\ref{eq:reservoir} assuming homogeneous pumping, adiabatic reservoir dynamics and low excitation power. In this simplified framework, the reservoir density is clamped to $P/\gamma_{R}$ and the dynamics can be solved exactly in the Fourier domain upon neglecting the polariton blueshift ($g \!=\! 0$). A detailed derivation of the first-order coherence function in this context is given in SI. We show that this model is fully determined by three experimentally accessible parameters: $m$, $\gamma_{0}$ and $\gamma_{2}$ (See SI for their estimation). At low pump power, excellent agreement is found between the experimental data (blue diamonds) and the linearized model up to $\sim 30\,\mathrm{ps}$, where the measured coherence reaches its minimum, that is, when interference fringes become indistinguishable from background noise.  This analysis is particularly reliable because of the absence of any fitting parameter. Conversely, at large pump powers, the linearized model only reproduces the experimental data (purple diamonds) at very short time delays, while largely underestimating the coherence elsewhere. In order to quantify the deviation of the experimental data from the linear model, we compute the symmetric mean absolute percentage error (or sMAPE) between data and the linear model predictions over the first $30 \, \mathrm{ps}$, as a function of the pumping power (see Fig.~\ref{fig:AF:comp_lin_mod}b. The sMAPE values continuously increase above a well-defined threshold power $P_{\rm th}$ (red star), which we identify as the condensation threshold whose value is subsequently reported in Fig.~\ref{fig:condensation}b. Indeed this marks the onset of the gain saturation nonlinearity, and thus the departure from the low-power linear photoluminescence regime toward genuine nonlinear condensate dynamics. The linear model also enables to determine the transparency threshold $P_{\rm trp}$ (blue star in Fig.~\ref{fig:AF:comp_lin_mod}b), which is achieved when stimulated scattering into the condensate compensates for the loss rate $\gamma_{0}$.

\subsection*{Experimental set-up}

We use a cw laser tuned to $735~\mathrm{nm}$ to excite the microcavity sample positioned in a closed-cycle cryostat and maintained at a temperature of $4 \, \mathrm{K}$. A top-hat beam shaper converts the Gaussian excitation beam into a nearly collimated flat-top beam, which is subsequently demagnified to uniformly illuminate the sample with a $60~\mathrm{\mu m}$-diameter area. The microcavity emission is collected in reflection. Above condensation threshold, we select a condensate polarization mode using a set of waveplates and polarizers. To image the polariton dispersion we project the Fourier plane of the collection lens on the entrance slit of a spectrometer coupled to a CCD camera. For interferometry experiments, we send the real space image of the emission into a Michelson interferometer and project the interferometer output onto a CCD camera (see sketch of the interferometer in Fig.~\ref{fig:AF:setup}). In one arm of the interferometer we insert a retroreflector so that we interfere the real space image with its $\pi$ rotated copy. The retroreflector is mounted on a motorized translation stage so that we can control the path-length difference between both arms. More technical details on the experimental setup are given in SI.

\subsection*{Detailed description of the $g^{(1)}$ heat map and origin of the dispersive branches}

The heat map in Fig.~\ref{fig:condensation}d reveals dispersive modes in the space-time correlations, visible as alternating regions of high and low coherence. The presence of these modes is a direct consequence of the fact that the first-order coherence function $g^{(1)}$ and the power spectrum are related through a Fourier transform. The residual population of the parabolic polariton dispersion near $k=0$ (seen for instance in the upper inset of Fig.~\ref{fig:condensation}b) produces dispersive branches whose shape in the $(\Delta r, \Delta t)$ plane only depends on the polariton mass $m$: $\Delta r = \sqrt{2 n \pi \hbar \Delta t/m}$, where $n$ is an integer. The corresponding branches obtained for $n = 2$, $n = 4$ and $n = 6$ are plotted as black dotted lines in Fig.~\ref{fig:condensation}d, and agree well with the three observed low-coherence branches. Note that at low powers where a significant portion of the polaritons are uncondensed, the dispersive modes produce strong modulations in the correlations. In order to reveal the intrinsic spatiotemporal decay of the condensate coherence, we thus realize a careful filtering of the dispersive modes (see SI). At higher pump powers where the condensed portion of polaritons dominates over uncondensed polaritons, the dispersive branches only weakly modulate the spatio-temporal coherence, dominated by the condensate contribution.

\subsection*{Statistical analysis of the coherence decay scalings}

The phase diagram shown in Fig.~\ref{fig:Phase_Diagram} is obtained by comparing the chi-square values of power-law and stretched-exponential fits to the temporal decay of $|g^{(1)}(0,\Delta t)|$ over the same fitting window. As detailed in the SI, we evaluate, for each pump power and detuning, the normalized chi-square difference $\Delta\chi^{2}/\delta\chi^{2}$, where $\Delta\chi^{2}$ denotes the difference between the chi-square values of the two fits (EW and KPZ) and $\delta\chi^{2}$ its associated $2 \sigma$ uncertainty. 

\section*{acknowledgments}
We thank  Iacopo Carusotto, Markus Holzmann, Paolo Comaron, Marzena Szyma{\'n}ska and Guillaume Malpuech for stimulating discussions. This work was supported by the European Research Council (ERC) under the Horizon Europe programme (project
ANAPOLIS, grant agreement no. 101054448) and the Union’s Horizon 2020 research and innovation programme through the Starting Grant ARQADIA (grant agreement no. 949730), by the RENATECH network and the General
Council of Essonne, by the Paris Ile-de-France Région via DIM QUANTIP, by the Plan France 2030 through the project QUTISYM ANR23-PETQ-0002, by the ANR projects Ngauge (ANR-24-CE92-0011) and HAWQ (ANR-25-CE47-7323). A.~M. and L.~C. acknowledge support from HQI (www.hqi.fr) initiative and is supported by France 2030 under the French National Research Agency grant number ANR-22-PNCQ-0002. L.~C. acknowledges support from ANR (grant ANR-18-
CE92-0019) and from Institut Universitaire de France. M.~W. acknowledges support from FWO-Vlaanderen (G0B2923N).

\section*{Author contributions}

Q.~F. built the experimental, performed the experiments and analyzed the data. D.~P.~D. contributed to the initial phase of the experiment building. M.~E. contributed to the measurement and data analysis. F.~H. realized the theoretical calculations and numerical simulations. S.~R. designed the sample. A.~L., M.~M. grew the sample by molecular beam epitaxy. Q.~F., F.~H., M.~E., M.~W., A.~M., L.~C., S.~R. and J.~B. participated to the scientific discussions about all aspects of the work. Q.~F., F.~H., L.~C., A.~M, S.~R. and J.B. wrote the manuscript. S.~R. and J.~B.  supervised the work.

\newpage

\setcounter{figure}{0}
\makeatletter
\renewcommand{\fnum@figure}{Extended Data Fig.~\thefigure}
\makeatother
\clearpage

\begin{figure}[h!]
    \centering
    \includegraphics[scale=0.5]{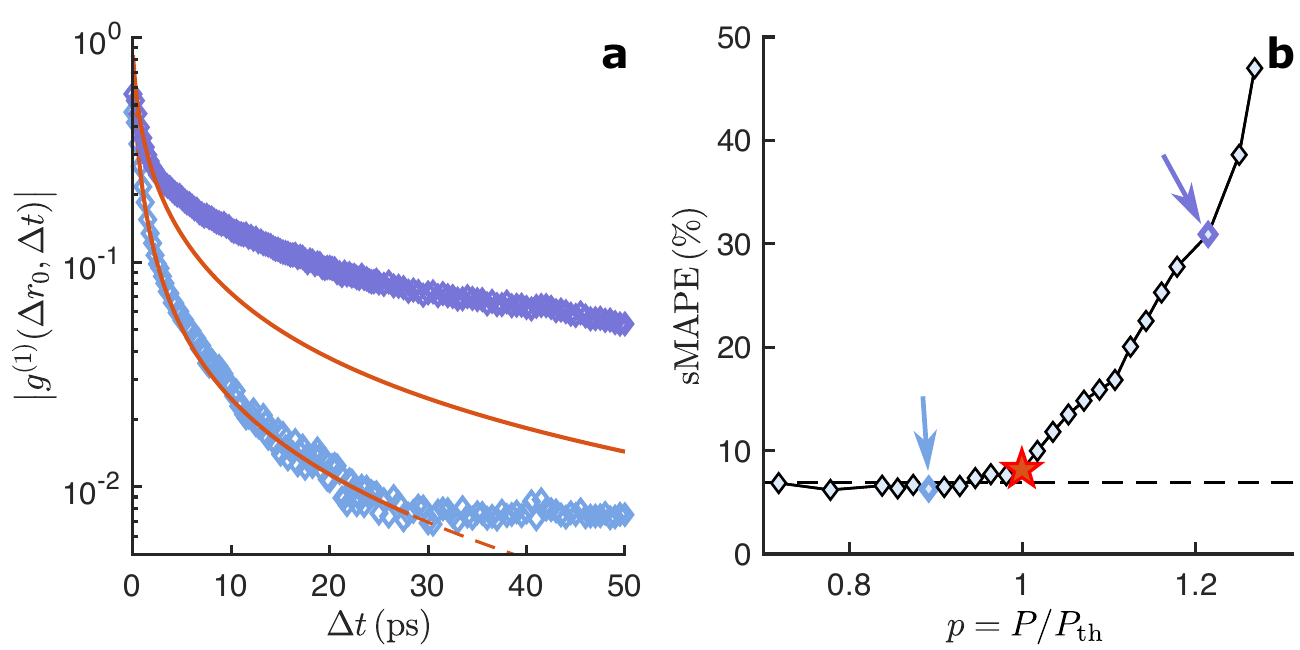}
    \caption{
    \textbf{Characterization of the condensation threshold from the onset of temporal coherence.}
    \textbf{a.}~Measured $|g^1(\Delta r_{0},\Delta t)|$ as a function of $\Delta t$ for $\Delta r_0 = 1~{\rm \mu m}$, $p = 0.89$ (light blue) and $p= 1.22$ (purple). The red solid lines represent predictions from the linear model described in Methods and SI.
    \textbf{b}~Deviation of the measured coherence decay from the linear model prediction. The corresponding sMAPE (diamond symbols) is plotted as a function of $p$. The colored diamonds and arrows highlight the two $p$ values chosen in \textbf{a}. The red star identifies the condensation threshold.
    }
    \label{fig:AF:comp_lin_mod}
\end{figure}

\newpage

\begin{figure}[h!]
    \centering
    \includegraphics[scale=0.8]{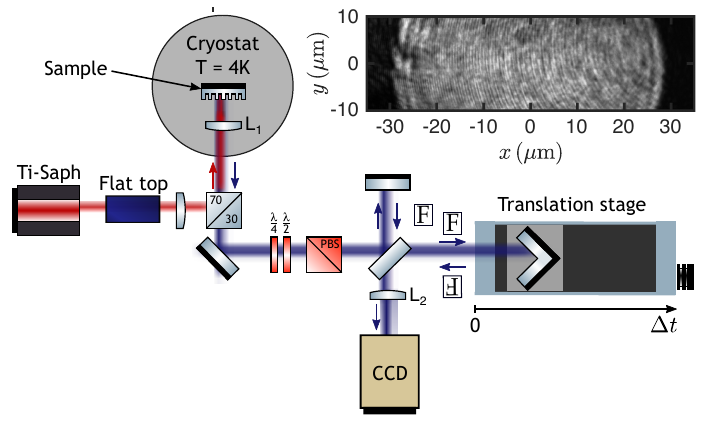}
    \caption{
    \textbf{Sketch of the interferometry setup.} The sample located in the cryostat is illuminated using a laser beam with a flat top profile (see measured intensity in top right inset). The reflected laser light is filtered out using a quarter and half waveplates, eanbling to collect and send the real space emission through a Michelson interferometer. In one of the two interferometer arms we insert a retroreflector mounted on a translation stage. The retroreflector rotates the image by $\pi$ and the translation stage changes the relative length between the two arms. The images from both arms are superimposed in the interferometer output and the resulting image is projected onto a CCD camera.}
    \label{fig:AF:setup}
\end{figure}

\newpage

\begin{figure}[h!]
    \centering
    \includegraphics[scale=0.5]{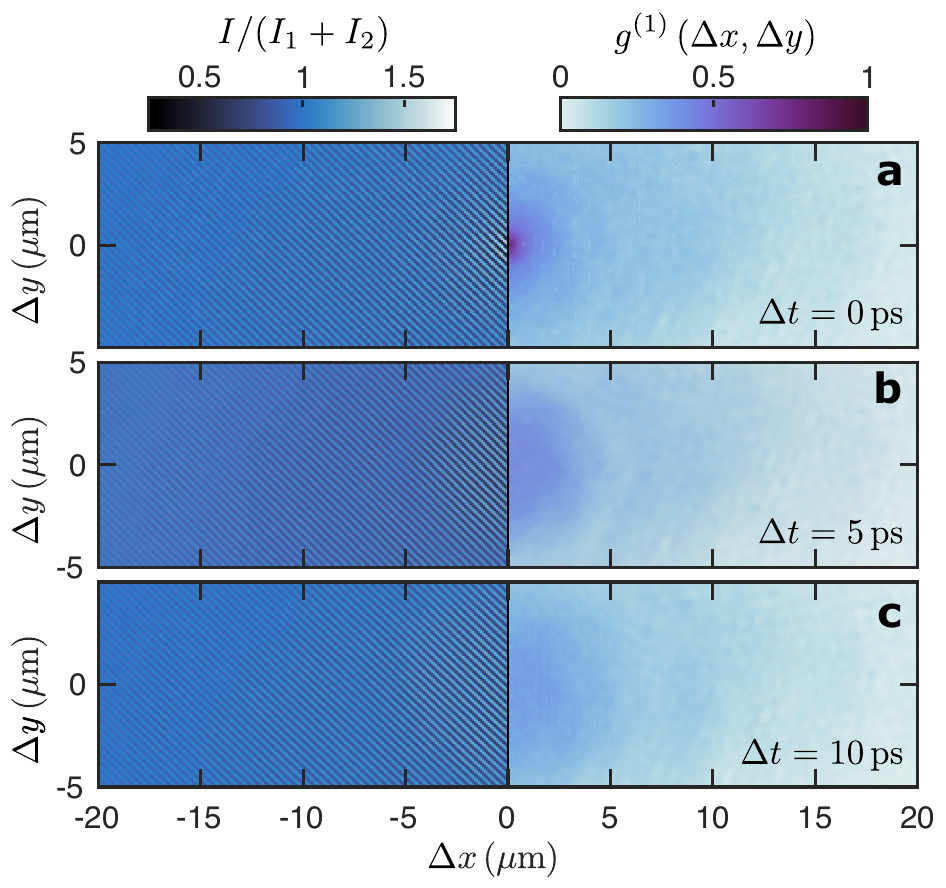}
    \caption{
    \textbf{Spatial heatmaps of the interferogram intensity pattern (left column) and of the corresponding measured $|g^1(\Delta x, \Delta y)|$ (right column).} The data is shown for
    \textbf{a.}~$\Delta t = 0~{\rm ps}$, \textbf{b.}~$\Delta t = 5~{\rm ps}$, and \textbf{c.}~$\Delta t = 10~{\rm ps}$.
    In the three columns, fringes covering the whole figure indicate the condensate extended spatial coherence. In the right column, the observed sharp peak of fast decaying coherence close to $\Delta x = \Delta y = 0$ is due to the contribution of uncondensed polaritons at short temporal and spatial scales.}
    \label{fig:AF:interfero}
\end{figure}

\newpage

\begin{figure}[h!]
    \centering
    \includegraphics[scale=0.5]{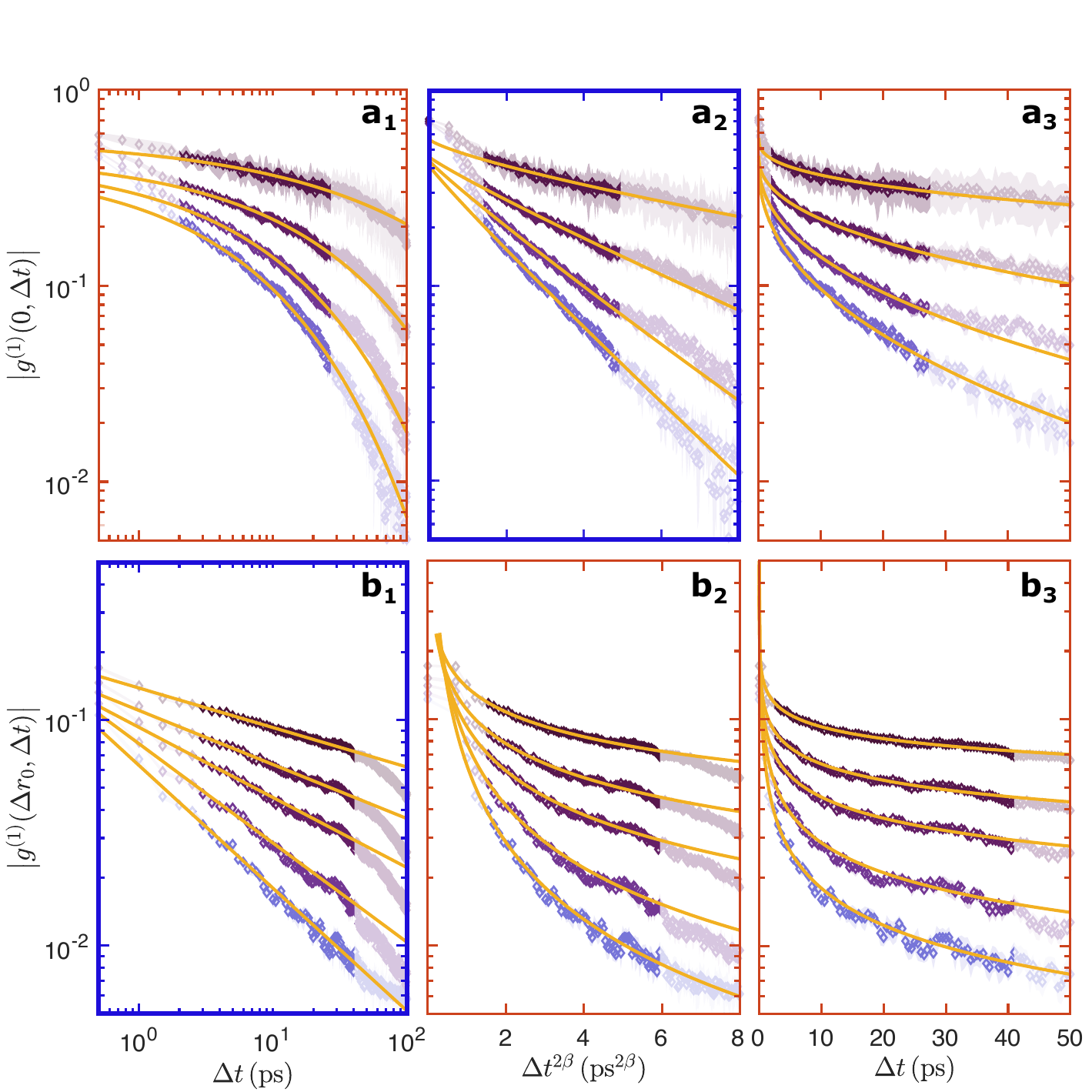}
    \caption{
    \textbf{Measured decays of $| g^1(\Delta r, \Delta t) |$ displayed using different sets of axes coordinates.}
    The first line (panels {$\bf a_1- a_3$}) shows plots of the data presented in Fig.~\ref{fig:scalingKPZ}.a (symbols) with a mask applied to highlight the data points found to be inside the KPZ window.
    The yellow solid lines show stretched exponential fits to the data points within the KPZ window. 
    The second line (panels \textbf{$\bf b_1-b_3$}) shows plots of the data presented in Fig.~\ref{fig:scalingEW}.a (symbols) with a mask applied to highlight the data points found to be inside the EW window.
    The yellow solid lines show power law fits to the data points within the EW window.
    In the first column, the graphs are displayed in logarithmical scale, where power-law decays are expected to show as straight lines.
    In the second column, the graphs are displayed in semi-logarithmic scale with rescaled horizontal axis $\Delta t ^{2 \beta}$ ($\beta=0.24$), where stretched exponential KPZ decays are expected to show as straight lines.
    In the third column, the graphs are displayed in semilogarithmic scales, where exponential decays are expected to show as straight lines.
    The plots clearly highlight that the correct scalings are KPZ stretched exponential in $\bf a_2$ (thick blue contour), and power-law (EW regime) in $\bf b_1$ (thick blue contour). }
    \label{fig:AF:KPZ_EW_all_scales}
\end{figure}

\clearpage

%

\clearpage
\onecolumngrid

\begin{center}
{\huge Supplementary Information}
\end{center}

\begin{center}
\vspace{1cm}
{\centering\Large\bfseries \papertitle \par}

\vspace{0.5cm}

Q. Fontaine, F. Helluin, M. Escalera, D. Pinto Dias,
A. Lemaître, M. Morassi, M. Wouters,
A. Minguzzi, L. Canet, S. Ravets, J. Bloch
\end{center}

\section*{Contents}

\bigskip

\supplementarytocentry{sec:Overview}{S1 Overview}

\supplementarytocentry{sec:supmat_Theory}{S2 Theoretical Approach}
\quad\supplementarytocentry{sec:supmat_model}{1 Model}
\quad\supplementarytocentry{sec:Mapping}{2 Effective phase dynamics}
\quad\supplementarytocentry{sec:supmat_properties_EPBEC_coherence}{3 Properties of the condensate coherence}
\quad\supplementarytocentry{sec:supmat_Linearized_model}{4 Linearized model}

\supplementarytocentry{sec:supmat_Experiment}{S3 Experiments - Additional Information and Data}
\quad\supplementarytocentry{sec:supmat_Experiment_Sample}{1 Sample structure and initial characterization}
\quad\supplementarytocentry{sec:supmat_Experiment_Setup}{2 Experimental setup and data processing}
\qquad\supplementarytocentry{sec:supmat_Experiment_Setup_Desciption}{1 Experimental setup}
\qquad\supplementarytocentry{sec:supmat_Experiment_Setup_Experimental_Sequence}{2 Experimental sequence}
\qquad\supplementarytocentry{sec:supmat_Experiment_Setup_Interferometry}{3 Michelson interferometry for coherence measurements}
\quad\supplementarytocentry{sec:supmat_micro_params_condensate}{3 Experimental determination of the microscopic parameters}
\qquad\supplementarytocentry{sec:supmat_micro_params_condensate_mass}{1 Determination of the polariton mass $m$}
\qquad\supplementarytocentry{sec:supmat_micro_params_condensate_linewidth}{2 Determination of the zero-momentum linewidth $\gamma_0$}
\qquad\supplementarytocentry{sec:supmat_micro_params_condensate_gamma2}{3 Determination of the gain-curvature coefficient $\gamma_2$}
\qquad\supplementarytocentry{sec:supmat_micro_params_reservoir}{4 Determination of the reservoir coefficient $\gamma_R/R$}
\quad\supplementarytocentry{sec:supmat_condensation_vs_transparency}{4 Determination of the condensation threshold $P_{\rm th}$}
\quad\supplementarytocentry{sec:supmat_filtering_dispersive_branches}{5 Filtering of the dispersive branches}
\quad\supplementarytocentry{sec:supmat_normalization}{6 Normalization of the coherence}
\quad\supplementarytocentry{sec:supmat_phase_diagram}{7 Computation of the phase diagram in Fig.~\ref{fig:Phase_Diagram}}
\quad\supplementarytocentry{sec:additional_data}{8 Additional data demonstrating 2D KPZ scaling in polariton condensates}

\supplementarytocentry{sec:supmat_Numerics_method}{S4 Numerical Simulations - Method and Parameters}
\quad\supplementarytocentry{sec:supmat_Numerics_method_parameters}{1 Numerical scheme and parameters}
\quad\supplementarytocentry{sec:supmat_Numerics_method_vortex_tracking}{2 Vortex tracking}

\supplementarytocentry{sec:supmat_Numerics_discussion}{S5 Numerical simulations - Discussion}
\quad\supplementarytocentry{sec:supmat_condensate_stability}{1 Stability of the condensate}
\quad\supplementarytocentry{sec:supmat_accessing_EW_KPZ}{2 Accessing KPZ and EW regimes}
\quad\supplementarytocentry{sec:supmat_robustness_EW_KPZ}{3 Robustness of the KPZ and EW regimes}
\quad\supplementarytocentry{sec:supmat_Nature_vortices}{4 Nature of the vortices}
\qquad\supplementarytocentry{sec:supmat_vortex_clustering}{1 Vortex clustering}
\qquad\supplementarytocentry{sec:supmat_coarse_graining}{2 Spatial coarse-graining}
\qquad\supplementarytocentry{sec:supmat_lifetime}{3 Defect lifetime}
\qquad\supplementarytocentry{sec:supmat_pair_correlation}{4 Defect pair correlation and discussion of the KPZ mapping}

\clearpage

\setcounter{secnumdepth}{3} 
\setcounter{section}{0}
\renewcommand{\thesection}{S\arabic{section}}
\renewcommand{\thesubsection}{\arabic{subsection}}
\renewcommand{\thesubsubsection}{\arabic{subsubsection}}

\setcounter{figure}{0}
\makeatletter
\renewcommand{\fnum@figure}{Fig.~\thefigure}
\makeatother
\clearpage

\setcounter{figure}{0}
\renewcommand{\thefigure}{S\arabic{figure}}


\section{\label{sec:Overview}Overview}

In this Supplementary Information, we provide additional information on the experiments and on the numerical simulations. The theoretical model is introduced in Sec.~\ref{sec:supmat_Theory}, where we recall the derivation of the mapping to the KPZ equation for the effective phase dynamics, highlighting its implications for the condensate coherence. We also analyze the regime far below the transparency threshold, where the linearization of the model yields analytical expressions for both the first-order correlation function and the momentum distribution. In Sec.~\ref{sec:supmat_Experiment}, we describe experimental procedures used for data measurement and data analysis. Importantly, we show additional datasets demonstrating KPZ scaling of 2D polariton condensates for different values of the excitation power and of the detuning $\delta$. Sec.~\ref{sec:supmat_Numerics_method} outlines the numerical methods and simulation parameters. Lastly, Sec.~\ref{sec:supmat_Numerics_discussion} presents a detailed discussion of the numerical results.

\medskip

The subsequent sections provide a comprehensive account of the experimental and numerical work. The main results reported in this Supplementary Information are highlighted below:

\begin{itemize}

\item We provide a detailed description of the experimental setup, and of the procedures used to experimentally determine the microscopic parameters entering the theoretical model in both the EW and KPZ regimes (see Sec.~\ref{sec:supmat_micro_params_condensate}).

\item We properly define and determine experimentally the threshold $P_{\rm th}$ and transparency threshold $P_{\rm trp}$ (see Sec.~\ref{sec:supmat_condensation_vs_transparency}).

\item We describe in details the procedure employed to filter the dispersive branches from the space-time heat maps of the first-order coherence (see Sec.~\ref{sec:supmat_filtering_dispersive_branches}).

\item We provide a detailed description of the procedure used to construct the experimental phase diagram showing the crossover from EW to KPZ universality classes (see Sec.~\ref{sec:supmat_phase_diagram}).

\item We consolidate the validity of our experimental findings by providing additional data demonstrating KPZ universal scaling in polariton condensates (see Sec.~\ref{sec:additional_data}).

\item We numerically reproduce the scaling behavior and data collapse of the $g^{(1)}$ function observed in experiments, and evidence that it directly reflects the KPZ and EW scaling of phase correlations, {\it i.e.} we verify that the assumptions underlying the KPZ mapping are fulfilled (See Sec.~\ref{sec:supmat_accessing_EW_KPZ}).

\item We point out the presence of vortices in numerical simulations of both the EW and KPZ regimes, and characterize their statistical properties, which allow us to explain why they do not suppress quasi-order (See Sec.~\ref{sec:supmat_Nature_vortices}).

\end{itemize}


\newpage

\section{\label{sec:supmat_Theory}Theoretical Approach}


\subsection{\label{sec:supmat_model}Model}

We consider a polariton condensate formed in a two-dimensional optical microcavity. Using a low energy effective description, we consider  the dynamics of the polariton condensate wavefunction $\psi(\boldsymbol{r},t)$ at position $\boldsymbol{r}$ and time $t$ in the lower polariton branch (LPB), neglecting the effect of other branches \cite{carusotto2013}. The dynamics of the condensate wavefunction is described by the two following coupled equations:

\begin{eqnarray}
i\hbar \, \partial_t\psi(\boldsymbol{r},t) & = & \Bigg[ \epsilon(\hat{\boldsymbol{k}}) + \dfrac{i\hbar}{2}\left(Rn_R -\gamma(\hat{\boldsymbol{k}})\right) + g|\psi|^2 + 2 g_Rn_R  \Bigg]\psi(\boldsymbol{r},t) + \xi(\boldsymbol{r},t) \, , \label{eq:supmat_gGPE} \\
\partial_tn_R(\boldsymbol{r},t) & = & P(\boldsymbol{r}) - \left( \gamma_R n_R(\boldsymbol{r},t)  - R|\psi^2|\right) n_R^2 \label{eq:supmat_xreservoir} \,.
\end{eqnarray}

\noindent
Equation \eqref{eq:supmat_gGPE} is a generalized Gross-Pitaevskii equation (gGPE), where $\epsilon(\hat{\boldsymbol{k}})$ represents the dispersion relation of the LPB, with $\hat{\boldsymbol{k}} = -i\boldsymbol{\nabla}$ the momentum operator. In the following, the polariton dispersion relation is approximated close to the bottom of the LPB by the parabola $\epsilon(\hat{\boldsymbol{k}}) = \hbar^2 \boldsymbol{\hat{k}}^2 / (2m)$ with $m$ the polariton mass. The $k$-dependent effective loss rate is defined by $\gamma(\hat{\boldsymbol{k}}) = \gamma_0 + \gamma_2 \hat{\boldsymbol{k}}^2$ \cite{Fontaine2022, deligiannis2022}, where $\gamma_0$ is the bare polariton linewidth while $\gamma_2$ emerges effectively from the LPB curvature \cite{Porras2002, Wouters2009}. Reservoir excitons interact with condensed polaritons with an amplitude $g_R$, while polariton-polariton interactions occur with amplitude $g$. The term $\xi$ is a complex stochastic noise with Gaussian statistics, characterized by a zero average $\langle \xi(\boldsymbol{r},t) \rangle = 0$ and delta correlations in space and time $\langle \xi(\boldsymbol{r},t)\xi^*(\boldsymbol{r'},t') \rangle = 2\hbar^2\sigma\delta(\boldsymbol{r}-\boldsymbol{r'})\delta(t-t')$, $\langle \xi(\boldsymbol{r},t)\xi(\boldsymbol{r'},t') \rangle=\langle \xi^*(\boldsymbol{r},t)\xi^*(\boldsymbol{r'},t') \rangle=0$ \cite{Gardiner2003, Wouters2009, Sieberer2016}. The strength of this stochastic contribution is given by $\sigma = Rn_R/2$. Equation~\eqref{eq:supmat_xreservoir} is a rate equation for the density of the incoherent exciton reservoir $n_R(\boldsymbol{r},t)$. It is driven by an external pump at rate $P(\boldsymbol{r})$. Reservoir excitons either relax into the polariton condensate by scattering with condensed polaritons with a rate $R$, capturing phonon-assisted and Coulomb-mediated processes between the exciton reservoir and the LPB \cite{carusotto2013}, or decay via other channels with a rate $\gamma_R$.


\clearpage

\subsection{\label{sec:Mapping}Effective phase dynamics}

The condensate wavefunction can be written in the density-phase representation as $\psi(\boldsymbol{r},t) = \sqrt{n(\boldsymbol{r},t)} e^{-i\mu t/\hbar + i\theta(\boldsymbol{r},t)}$, with $\mu$ the polariton blueshift. Assuming that the condensate density $n$ weakly fluctuates around its average value, it can be shown that phase fluctuations of the condensate wavefunction are associated with low-energy gapless excitations. These correspond to a dissipative Goldstone mode, which dominates the condensate dynamics at large scales \cite{Altman_PRX2015_2D_superfluidity_anisotropy}. In particular, the large scale properties of the condensate are controlled by the effective phase dynamics, which is mapped onto a Kardar--Parisi--Zhang equation
\begin{equation}
    \partial_t \theta = \nu \nabla^2 \theta + \dfrac{\lambda}{2}(\boldsymbol{\nabla}\theta)^2 + \sqrt{D}\eta \,, \label{eq:supmap_KPZ_mapping}
\end{equation}
\noindent
where  the effective parameters are given in terms of the microscopic parameters from Eqs.~\ref{eq:supmat_gGPE}-\ref{eq:supmat_xreservoir} as:
\begin{equation}
 \nu = \frac{\gamma_2}{2} + \frac{\hbar g_{e}}{2\hbar mg_{i}} \,, \;\;\; \lambda = -\frac{\hbar}{m} + \gamma_2 \frac{g_{e}}{\hbar g_{i}} \,, \;\;\; D = \frac{\sigma}{2n_0}\left[ 1+ \left(\frac{g_{e}}{\hbar g_{i}}\right)^2 \right] \,, \label{eq:supmat_effective_KPZ_params}
\end{equation}
\noindent
with the effective interaction strength $g_{e}=g - 4 g_Rg_{i} /R $ and the effective loss rate $g_{i}=\gamma_0^2/2P$. The noise $\eta$ is real and Gaussian,  with $\langle \eta \rangle = 0$ and $\langle \eta(\boldsymbol{r},t)\eta(\boldsymbol{r'},t') \rangle = 2\delta(\boldsymbol{r}-\boldsymbol{r'})\delta(t-t')$. The effective nonlinearity of the KPZ equation \eqref{eq:supmap_KPZ_mapping} is defined by $g_{\rm KPZ}= \lambda^2D / \nu^3$. A detailed derivation and discussion of this mapping can be found in Ref.~\cite{Fontaine2022}.

\medskip

If the phase dynamics follows a KPZ equation, its two-point correlation $C_{\theta\theta}(\Delta \boldsymbol{r}, \Delta t) = \langle \left[ \theta(\Delta \boldsymbol{r}, \Delta t) - \theta(\boldsymbol{0},0)\right]^2 \rangle$ is given at large distances and long times by the scaling form
\begin{equation}
C_{\theta\theta}(\Delta \boldsymbol{r}, \Delta t) = C_0 \Delta t^{2\beta}F_{\rm KPZ}(y_0|\Delta\boldsymbol{r}|/\Delta t^{1/z}) \sim \left\{\begin{array}{l l}
\Delta t^{2\beta}, \qquad&  {\rm for}\;|\Delta \boldsymbol{r}|=0\\
|\Delta \boldsymbol{r}|^{2\chi}, & {\rm for}\; \Delta t=0
\end{array}\right. \,,
\label{eq:supmat_KPZscaling}
\end{equation}

\noindent
where $\beta$, $\chi$ are universal exponents and $F_{\rm KPZ}$ is the KPZ universal scaling function. The dynamical exponent $z$ is defined according to $z = \chi/\beta$, and $C_0$, $y_0$ are non-universal constants. In dimension $d=2$, the values of the universal exponents have been estimated numerically and are approximately given by $\beta\approx0.24$, $\chi\approx0.39$, $z\approx1.62$ \cite{Pagnani2015}. The  KPZ universal scaling function $F_{\rm KPZ}$ in  $d=2$ has been determined  within the functional renormalization group approach \cite{Canet2012_Scaling_fct_amplitude_ratios_1d_2d_3d}.

\medskip

When the non-linearity $\lambda$ vanishes (yielding $g_{\rm KPZ}=0$), the KPZ equation becomes a stochastic diffusion equation called the Edwards--Wilkinson (EW) equation \cite{edwards1982surface}. The two-point correlation function of a EW interface can be written under a scaling form similar to Eq.~\eqref{eq:supmat_KPZscaling}, with the exact exponents $\chi  =\frac{2-d}{2}$, $\beta  =\frac{2-d}{4}$, $z=2$ and a universal scaling function $F_{\rm EW}$ determined in \cite{Nattermann_PRA1992_solutionEW}. In dimension $d=2$, both the spatial and the temporal exponents vanish. As a consequence, the two-point  correlations do not grow as  power laws, but logarithmically:
\begin{equation}
C_{\theta\theta}(\Delta \boldsymbol{r}, \Delta t) = \dfrac{D}{\pi\nu}F_{\rm EW}\left(\frac{|\Delta \boldsymbol{r}|}{2\sqrt{\nu \Delta t} };\;2\Lambda\sqrt{\nu \Delta t}\right)\sim\left\{\begin{array}{l l}
D \ln(\Delta t\Lambda^2) / (2\pi\nu), \qquad&  {\rm for}\;|\Delta \boldsymbol{r}|=0\\
D \ln(|\Delta \boldsymbol{r}|\Lambda) / (\pi\nu), &  {\rm for}\;\Delta t=0
\end{array}\right. \, ,
\label{eq:supmat_EWscaling}
\end{equation}
\noindent
with the function:
\begin{equation}
    F_{\rm EW}\left(y;\; 2\Lambda\sqrt{\nu \Delta t}\right) = \int_0^{2\Lambda\sqrt{\nu \Delta t}} dp \dfrac{1 - J_0(py)e^{-p^2/4}}{p},
    \label{eq:supmat_EW_function}
\end{equation}
\noindent
where $y=|\Delta \boldsymbol{r}|/2\sqrt{\nu\Delta t}$ is the dimensionless scaling parameter and where $\Lambda$ is a UV cutoff \cite{Nattermann_PRA1992_solutionEW}. 

\medskip

In the weakly non-linear regime, \textit{i.e.} when the effective nonlinearity is very small but non-zero ($g_{\rm KPZ}\ll1$), the two-point  correlation function can grow logarithmically according to Eq.~\eqref{eq:supmat_EWscaling} for extended spatiotemporal scales. The KPZ power-law growth Eq.~\eqref{eq:supmat_KPZscaling} is eventually reached after a crossover time $t^*$. As the effective nonlinearity $g_{\rm KPZ}$ increases, $t^*$ decreases, and the KPZ regime is accessed at shorter scales. In a finite observation window, typically induced by finite sizes, a crossover from Eq.~\eqref{eq:supmat_EWscaling} to Eq.~\eqref{eq:supmat_KPZscaling} is thus expected, as $g_{\rm KPZ}$ is increased. This crossover has been reported in numerical simulations performed in multiple contexts: discrete models in the KPZ universality class \cite{Tang_PRA1992_hybercube_stacking_EW-KPZ_crossover}, the two-dimensional noisy Kuramoto-Sivashinksy equation \cite{Cuerno_KPZasympotics_2DKS}, and more recently in the dynamics of two-dimensional exciton-polariton condensates \cite{Helluin2025}.


\clearpage

\subsection{\label{sec:supmat_properties_EPBEC_coherence}Properties of the condensate coherence}

The first-order correlation function of the polariton condensate is defined by:
\begin{equation}
    g^{(1)}(\Delta \boldsymbol{r}, \Delta t) = \dfrac{\big\langle \psi(\Delta \boldsymbol{r} + \boldsymbol{r_0}, \Delta t + t_0) \psi^*(\boldsymbol{r_0}, t_0)\big\rangle}{\sqrt{ \big\langle|\psi(\Delta \boldsymbol{r} + \boldsymbol{r_0}, \Delta t + t_0)|^2\big\rangle \big\langle |\psi(\boldsymbol{r_0}, + t_0)|^2 \big\rangle}} \, , \label{eq:supmat_g1_def}
\end{equation}
\noindent
where $t_0$ is a reference time chosen in the non-equilibrium steady state of the condensate dynamics, and $\boldsymbol{r_0}$ a reference point chosen at the center of the condensate profile. The average $\langle \cdot \rangle$ denotes both an average over independent realizations of the dynamics and an angular average over all directions at fixed distance $\Delta r = |\Delta \boldsymbol{r}|$. Following the hypotheses of the mapping Eq.~\eqref{eq:supmap_KPZ_mapping}, the first-order correlation function of the condensate can be approximated by:
\begin{equation}
    g^{(1)}(\Delta \boldsymbol{r}, \Delta t) \approx g^{(1)}_{n}(\Delta \boldsymbol{r}, \Delta t) \big\langle e^{i\Delta\theta(\Delta \boldsymbol{r}, \Delta t)} \big\rangle
    \label{eq:supmat_g1_decoupling_density_phase}
\end{equation}
\noindent
where density-phase correlations have been neglected and we have defined density correlations as
\begin{equation}
    g^{(1)}_{n}(\Delta \boldsymbol{r}, \Delta t) = \dfrac{\big\langle \sqrt{n(\Delta \boldsymbol{r} + \boldsymbol{r_0}, \Delta t + t_0)n(\boldsymbol{r_0}, t_0)}\big\rangle}{\sqrt{\big\langle n(\Delta \boldsymbol{r} + \boldsymbol{r_0}, \Delta t + t_0) \big\rangle\big\langle n(\boldsymbol{r_0}, t_0) \big\rangle}}.
\end{equation}
\noindent
When density correlations weakly depend on temporal and spatial delays, the decay of the first-order correlation function is dominated by the growth of phase correlations, which can be further approximated by $\langle e^{i\Delta\theta(\Delta \boldsymbol{r}, \Delta t)}\rangle \approx e^{-\frac{1}{2}C_{\theta\theta}(\Delta\boldsymbol{r}, \Delta t)}$. In the KPZ regime, where $C_{\theta\theta}$ follows the power-laws of Eq.~\eqref{eq:supmat_KPZscaling}, the condensate coherence is thus expected to decay following a stretched exponential behavior
\begin{equation}
g^{(1)}(\Delta r, \Delta t)\propto  \left\{\begin{array}{l l}
e^{-A_t\Delta t^{2\beta}}, \qquad&  {\rm for}\;\Delta r=0\\
e^{- A_r\Delta r^{2\chi}}, &  {\rm for}\;\Delta t=0
\end{array}\right. \,,
\label{eq:supmat_g1_KPZscaling}
\end{equation}
\noindent
where $A_t$ and $A_r$ are non-universal constant amplitudes. In the EW regime, where $C_{\theta\theta}$ follows the logarithmic growth of Eq.~\eqref{eq:supmat_EWscaling}, the condensate coherence is instead expected to display algebraic decays in both space and time,

\begin{equation}
g^{(1)}(\Delta \boldsymbol{r}, \Delta t)\propto \left\{\begin{array}{l l}
\Delta t^{-a_t}, \qquad&  {\rm for}\;|\Delta \boldsymbol{r}|=0\\
|\Delta \boldsymbol{r}|^{-a_s}, &  {\rm for}\;\Delta t=0
\end{array}\right. \,,
\label{eq:supmat_g1_EWscaling}
\end{equation}

\noindent
with $a_s = D/2\pi\nu$ and $a_s/a_t=2$ \cite{Szymanska2006, Helluin2025}. An effective temperature $T_{\rm eff}$ can be introduced as $k_BT_{\rm eff} = 2D/\nu \times \hbar^2n_0/(2m)$, such that $a_s$ can be expressed as  $a_s = 1/n_0\lambda_{T_{\rm eff}}^2$, in analogy with 2D quasi-BECs in their quasi-ordered phase. Note however that $T_{\rm eff}$ should not be interpreted as the cryostat temperature, which may instead be taken into account upon including a phonon bath to the microscopic model, rescaling the noise strength $\sigma$ \cite{Shelykh_PRL2013_stochastic_GPE_dynamical_thermalization, Frerot_Richard_PRX2023_Bogo_excitations_thermal_phonons}. The non-equilibrium nature of the weakly non-linear polariton condensate is perfectly explicit, since $T_{\rm eff}$ depends on drive and dissipation, while the phase dynamics is not constrained by any fluctuation-dissipation type of relation. In addition, contrary to the equilibrium case, $a_s$ is not bounded to $a_s\leq1/4$ \cite{chiocchetta2013}, while the exponent ratio $a_s/a_t=2$ reflects the dissipative nature of the Goldstone mode \cite{Szymanska2006, Comaron2021}.

\medskip

We finally point out that, in a finite-size system of linear size $L$, the long-time behavior of the first-order correlation function departs from both the KPZ and EW predictions given by Eqs.\eqref{eq:supmat_g1_KPZscaling} and \eqref{eq:supmat_g1_EWscaling} respectively. Indeed for $\Delta t$ exceeding a crossover time $\propto L^z$, the coherence decays exponentially~\cite{Keeling2010}, namely:
\begin{equation}
g^{(1)}(\Delta \boldsymbol{r}=\boldsymbol{0}, \Delta t) \, {\sim} \, e^{-\Delta t/\tau_r} \, ,
\label{eq:supmat_ST}
\end{equation}
\noindent
where the relaxation time $\tau_r$ depends on both the system size and its microscopic parameters. This behavior is general, and was confirmed in numerical simulations of both one and two dimensional polariton condensates \cite{Amelio2024, Helluin2025}. It is analogous to the Schawlow--Townes regime predicted for laser systems \cite{Schawlow1958, Fabre2010}.


\clearpage

\subsection{\label{sec:supmat_Linearized_model}Linearized model}

In this section, we derive analytical results describing the emission of uncondensed polaritons, valid at low pumping power, \textit{i.e.}, far below threshold. Indeed, under this condition, the microscopic model Eqs.~\eqref{eq:supmat_gGPE}-\eqref{eq:supmat_xreservoir} can be linearized assuming that:

\begin{enumerate}
    \item The pump is spatially homogeneous,
    \item The polariton density is low enough so that interactions can be neglected $g|\psi|^2\approx0$,
    \item The reservoir dynamics is adiabatically eliminated using $R|\psi|^2\ll\gamma_R$, yielding $n_R \simeq \frac{P}{\gamma_R+R|\psi|^2} \simeq \frac{P}{\gamma_R}$.
\end{enumerate}

\noindent 
The polariton field in Fourier space then reads

\begin{equation}
    \psi(\boldsymbol{k}, \omega) = \frac{ \xi(\boldsymbol{k}, \omega)/\hbar }{\left[ \omega - \overline{\omega}(k) \right] + \frac{i\hbar}{2} \left[ \gamma_{0} (1-p) +\gamma_{2} k^{2} \right]},
    \label{eq:supmat_2D_psi_k}
\end{equation}

\noindent 
where $\langle \xi(\boldsymbol{k}, \omega) \xi^{*}(\boldsymbol{k'}, \omega') \rangle = 2\hbar^2\sigma\times (2 \pi)^{2+1} \delta^{(2)}(\boldsymbol{k}-\boldsymbol{k'}) \delta(\omega-\omega')$ and $\overline{\omega}(k) = 2 g_{R} n_{R} + \hbar^{2} k^{2}/ 2 m$ is the dispersion in the vicinity of the ground state. We also introduced the reduced pump $p_{\rm trp} = P/P_{\mathrm{trp}}$ where $P_{\mathrm{trp}} = \gamma_{0} \gamma_{R}/R$ is the mean-field transparency threshold. The results derived below are valid for $p_{\rm trp}\ll1$.

\medbreak

The spectral density $\rho(k, \omega)$ is defined according to

\begin{equation}
    \rho(k, \omega) \, = \,  \int \frac{\boldsymbol{dk'} d\omega'}{(2\pi)^{2+1}} \big\langle \psi(\boldsymbol{k}, \omega) \psi^{*}((\boldsymbol{k'}, \omega')) \big\rangle = \frac{2 \sigma}{\left[ \omega - \overline{\omega}(k)\right]^{2}+\left[ \hbar\Gamma(k)/2 \right]^{2}}, \label{eq:supmat_rho_k_omega}
\end{equation}

\noindent 
where  $\Gamma(k) = \gamma_{0} (1-p_{\rm trp}) +\gamma_{2} k^{2}$ is the momentum-dependent linewidth. The momentum distribution of the polariton photo-emission is obtained by integrating Eq.~\eqref{eq:supmat_rho_k_omega} over frequencies:

\begin{equation}
    n(k) \, = \, \int \dfrac{d\omega}{2\pi} \, \rho(k, \omega) = \frac{2\sigma/\hbar}{\gamma_{0}(1-p_{\rm trp})+\gamma_{2} k^{2}}.
\end{equation}

\noindent 
Using $\sigma = R n_R/2$, the momentum distribution can be expressed as a function of the ratio $\gamma_{2}/\gamma_{0}$:

\begin{equation}
    \frac{n(0)}{n(k)} = 1+ \frac{\gamma_{2}}{\gamma_{0}} \frac{1}{1-p_{\rm trp}} k^{2}.
    \label{eq:n_k}
\end{equation}

\noindent
Equation \eqref{eq:n_k} is used in Sec.~\ref{sec:supmat_micro_params_condensate} to estimate the value of $\gamma_2$ in experiments.

\medskip

Finally, we derive the first-order coherence function $g^{(1)}( \Delta\boldsymbol{r}, \Delta t)$ induced by the emission of uncondensed polaritons. It is obtained, up to a normalization constant, calculating the Fourier transform of the spectral density as follows:

\begin{align}
    g^{(1)}(\Delta\boldsymbol{ r}, \Delta t) = & \;  \int \frac{\boldsymbol{dk} d\omega}{(2 \pi)^{2+1}} \rho(k, \omega) \, e^{i(\omega \Delta t - \boldsymbol{k} \cdot \Delta\boldsymbol{ r})} \nonumber \\
    =& \; \frac{2 \sigma}{(2 \pi)^{1+1}} \int_{0}^{\infty} \frac{k J_{0}(k \Delta r) dk}{\Gamma(k) / 2 \pi} \left\{ \int_{-\infty}^{+ \infty} \frac{\Gamma(k)/2\pi}{\left[ \omega - \overline{\omega}(k) \right]^{2} +\left[\hbar \Gamma(k)/2 \right]^{2}} e^{i \omega \Delta t} d\omega \right\} \nonumber \\ 
    = & \; \frac{\sigma}{2 \pi \gamma_{2}} e^{-\frac{\gamma_{0}}{2}(1-p_{\rm trp}) \Delta t} \, e^{i \mu \Delta t/\hbar} \int_{0}^{\infty} \frac{k J_{0}(k \Delta r) dk}{\beta+k^{2}} \, e^{- \eta\Delta t k^{2}},
    \label{eq:supmat_2D_g1_dr_dt}
\end{align}

\noindent
where we defined the blueshift  $\mu = 2 g_{R} n_{R}$, $\beta = \gamma_{0} (1-p_{\rm trp})  / \gamma_{2}$ and $\eta = \gamma_{2}/2 - i \hbar / (2m)$. Notice that, in the low-power regime where emission of uncondensed polaritons dominates, the first-order coherence function Eq.~\eqref{eq:supmat_2D_g1_dr_dt} is fully determined by the polariton mass ($m$), the linewidth at $\boldsymbol{k = 0}$ ($\gamma_{0}$) and the effective quadratic curvature of the linewidth ($\gamma_{2}$) in momentum space. The reservoir parameters also enter Eq.~\eqref{eq:supmat_2D_g1_dr_dt} through the blueshift $\mu$, but only as a global phase factor that drops out when considering the modulus. The exponential damping prefactor reflects the Schawlow--Townes contribution to the coherence decay. It accounts for the increase of the coherence time with the ground-state population as $P$ approaches $P_{\rm trp}$. At zero spatial separation ($\Delta\boldsymbol{ r}=\boldsymbol{0}$), the first-order coherence function follows the simple analytical form:

\begin{equation}
    \left| g^{(1)}(\Delta \boldsymbol{r} = \boldsymbol{0}, \Delta t) \right| = \frac{\sigma}{2 \pi \gamma_{2}} |E_{1}\left[\eta\Delta t \beta \right] | \, ,
    \label{eq:supmat_2D_g1_dr_0_dt}
\end{equation}

\noindent
where $E_{1}$ is the exponential integral function. Equation \eqref{eq:supmat_2D_g1_dr_0_dt} is used in Sec.~\ref{sec:supmat_condensation_vs_transparency} to determine the condensation threshold in experiments.

\newpage


\section{\label{sec:supmat_Experiment}Experiments - Additional Information and Data}


\subsection{\label{sec:supmat_Experiment_Sample}Sample structure and initial characterization}

The sample is a semiconductor heterostructure grown by molecular beam epitaxy. It consists of a $5 \lambda/2$ Ga$_{0.05}$Al${_0.95}$As microcavity surrounded by two Al$_{0.20}$Ga$_{0.80}$As/Al$_{0.05}$Ga$_{0.95}$As distributed Bragg reflectors with 26 (38) pairs in the top (bottom) mirror. Five stacks of pairs of $20~\mathrm{nm}$ thick GaAs quantum wells (QWs) are distributed at the anti-nodes of the electromagnetic field in the microcavity spacer. The sample is single-side polished for reflectivity measurement only. A small gradient in the cavity spacer provides spatial tuning of the zero-momentum cavity photon energy.

\medskip

\begin{figure}[h!]
    \centering
    \includegraphics[scale=0.75]{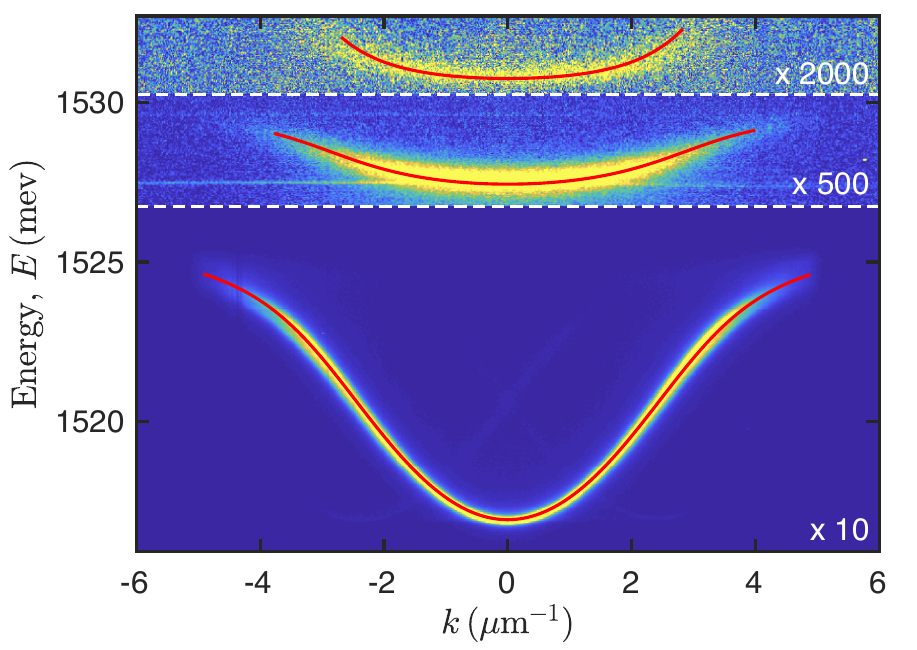}
    \caption{
    \textbf{Low-power photoluminescence spectrum} showing the three polariton branches together with the fitted polariton dispersion (solid red lines). Different saturation levels are applied across the photoluminescence image to enhance the visibility of the two upper polariton branches.
    }
    \label{fig:sup_map_3bandsmodel}
\end{figure}

In contrast to earlier samples~\cite{baboux2018, Fontaine2022}, which contained $7~\mathrm{nm}$ QWs, the increased QW thickness in the present sample lowers the light-hole (lh) exciton energy splitting with respect to heavy-hole (hh) excitons, leading to the observation of three polariton branches in photoluminescence measurements (see Fig.~\ref{fig:sup_map_3bandsmodel}). The low-power photo-emission dispersion is well captured by a three-band model. This model is parametrized by \textit{(i)} the lh and hh exciton detunings from the zero-momentum cavity mode, $\delta_{\mathrm{lh}}$ and $\delta_{\mathrm{hh}}$, \textit{(ii)} the exciton-photon couplings $\Omega_{\mathrm{lh}}$ and $\Omega_{\mathrm{hh}}$ and \textit{(iii)} the photon effective mass $m_{\mathrm{ph}}$. A fit of the measured dispersion (shown as red solid lines in Fig.~\ref{fig:sup_map_3bandsmodel}) yields $\hbar \Omega_{\mathrm{lh}} = 5.2 \pm 0.1~\mathrm{meV}$, $\hbar \Omega_{\mathrm{hh}} = 8.6 \pm 0.2~\mathrm{meV}$ and $m_{\mathrm{ph}} = 4.2 \times 10^{-5} \, m_{e}$ (with $m_{e}$ the electron mass). Throughout the main text and the SI, the exciton-photon detuning $\delta$ is identified with the heavy-hole exciton detuning $\delta = \delta_{\mathrm{hh}}$.

\clearpage

\subsection{\label{sec:supmat_Experiment_Setup}Experimental setup and data processing}

In this section, we describe the experimental techniques employed in our experiments to measure the coherence of polariton condensates.

\subsubsection{\label{sec:supmat_Experiment_Setup_Desciption}Experimental setup}

\begin{figure}[h!]
    \centering
    \includegraphics[scale=0.89]{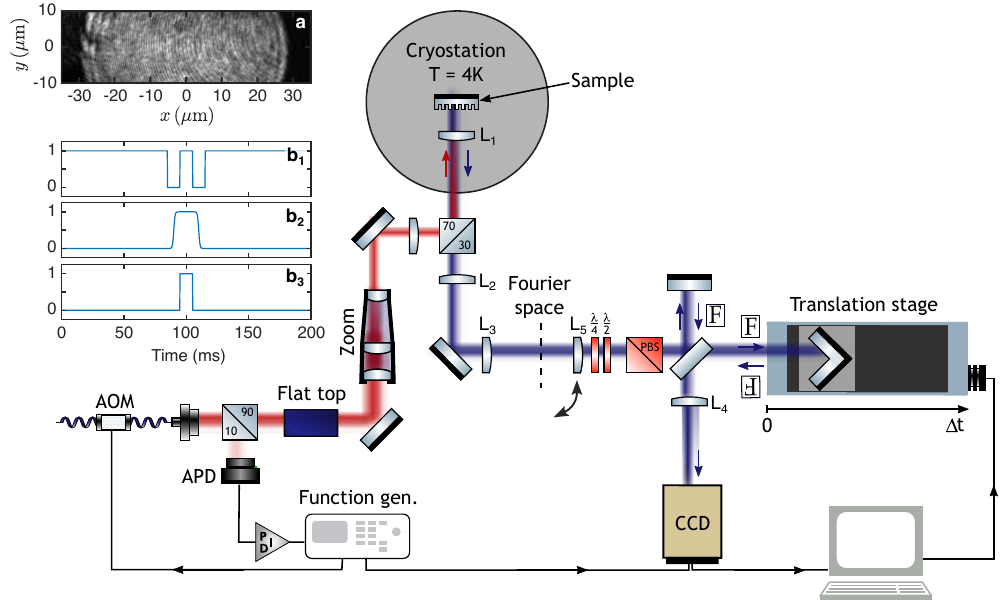}
    \caption{
    \textbf{Sketch of the experimental setup.}}
    \label{fig:sup_map_schematics_setup}
\end{figure}

The sample described previously is kept at $4~\mathrm{K}$ in a closed-cycle cryostat. A sketch of the optical setup is shown in Fig.~\ref{fig:sup_map_schematics_setup}. Polaritons are generated by off-resonant, quasi-cw excitation of the microcavity at $735~\mathrm{nm}$. A top-hat beam shaper converts the excitation into a nearly collimated flat-top beam, which is subsequently de-magnified to uniformly illuminate a $60~\mathrm{\mu m}$-diameter area on the sample (see measured profile of the excitation beam in inset~\textbf{a}). The microcavity emission at $780~\mathrm{nm}$ is collected in reflection through a 70/30 (R/T) beam splitter, while the reflected part of the pump is rejected using a high-extinction edge filter. A combination of a half-waveplate ($\lambda / 2$), a quarter-waveplate ($\lambda / 4$) and a polarizing beam splitter (PBS) enables precise selection of a given polarization. The sample plane is relayed by a 4-f imaging system formed by lenses $L_{1}$ and $L_{2}$, and subsequently imaged onto the CCD through lenses $L_{3}$ and $L_{4}$. The same configuration enables imaging of the back focal plane (Fourier space) of the excitation lens, which is relayed to infinity through $L_{5}$. 

\subsubsection{\label{sec:supmat_Experiment_Setup_Experimental_Sequence}Experimental sequence}

An acousto-optic modulator (AOM) driven by a waveform generator (WG) shapes the temporal intensity profile of the excitation beam into the sequence shown in the inset \textbf{b1}: alternating $170~\mathrm{ms}$ and $10~\mathrm{ms}$ pulses separated by $10~\mathrm{ms}$ dark intervals, with a $5~\mathrm{Hz}$ repetition rate. A PID controller stabilizes the intensity at the desired level for both long and short pulses. A homemade mechanical shutter with a rise time of a few milliseconds (inset \textbf{b2}), positioned at the focus of a magnifying telescope, isolates the short pulses (inset \textbf{b3}) before they propagate through the excitation setup. For image acquisition on the CCD, the camera is triggered by the excitation pulse front edge with an exposure time of $200~\mathrm{ms}$.

\subsubsection{\label{sec:supmat_Experiment_Setup_Interferometry}Michelson interferometry for coherence measurements}
\label{sec:interferometry}

The spatiotemporal coherence of the microcavity emission is measured by inserting a Michelson interferometer between $L_{3}$ and $L_{4}$, where the light is collimated. A retroreflector mounted on a high-precision motorized translation stage controls the path-length difference between arms. To access the temporal coherence decay, the retroreflector is scanned over $45~\mathrm{mm}$ ($\sim 300~\mathrm{ps}$) at low power ($P \lesssim 0.8 P_{\mathrm{trp}}$) and $75~\mathrm{mm}$ ($\sim 500~\mathrm{ps}$) at high power ($P > 0.8 P_{\mathrm{trp}}$), where the coherence increases. The step size is $75~\mathrm{\mu m}$ ($\sim 0.5~\mathrm{ps}$) or even $37.5~\mathrm{\mu m}$ ($\sim 0.25~\mathrm{ps}$) when high resolution is needed. At each delay, 10 images are recorded. Spatial coherence is extracted from the fringe contrast (see below), radially averaged, and its uncertainty estimated from the $2 \sigma$ shot-to-shot variations.

\medskip

In our experiments, the field amplitude emitted by the  at time $t_{0}$, $E_{1}(\boldsymbol{r/2}, t_{0}) \propto \psi(\boldsymbol{r/2}, t_{0})$ interferes with that emitted at the symmetric point $-\boldsymbol{r/2}$ at time $t_{0} + \Delta t$ , $E_{2}(-\boldsymbol{r/2}, t_{0} + \Delta t) \propto \psi(-\boldsymbol{r/2}, t_{0} + \Delta t)$, where $\Delta t$ is the delay between the two interferometer arms.
The resulting intensity measured by the CCD reads:
\begin{equation}
    I(\Delta\boldsymbol{ r}, \Delta t) = \frac{1}{4} \left[ I_{1}(\boldsymbol{r}) +I_{2}(-\boldsymbol{r}) + 2\sqrt{I_{1}(\boldsymbol{r}) I_{2}(-\boldsymbol{r})} \left| g^{(1)} (\Delta\boldsymbol{ r}, \Delta t) \right| \mathrm{cos} (\Delta \Phi)\right],
\end{equation}
\noindent where $I_{i}(\boldsymbol{r}) = \langle \left| E_{i}(\boldsymbol{r}, t)\right|^{2} \rangle_{\tau}$ is the intensity distribution coming from arm $i$, averaged over the camera exposure time $\tau$, and $\Delta \Phi \!=\! \boldsymbol{\delta q} \cdot \boldsymbol{r}$ is the relative geometric phase arising from the nonzero transverse wave-vector mismatch $\boldsymbol{\delta q}$ between the condensate field and its spatially inverted replica. 
The first-order coherence function $g^{(1)}(\Delta\boldsymbol{ r}, \Delta t)$, defined in Eq.~\eqref{eq:supmat_g1_def}, is retrieved by measuring the fringe visibility $V \!=\! (I_{+} - I_{-})/(I_{+} + I_{-})$, where $I_{+}$ and $I_{-}$ are the upper and lower envelopes of $I$, respectively. 
At each time delay $\Delta t$,  $I_{\pm}$ are extracted from the interferogram using Fourier analysis. The fringe visibility is then related to the first-order coherence through:
\begin{equation}
    V(\Delta r, \Delta t) = \frac{2 \sqrt{I_{1}(\boldsymbol{r}) I_{2}(-\boldsymbol{r})}}{I_{1}(\boldsymbol{r}) + I_{2}(-\boldsymbol{r})} \left| g^{(1)}(\Delta r, \Delta t) \right| = K(\Delta r) \left| g^{(1)}(\Delta r, \Delta t) \right|,
\end{equation}
\noindent where $K(\Delta r)$ is a normalization factor taking into account possible imbalance between $I_{1}$ and $I_{2}$. 
In our case,$\;$this factor always remains close to 1.


\newpage

\subsection{\label{sec:supmat_micro_params_condensate}Experimental determination of the microscopic parameters}

In this section, we experimentally determine the microscopic parameters characterizing the polariton condensate, and entering Eq.~\eqref{eq:supmat_gGPE}. Namely we provide estimated values for the polariton mass $m$, the bare polariton linewidth $\gamma_0$, and the gain-curvature coefficient $\gamma_2$.

\subsubsection{\label{sec:supmat_micro_params_condensate_mass}Determination of the polariton mass $m$}

The polariton effective mass is extracted by fitting the bottom of the lower polariton branch with a parabola. It slightly varies with the exciton-photon detuning: $\boldsymbol{m \!=\! 5.9 \!\times\! 10^{-5} \, m_e}$ for $\hbar \delta \!=\! -5.2 \, \mathrm{meV}$ (Fig.~2 of the main text) and $\boldsymbol{m \!=\! 4.7 \!\times\! 10^{-5} \, m_e}$ for $\hbar \delta = -12.6 \, \mathrm{meV}$ (Fig.~3 of the main text), where $m_e$ is the electron mass.

\subsubsection{\label{sec:supmat_micro_params_condensate_linewidth}Determination of the zero-momentum linewidth $\gamma_{0}$}

\begin{figure}[t!]
    \centering
    \includegraphics[width=\linewidth]{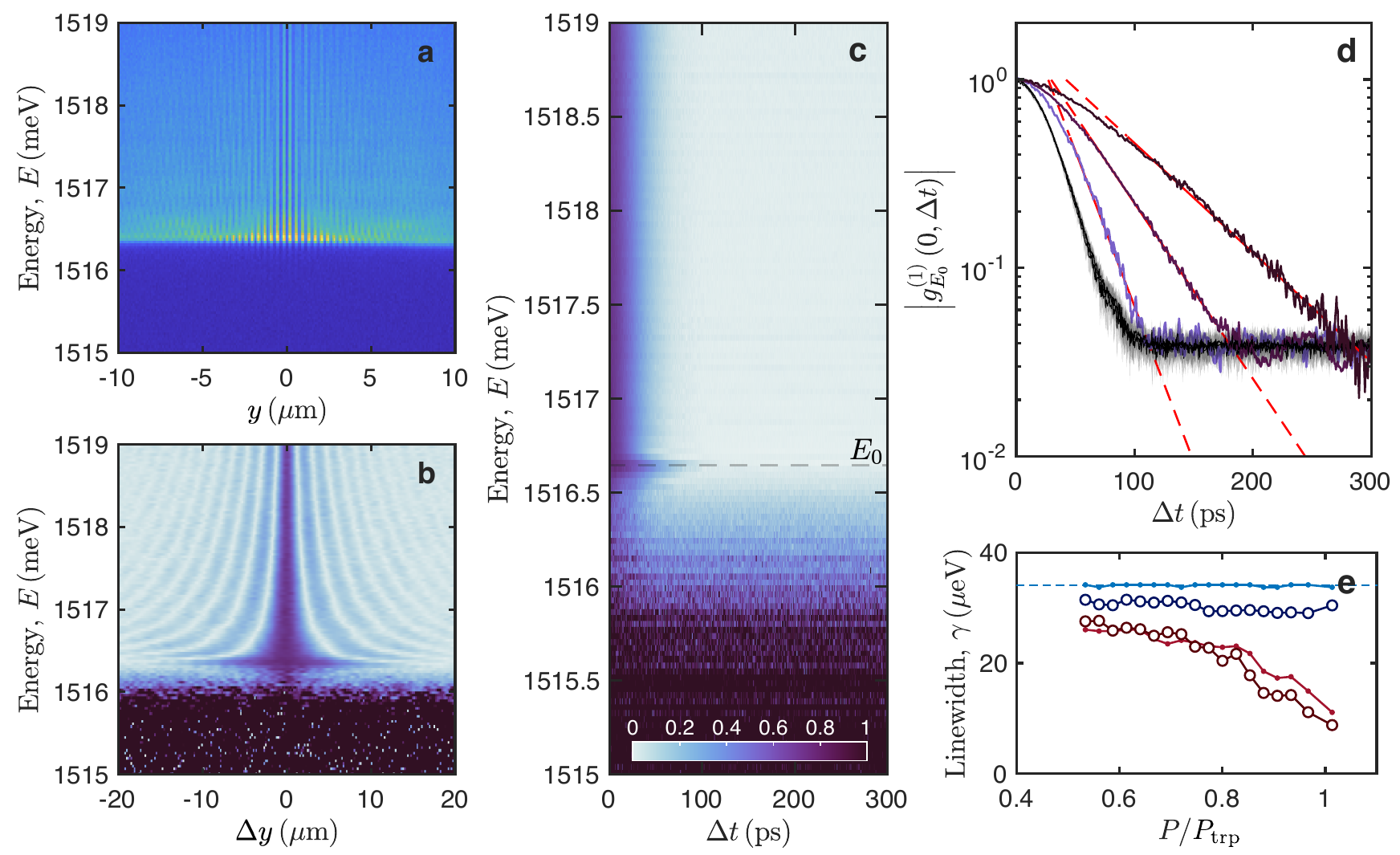}
    \caption{\textbf{Zero-momentum linewidth estimation from energy-resolved coherence measurement}. 
    \textbf{a} Zero-delay energy resolved interferogram. 
    \textbf{b} Zero-delay energy resolved coherence heat map. 
    \textbf{c} Energy-resolved coherence as a function of the time delay. 
    \textbf{d} Temporal coherence decay at $E = E_{0}$ (colored curves) for $P/P_{\mathrm{trp}} \approx 0.7$, 0.9 and 1. The red lines correspond to exponential fits of the long-delay tails of the ground-state coherence. For reference the black curves show the temporal decay at $E \!=\! 1518 \, \mathrm{meV}$ for the same powers (three overlapping curves).
    \textbf{e} Linewidth dependence on the reduced pump power $p$. 
    Red ($E=E_{0}$) and blue ($E=1518~\mathrm{meV}$) circles indicate the linewidths extracted from the fits, while the corresponding dots show the effective linewidths obtained from the $1/e$ coherence times.
    For all graphs, $\hbar \delta = -7.3 \, \mathrm{meV}$.
    } 
    \label{fig:sup_map_meas_gam0}
\end{figure}

As detailed in Sec.~\ref{sec:interferometry}, the first-order coherence function, $\left|g^{(1)}(\Delta x,\Delta y,\Delta t)\right|$, is extracted at each time delay $\Delta t$ from the fringe visibility of the corresponding interferogram. By closing the spectrometer slit around $x=0$ and imaging the first diffraction order onto the CCD, we additionally obtain energy-resolved interferograms. Applying the same analysis as in Sec.~\ref{sec:interferometry} yields the energy-resolved first-order coherence, denoted $\left| \vphantom{g^{(1)}} \right. \!\! g_{E}^{(1)}(\Delta y,\Delta t) \!\! \left. \vphantom{g^{(1)}}\right|$. The zero-delay interferogram and the corresponding coherence heat map, $\left| \vphantom{g^{(1)}} \right. \!\! g_{E}^{(1)}(\Delta y,0) \!\! \left. \vphantom{g^{(1)}}\right|$, are shown in Figs.~\ref{fig:sup_map_meas_gam0}\textbf{a} and~\ref{fig:sup_map_meas_gam0}\textbf{b}, for $P/P_{\mathrm{trp}} = 0.5$. Only the ground-state coherence remains uniform along $\Delta y$, as it originates from the $k_y = 0$ state. At higher energies, each spectral slice is related to polaritons with finite in-plane momentum. Consequently, the coherence is modulated along $\Delta y$, with the oscillations arising from the interference between the emission components at $\pm k_y$, as observed in panel~\textbf{b}. Additionally, Fig.~\ref{fig:sup_map_meas_gam0}\textbf{c} displays the energy-resolved first-order coherence at $P/P_{\mathrm{trp}}=0.8$ as a function of delay time for $\Delta x = \Delta y=0$, $\left| \vphantom{g^{(1)}} \right. \!\! g_{E}^{(1)}(0, \Delta t) \!\! \left. \vphantom{g^{(1)}}\right|$. The temporal coherence is markedly enhanced at the ground-state energy $E_{0}$ (black dotted line), compared to that of higher-energy states. In contrast, all states with $E>E_{0}$ display nearly identical coherence decay. Figure~\ref{fig:sup_map_meas_gam0}\textbf{d} shows cuts of $\left| \vphantom{g^{(1)}} \right. \!\! g_{E}^{(1)}(0, \Delta t) \!\! \left. \vphantom{g^{(1)}}\right|$ at $E \!=\! E_{0}$ (colored curves) and $E \!=\! 1518 \, \mathrm{meV}$ (black curves, averaged over 10 pixels), for $P/P_{\mathrm{trp}} \!\approx \!0.7$, 0.9 and 1. The exponential tails ($0.05 \leq \left| \vphantom{g^{(1)}} \right. \!\! g_{E}^{(1)}(0, \Delta t) \!\! \left. \vphantom{g^{(1)}}\right| \leq 0.5$) are fitted to extract the linewidth $\gamma_0$, yielding the red ($E\!=\!E_{0}$) and blue ($E \!=\! 1518 \, \mathrm{meV}$) circles in Fig.~\ref{fig:sup_map_meas_gam0}\textbf{e}. For comparison, the $1/e$ coherence time is also converted into an effective linewidth, shown as the corresponding red and blue dots on the same panel. This analysis highlights the linewidth narrowing of the ground state with increasing polariton population. At low excitation power, the ground-state linewidth is expected to coincide with that of higher-energy states, providing an estimate of $\boldsymbol{\gamma_{0}=35\pm5~\mathrm{\mu eV}}$.

\subsubsection{\label{sec:supmat_micro_params_condensate_gamma2}Determination of the gain-curvature coefficient $\gamma_{2}$}

\noindent The gain-curvature coefficient $\gamma_2$ is measured experimentally through the fit of the condensate momentum distribution, $n(k)$, below transparency threshold to the analytical prediction Eq. \eqref{eq:n_k}. Experimentally, we do not directly access $n(k)$, but rather the cavity emission intensity in $k$-space $I(k) \propto \gamma_0 n(k)$. The latter is extracted either by integrating the energy-resolved spectrum or by radially averaging a direct CCD image of $I(\boldsymbol{k})$ about $\boldsymbol{k}=\boldsymbol{0}$. Careful background subtraction is essential, as residual counts can bias the zero-momentum density and compromise the normalization of $n(k)$. This effect is stronger at large negative detunings where high-momentum  luminescence from the relaxation bottleneck produces a strong momentum$\text{-}$dependent background beneath the low-$k$ population.

\begin{figure}[t!]
    \centering
    \includegraphics[width=\linewidth]{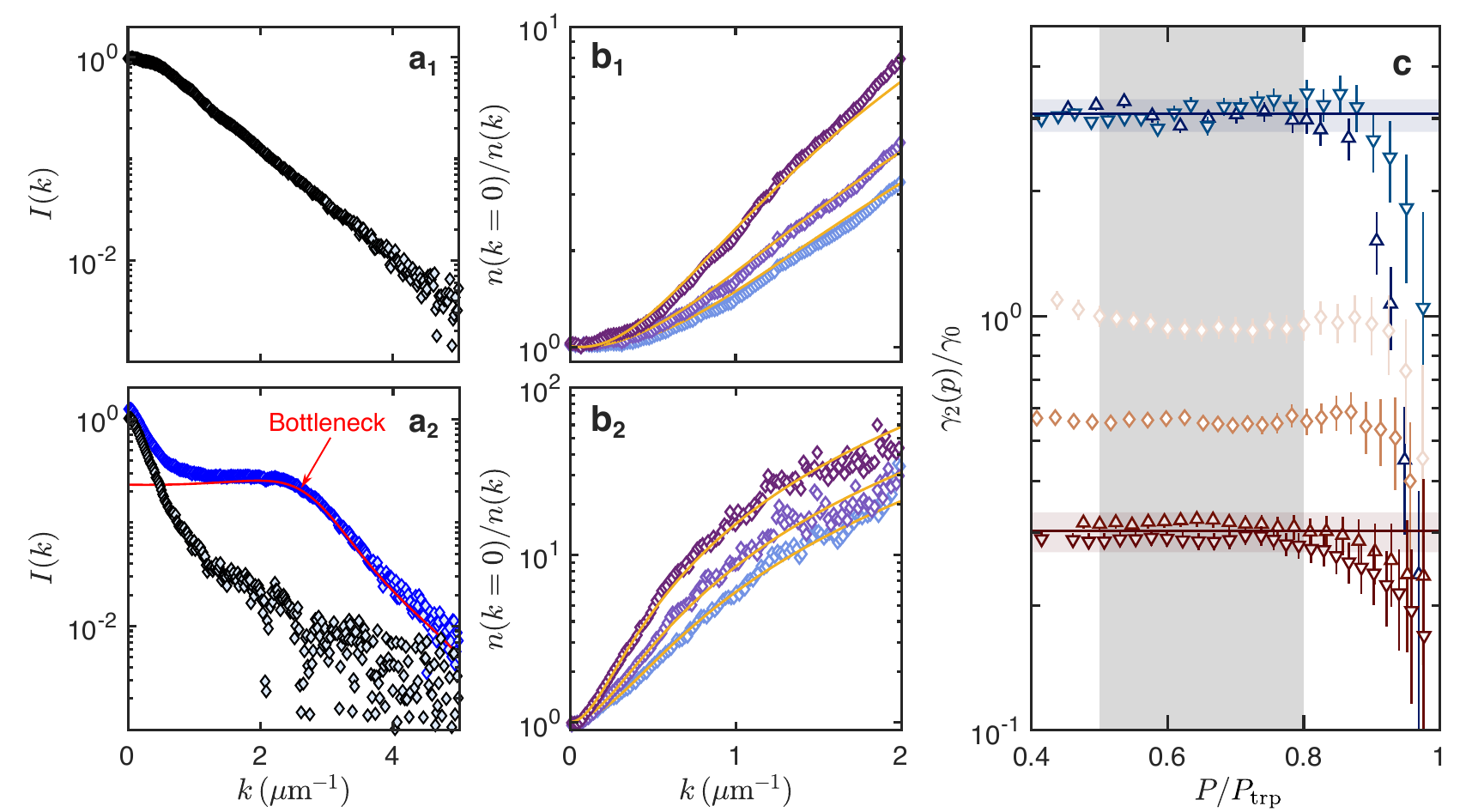}
    \caption{\textbf{Measurement of the gain-curvature coefficient $\boldsymbol{\gamma_{2}}$.}
    $\bf{a_1}$-$\bf{a_2 .}$ Radially averaged emission intensity $I(k)$ at $P/P_{\mathrm{trp}}\!=\!0.8$ for $\hbar \delta=-5.2~\mathrm{meV}$ (\textbf{a$\boldsymbol{_{1}}$}) and $-12.6~\mathrm{meV}$ (\textbf{a$\boldsymbol{_{2}}$}). For \textbf{a$\boldsymbol{_{2}}$}, the photo-emission background arising from the relaxation bottleneck is fitted (red curve) and subtracted from the raw data (blue circles), yielding the background-corrected profile (black circles). 
    $\bf{b_1}$-$\bf{b_2 .}$ Normalized momentum distribution, $n(0)/n(k)$, measured at $P/P_{\mathrm{trp}}=0.4$, $0.6$, and $0.8$. Fits to the data with Eq.~\eqref{eq:n_k} are shown as orange lines.
    \textbf{c} Ratio $\gamma_{2}/\gamma_{0}$ as a function of the reduced pump power $P/P_{\mathrm{trp}}$ for the different cavity-exciton detunings.}
    \label{fig:sup_map_meas_gam2}
\end{figure}

\medskip

Figure~\ref{fig:sup_map_meas_gam2} highlights the procedure used to extract the ratio $\gamma_{2}/\gamma_{0}$ for $\hbar \delta = -5.2~\mathrm{meV}$ (upper panels $\bf{a_1}$ and $\bf{b_1}$) and $\hbar \delta =-12.6~\mathrm{meV}$ (lower panels $\bf{a_2}$ and $\bf{b_2}$). Figure~\ref{fig:sup_map_meas_gam2}\textbf{a} shows the radial average of the emission intensity, $I(k)$, for $P/P_{\mathrm{trp}} \approx 0.8$. While no background subtraction is needed in panel~$\bf{a_1}$, the relaxation bottleneck gives rise to a large photo-luminescence background in panel~$\bf{a_2}$. This contribution is fitted and subtracted from the raw data (blue symbols), revealing the background$\text{-}$corrected low$\text{-}k$ emission intensity (black symbols). The normalized profiles $n(k)/n(0)$ are fitted at all pump powers below transparency using $f_{a}(k)=1/[1+a(p_{\rm trp})k^{2}]$ (see Fig.~\ref{fig:sup_map_meas_gam2}\textbf{b}), for $k \leq 2 \, \mathrm{\mu m^{-1}}$. According to Eq.~\eqref{eq:n_k}, $a(p_{\rm trp}) = \gamma_{2}/\!\left(\gamma_{0} (1-p_{\rm trp}) \right)$. Consequently, plotting $a(p_{\rm trp}) (1-p_{\rm trp})$ directly provides the ratio $\gamma_{2}/\gamma_{0}$, as shown in Fig.~\ref{fig:sup_map_meas_gam2}\textbf{c}. Besides the KPZ (red downward triangles) and EW datasets (blue downward triangles), measurements at $\hbar \delta= -13.9, \, -9.2, \, -7.6$ and $-5.5 \, \mathrm{meV}$ are also included in this plot (blue to red). In Fig.~\ref{fig:sup_map_meas_gam2}\textbf{c}, the errorbars include the uncertainties on both $a(p_{\rm trp})$ and the threshold power $P_{\mathrm{trp}}$. For all datasets, $\gamma_{2}/\gamma_{0}$ remains approximately constant over a finite pump power range.  Fitting the upper and lower groups with a constant yields $\boldsymbol{\gamma_{2}/\gamma_{0} = 0.30 \pm 0.05~\mathrm{\mu m}^{2}}$ in the KPZ phase and $\boldsymbol{\gamma_{2}/\gamma_{0} = 3.0 \pm 0.5~\mathrm{\mu m}^{2}}$ in the EW phase.

\subsubsection{\label{sec:supmat_micro_params_reservoir}Determination of the reservoir coefficient $\gamma_R / R$}

In this paragraph, we experimentally determine the microscopic parameter characterizing the exciton reservoir in Eq.~\eqref{eq:supmat_xreservoir}. Namely we provide estimated values for the ratio $\gamma_R/R$ through the I-P curve.

\medskip

\begin{figure}[t!]
    \centering
    \includegraphics[scale=0.75]{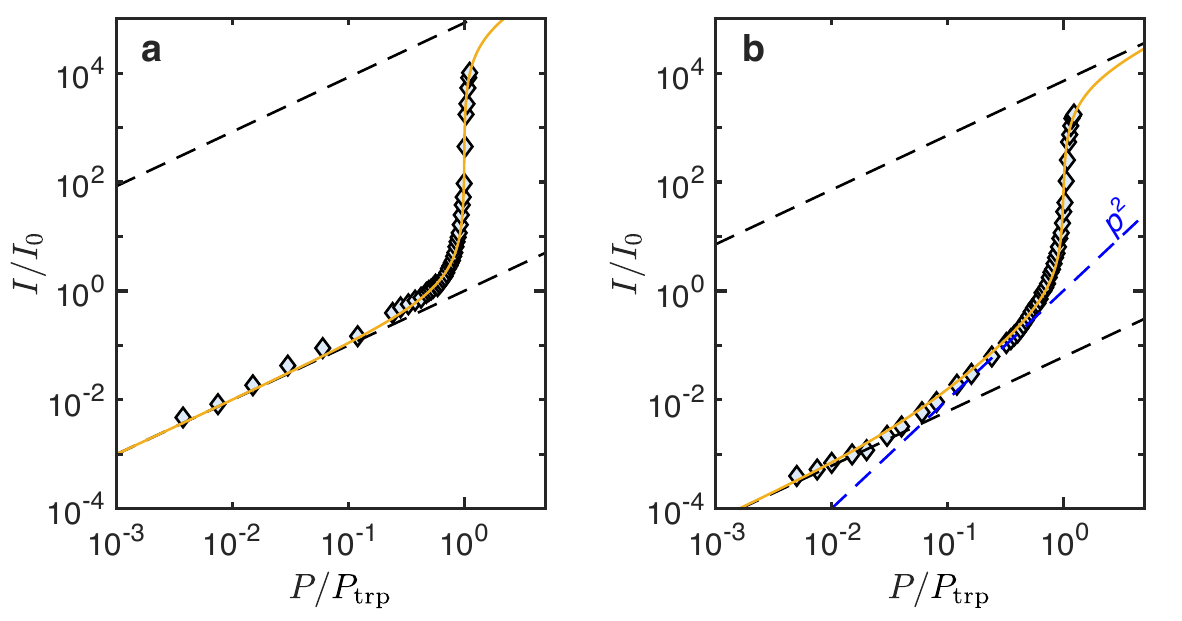}
    \caption{
    \textbf{I-P characteristics} obtained \textbf {a} in the KPZ regime ($\hbar \delta = -5.2 \, \mathrm{meV}$) and \textbf{b} in the EW regime ($\hbar \delta = -12.6 \, \mathrm{meV}$).
    }
    \label{fig:sup_mat_I-P_charac}
\end{figure}

A minimal kinetic model for a single-mode polariton condensate, depleted at rate $\gamma_{0}$ and fed by an exciton reservoir via (i) phonon-assisted relaxation at rate $R$ and (ii) exciton--polariton scattering at rate $R_{\mathrm{XP}}$, reads~\cite{Deng2010}:
\begin{align}
    \frac{d n}{d t} =& \,- \gamma_{0} n + R n_{R} (n+1) + R_{\mathrm{XP}} n_{R}^{2} (n+1) \label{eq:polariton_density} \\
    \frac{d n_{R}}{d t} =& \, P -\gamma_{R} n_{R} - R n_{R} (n+1) - R_{\mathrm{XP}} n_{R}^{2} (n+1) \label{eq:reservoir_density}
\end{align}
\noindent The steady state solution of this set of equations gives the relation between the pump power $P$ and the number $n$ of polaritons in the single mode condensate. From Eq.~\eqref{eq:polariton_density}, we get:
\begin{equation}
    n_{R} = \frac{1}{2} \frac{R}{R_{\mathrm{XP}}} \left( \sqrt{1+4 \frac{\gamma_{0} R_{\mathrm{XP}}}{R^{2}} \frac{n}{n+1}} - 1\right) = \frac{1}{2} \frac{R}{R_{\mathrm{XP}}} f_{\xi}(n) \, ,
    \label{eq:reservoir_density_SS}
\end{equation}
\noindent where $\xi = 2\gamma_{0} R_{\mathrm{XP}}/ R^{2}$. 
Substituting Eq.~\eqref{eq:reservoir_density_SS} into the steady state solution of Eq.~\eqref{eq:reservoir_density} yields:
\begin{equation}
    P = \frac{\gamma_{R} \gamma_{0}}{R} \frac{1}{\xi} \left( f_{\xi}(n) +  \frac{R}{\gamma_{R}} (n+1) f_{\xi}(n) + \frac{1}{2}  \frac{R}{\gamma_{R}} (n+1) f_{\xi}(n)^{2}\right) \, .
    \label{eq:final_P_VS_n}
\end{equation}
\noindent Below transparency, gain saturation is negligible and $n_{R} \simeq P/\gamma_{R}$. Assuming this relation remains valid at threshold, where gain balances losses, yields the following expression for the transparency threshold:
\begin{equation}
   P_{\mathrm{trp}} = \frac{\gamma_{0} \gamma_{R}}{R} \frac{1}{\xi} \left( \sqrt{1+2 \xi} - 1 \right).
\end{equation}
\noindent Finally, Eq.~\eqref{eq:final_P_VS_n}  can be rewritten as:
\begin{equation}
\boxed{
    \frac{P}{P_{\mathrm{trp}}} = \frac{f_{\xi}(n)}{\sqrt{1+2 \xi} - 1 } \left[1 + \frac{R}{\gamma_{R}} (n+1) + \frac{1}{2} \frac{R}{\gamma_{R}} (n+1) f_{\xi}(n) \right] \, .
    }
    \label{eq:IP_charac}
\end{equation}
\noindent In the limit $\xi \rightarrow 0$ (that is, when $\gamma_{0} R_{\mathrm{XP}} \ll R^{2}$), Eq.~\eqref{eq:IP_charac} simplifies to:
\begin{equation}
\boxed{
    \frac{P}{P_{\mathrm{trp}}} \underset{\xi \rightarrow 0}{=} \frac{R}{\gamma_{R}}n+\frac{n}{n+1} \, ,
    \label{eq:IP_charac_xi_0}
    }
\end{equation}
\noindent where $P_{\mathrm{trp}} = \gamma_{0} \gamma_{R}/R$. Equations~\eqref{eq:IP_charac} and~\eqref{eq:IP_charac_xi_0} provide a fit function to the I--P curve, from which estimates of $P_{\rm trp}$ and of the ratio $\gamma_{R}/R$ can, in principle, be extracted, thereby constraining the microscopic reservoir parameters used in simulations. In the main text, the cavity zero-momentum emission intensity $I$ is plotted in Fig.~1\textbf{b} as a function of the reduced pumping power $p = P/P_{\mathrm{th}}$ for $\hbar\delta = -5.2~\mathrm{meV}$. This figure has been reported in the panel \textbf{a} of Fig.~\ref{fig:sup_mat_I-P_charac} for convenience but as function of $p_{\rm trp} = P/P_{\mathrm{trp}}$ this time. The yellow solid line shows a fit of the data using Eq.~\eqref{eq:IP_charac_xi_0}. The normalization factor, $I_{0}$, is treated as a third fitting parameter in order to circumvent the conversion between the measured intensity and the polariton number. The fit curve exhibits a characteristic S-shape in double logarithmic scales, with linear regimes at low ($p_{\rm trp} \ll 1$) and high ($p_{\rm trp} \gg 1$) pump powers, separated by a strongly nonlinear transition around the transparency threshold. The ratio $\gamma_{R}/R$ is directly encoded in the vertical separation between the two linear asymptotes (black dashed lines) in this graphical representation. Figure~\ref{fig:sup_mat_I-P_charac}\textbf{b} shows the analogue of panel~\textbf{a} for a detuning of $\hbar \delta = -12.6 \, \mathrm{meV}$.$\;$The data are fitted with the full model of Eq.~\eqref{eq:IP_charac} this time, as it captures the quadratic departure of the emission intensity $I$ from the low-power linear asymptotic behavior observed at large negative detuning. 

\medskip

The values of $\gamma_{R}/R$ extracted from the fits are subject to large uncertainties in both cases. At high pump power ( $p_{\rm trp} \!\gtrsim\! 1.4$), modulation instability drives the condensate into a multimode regime, such that the zero-momentum intensity no longer directly reflects the condensate occupation $n$. The lack of data prevents a robust constraint of the fit in this regime and an accurate estimation of $\gamma_{R}/R$. We find $\boldsymbol{\gamma_{R}/R=8\pm2\times10^{4}}$ in the KPZ phase and $\boldsymbol{12\pm2\times10^{4}}$ in the EW phase. We point out that $\gamma_{R}/R$ is dimensionless here, whereas it has units of inverse area experimentally. This difference arises because our model is single-mode, whereas the experimental system is transversely multimode. Therefore, the extracted value of $\gamma_{R}/R$ must be rescaled by the mode area $\mathcal{A}$, approximated here by the pump area ($\sim 3 \times 10^{3} \, \mu\mathrm{m}^{2}$). Moreover, further corrections may arise from the redistribution of the pump among the several modes of the system. Accounting for these effects requires a multimode extension of the model, which is beyond the scope of this work.

\newpage

\subsection{\label{sec:supmat_condensation_vs_transparency}Determination of the condensation threshold $P_{\rm th}$}

Table~\ref{tab:micro_para_lin_mod} lists the experimentally extracted microscopic parameters entering the linearized model described in Eq.~\eqref{eq:supmat_2D_g1_dr_0_dt} of Sec.~\ref{sec:supmat_Linearized_model}. In the present section, we compare the model predictions with the measured spatiotemporal coherence decay and identify the condensation threshold as the pump power at which the linearized description first starts departing significantly from the experimental data. This departure from the linear model marks the onset of gain saturation, where stimulated scattering into the ground state drives its macroscopic occupation and the emergence of a polariton condensate.

\begin{table}[h]
\centering
\renewcommand{\arraystretch}{1.4}
\setlength{\tabcolsep}{10pt}
\begin{tabular}{c|c|c|c|c|}
\cline{2-5}
 & $\hbar \delta \, (\mathrm{meV})$
 & $m\,(m_e)$ 
 & $\hbar\gamma_0\,(\mu\text{eV})$ 
 & $\gamma_{2}/\gamma_{0} \,(\mu\text{m}^2)$  \\
\hline
\multicolumn{1}{|c|}{EW} 
& $-12.6$
& $4.7 \times 10^{-5}$ 
& $35 \pm 5$ 
& $3.0 \pm 0.5$ \\
\hline
\multicolumn{1}{|c|}{KPZ} 
& $-15.2$
& $5.9 \times 10^{-5}$ 
& $35 \pm 5$  
& $0.3 \pm 0.05$ \\
\hline
\end{tabular}
\caption{\label{tab:micro_para_lin_mod} Microscopic parameters entering the linearized model of Sec.~\ref{sec:supmat_Linearized_model} (experimental values).}
\end{table}

\begin{figure}[t!]
    \centering
    \includegraphics[scale=0.6]{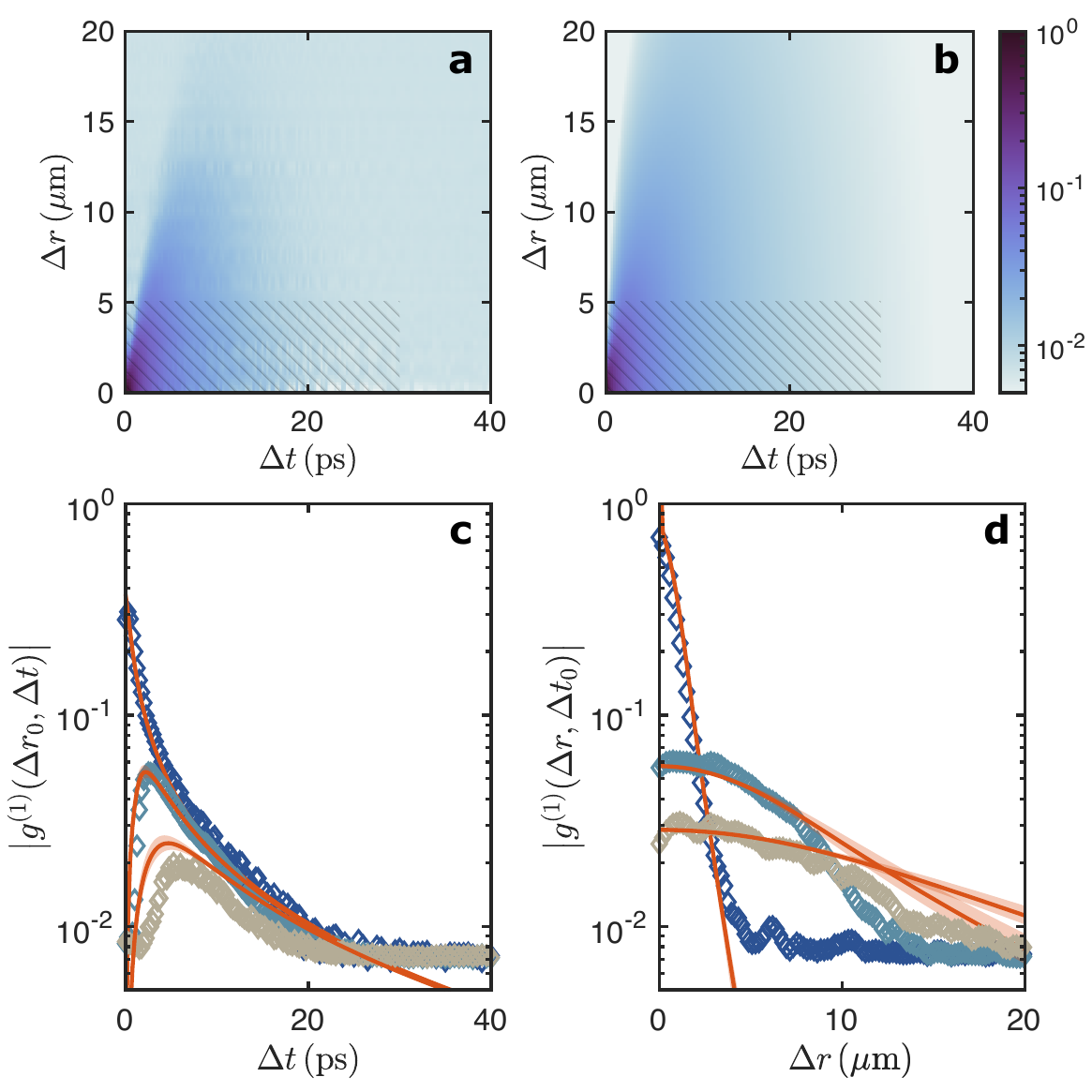}
    \caption{
    \textbf{Comparison of the low power coherence of polariton emission to the linearized model prediction.}
    \textbf{a} Measured values of $| g^{(1)} \left(\Delta r, \Delta t \right) |$ as a function of $\Delta r$ and $\Delta t$, for $P/P_{\mathrm{trp}} = 0.6$ and $\hbar \delta = -5.2 \, \mathrm{meV}$.
    \textbf{b} Corresponding coherence map computed using the linearized model Eq.~\eqref{eq:supmat_2D_g1_dr_dt} for $m = 5.9 \times 10^{-5} \, m_{e}$, $\hbar \gamma_{0} = 35 \, \mathrm{\mu eV}$, $\gamma_{2}/\gamma_{0} = 0.3 \, \mathrm{\mu m^{2}}$ , $P/P_{\mathrm{trp}} = 0.6$.
    The hatched regions in both panels indicate the $(\Delta r,\Delta t)$ domains over which the sMAPE is evaluated.
    \textbf{c}  Measured values of $| g^{(1)} \left(\Delta r_{0}, \Delta t \right) |$ versus $\Delta t$ for $\Delta r_{0} = 1 \, \mathrm{\mu m}$ (blue), $5 \, \mathrm{\mu m}$ (seafoam green) and $10 \, \mathrm{\mu m}$ (light brown).
    \textbf{d} Measured values of $| g^{(1)} \left(\Delta r, \Delta t_{0} \right) |$ as a function of $\Delta r$ for $\Delta t_{0} = 0 \, \mathrm{ps}$ (blue), $4 \, \mathrm{ps}$ (seafoam green) and $8 \, \mathrm{ps}$ (light brown).
    On both panels, the red curves show the corresponding$\;$cuts extracted from the theoretical coherence heat map \textbf{b}. 
    The red shaded area represents the variation of the model predictions between $\gamma_{2}/\gamma_{0}=0.25$ and $0.35 \, \mu\mathrm{m}^{2}$.
    }
    \label{fig:sup_map_lin_mod}
\end{figure}

Figures~\ref{fig:sup_map_lin_mod} compares the measured (panel~\textbf{a}) and calculated (panel~\textbf{b}) $\left|g^{(1)}(\Delta r,\Delta t)\right|$ heat maps at $P/P_{\mathrm{trp}} = 0.6$, far below threshold in the linear, low-density regime. The theoretical coherence is obtained from Eq.~\eqref{eq:supmat_2D_g1_dr_dt}, using a large UV cutoff $k_c = 10~\mathrm{\mu m^{-1}}$ to suppress spurious oscillations in the decay. The dispersive features are well resolved and closely reproduced by the model. Figures~\ref{fig:sup_map_lin_mod}\textbf{c} and \textbf{d} show temporal and spatial coherence decays obtained at fixed $\Delta r_0 = 1,5,10~\mathrm{\mu m}$ and $\Delta t_0 = 0,4,8~\mathrm{ps}$, respectively. In both panels, the red curves show the corresponding cuts from Fig.~\ref{fig:sup_map_lin_mod}\textbf{b}, surrounded by shaded areas accounting for the variation of the model prediction when $\gamma_{2}/\gamma_{0}$ is varied within its uncertainty range ($0.1 \, \mathrm{\mu m}^{2}$, see table~\ref{tab:micro_para_lin_mod}). The temporal decay is well reproduced by the model up to $\sim 30~\mathrm{ps}$, where the experimental signal reaches its noise floor. In particular, the position of the dispersive branch, only governed by the polariton mass, is well captured by the model. In contrast, the agreement for the spatial decay progressively worsens with increasing $\Delta t_{0}$, the experimental coherence decaying faster than the theoretical prediction. This deviation may result from finite-size effects, absent from the homogeneous model, which can accelerate coherence decay within the pumping area.

\medskip

By comparing the experimental and calculated coherence over the black hatched region in Figs.~\ref{fig:sup_map_lin_mod}\textbf{a}-\textbf{b}, we evaluate the symmetric mean absolute percentage error (sMAPE), defined as:
\begin{equation}
    \mathrm{sMAPE} = \frac{1}{N} \sum_{i, j} \left|\frac{g^{(1)}_{\scriptscriptstyle{\mathrm{exp}}}(\Delta r_{i}, \Delta t_{i}) - g^{(1)}_{\scriptscriptstyle{\mathrm{th}}}(\Delta r_{i}, \Delta t_{i}) }{g^{(1)}_{\scriptscriptstyle{\mathrm{exp}}}(\Delta r_{i}, \Delta t_{i}) +g^{(1)}_{\scriptscriptstyle{\mathrm {th}}}(\Delta r_{i}, \Delta t_{i}) }\right| \, ,
\end{equation}
\noindent where $g^{(1)}_{\scriptscriptstyle{\mathrm {exp}}}$ and $g^{(1)}_{\scriptscriptstyle{\mathrm {th}}}$ denote the experimental and theoretical coherence, respectively, and $N$ is the number of compared data points. Repeating this analysis when increasing $P/P_{\mathrm{trp}}$ gives the sMAPE evolution shown in Extended data~\ref{fig:AF:comp_lin_mod}\textbf{b}. A low-power plateau of small sMAPE values is followed by a sharp increase, which defines the condensation threshold, above which the macroscopic ground state occupation leads to coherence enhancement beyond the linear model description. Note that this transition occurs well bellow the transparency threshold ($P_{\mathrm{th}} \approx 0.6-0.7 \, P_{\mathrm{trp}}$).


\newpage

\subsection{\label{sec:supmat_filtering_dispersive_branches}Filtering of the dispersive branches}

\begin{figure}[h!]
    \centering
    \includegraphics[scale=0.65]{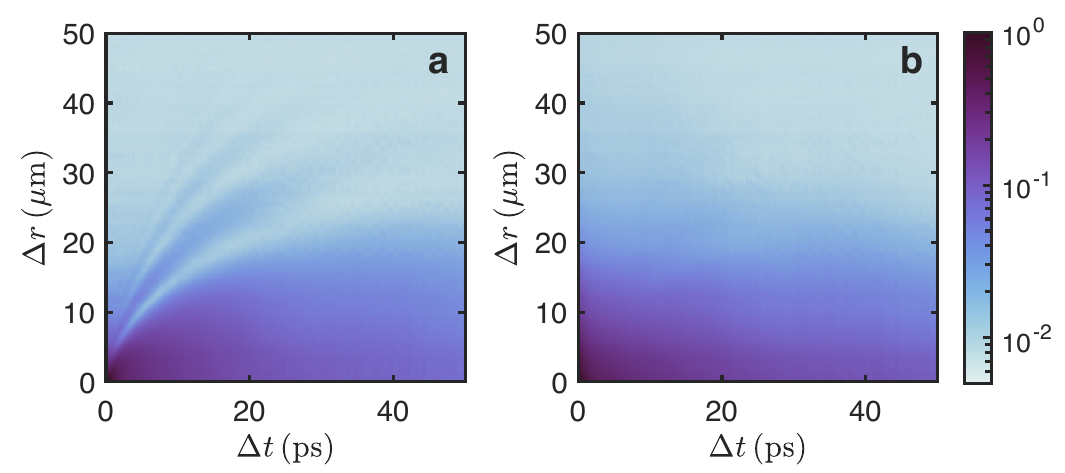}
    \caption{
    \textbf{Coherence heat maps} \textbf{a} before and \textbf{b} after filtering. 
    }
    \label{fig:sup_map_g1maps}
\end{figure}

As mentioned in the main text, the residual population of the polariton dispersion near the ground state gives rise to dispersive branches in the space-time correlations, appearing as alternating regions of enhanced/reduced coherence. These dispersive branches are most pronounced close to the condensation threshold, where the condensed fraction only marginally exceeds the non-condensed one. In this regime, filtering of the dispersive modes is required to isolate the condensate contribution in the emission coherence. In this section, we describe the procedure employed to perform this filtering step. The dispersion branches appear both in the phase and amplitude of the Fourier transform of $\left| g^{(1)} \right|$. We therefore independently filter the high-energy contributions in both components. The analysis presented below is done for an exciton-photon detuning of $\hbar \delta = -5.2~\mathrm{meV}$ (KPZ regime, Figs.~\ref{fig:condensation} and \ref{fig:scalingKPZ} of the main text) at $P/P_{\mathrm{trp}} = 0.92$. In the rest, $\mathrm{fft2}$ and $\mathrm{fftshift}$ denote the 2D fast Fourier transform and the operation that centers the zero-frequency component of the spectrum, respectively. The four steps of the filtering algorithm are detailed below. The coherence heat map is shown in Fig.~\ref{fig:sup_map_g1maps}\textbf{a} before filtering and in Fig.~\ref{fig:sup_map_g1maps}\textbf{b} after filtering.

\begin{figure}[h!]
    \centering
    \includegraphics[scale=0.65]{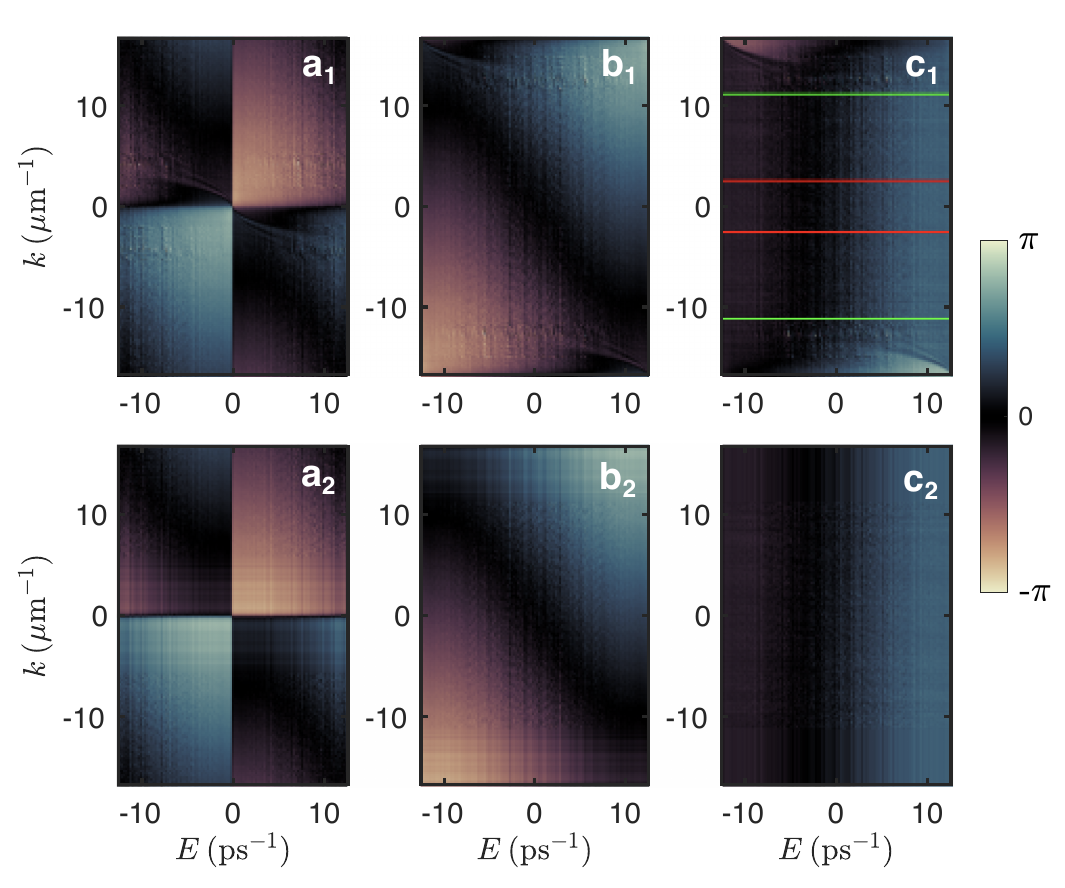}
    \caption{
    \textbf{Phase of} $\boldsymbol{\mathrm{FFT}(k, E)}$ the filtering process before removing the dispersive branches (panels $\textbf a_1$, $\textbf b_1$ and $\textbf c_1$) and after removing the dispersive branches (panels $\textbf a_2$ and $\textbf b_2$ and $\textbf c_2$).
    }
    \label{fig:sup_map_FFT_phase}
\end{figure}

\begin{figure}[h!]
    \centering
    \includegraphics[scale=0.65]{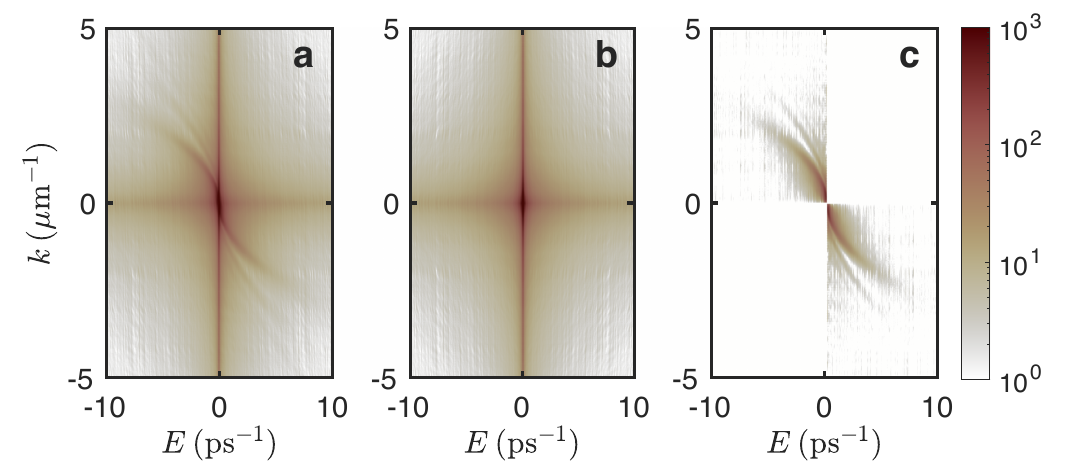}
    \caption{
    \textbf{Amplitude of} $\boldsymbol{\mathrm{FFT}(k, E)}$ \textbf{a} before and \textbf{b} after filtering. \textbf{c} Filtered signal.
    }
    \label{fig:sup_map_FFT_intensity}
\end{figure}

\begin{algobox}

\begin{itemize}
    \item [$\boldsymbol{\rightarrow}$] \textbf{STEP 1 - Fourier transform}
\end{itemize}

\begin{enumerate}
    \item The coherence heat map $\left|g^{(1)} (\Delta r, \Delta t)\right|$ is padded with zero such as $\mathrm{dim} \left(g_{p}^{(1)}\right) = 2  \,\mathrm{dim} \left(g^{(1)} \right) + 1$.
    \item We compute the fast Fourier transform (FFT) of $\left| g_{p}^{(1)} \right|$: $\mathrm{FFT}(k, E) = \mathrm{fftshift}\left( \mathrm{fft2} \left( \left| g_{p}^{(1)} (\Delta r, \Delta t) \right| \right) \right) $.
    \item We retrieve its phase $\theta_{0} = \mathrm{angle} \left( \mathrm{FFT} \right)$ (see Fig.~\ref{fig:sup_map_FFT_phase}\textbf{a}$\boldsymbol{_{1}}$) and amplitude $A_{0} = \left|  \mathrm{FFT} \right|$ (see Fig.~\ref{fig:sup_map_FFT_intensity}\textbf{a}).
\end{enumerate}

\end{algobox}

\begin{algobox}

\begin{itemize}
    \item [$\boldsymbol{\rightarrow}$] \textbf{STEP 2 - Phase filtering}
\end{itemize}

\begin{enumerate}
    \setcounter{enumi}{3}
    \item We start by computing $\theta = \mathrm{fftshift}\left(\theta_{0}\right)$ (Fig.~\ref{fig:sup_map_FFT_phase}\textbf{b}$\boldsymbol{_{1}}$).
    \item We remove the phase ramp $\theta(1, :)$ from the phase heat map $\theta$: $\Delta \theta = \theta-\theta(1, :)$.  
    The result is shown in Fig.~\ref{fig:sup_map_FFT_phase}\textbf{c}$\boldsymbol{_{1}}$.
    \item We average all rows in $\Delta \theta$ lying in between the red lines in Fig.~\ref{fig:sup_map_FFT_phase}\textbf{c}$\boldsymbol{_{1}}$). We call $\overline{\Delta \theta}$ the resulting mean vector.
    \item All rows in $\Delta \theta$ lying above (resp. bellow) the upper (resp. lower) green lines are replaced by the mean vector $\overline{\Delta \theta}$, thereby removing most of the dispersive features visible in the upper-left and lower-right corners of Fig.~\ref{fig:sup_map_FFT_phase}\textbf{c}$\boldsymbol{_{1}}$. The resulting heat map is shown in Fig.~\ref{fig:sup_map_FFT_phase}\textbf{c}$\boldsymbol{_{2}}$. We add back the ramp $\theta(1, :)$ to obtain the heat map of Fig.~\ref{fig:sup_map_FFT_phase}\textbf{b}$\boldsymbol{_{2}}$.
    \item We finally retrieve the corrected phase $\theta_{c}$ by performing another fftshift (see Fig.~\ref{fig:sup_map_FFT_phase}\textbf{a}$\boldsymbol{_{2}}$).
\end{enumerate}

\end{algobox}

\begin{algobox}

\begin{itemize}
    \item [$\boldsymbol{\rightarrow}$] \textbf{STEP 3 - Amplitude filtering}
\end{itemize}

\noindent \hspace{0.25cm} We now aim at filtering the fft amplitude $A_{0}$ (see Fig.~\ref{fig:sup_map_FFT_intensity}\textbf{a}).

\begin{enumerate}
    \setcounter{enumi}{8}
    \item The quadrant $\left\{k \! \geq \!0, E \! \geq \! 0 \right\}$ (as well as $\left\{k \!\leq \!0, E \! \leq 0 \right\}$) in $A_{0}$ does not show any dispersive feature. We replicate and flip this quadrant to replace $\left\{k \leq 0, E \geq 0 \right\}$ and $\left\{k \geq 0, E \leq 0 \right\}$ (see Fig.~\ref{fig:sup_map_FFT_intensity}\textbf{b}). The corrected amplitude is denoted $A_{c}$. Fig.~\ref{fig:sup_map_FFT_intensity}\textbf{c} shows the difference between $A_{0}$ and $A_{c}$, highlighting the filtered signal.
\end{enumerate}

\end{algobox}

\begin{algobox}

\begin{itemize}
    \item [$\boldsymbol{\rightarrow}$] \textbf{STEP 4 - Inverse Fourier transform}
\end{itemize}

\begin{enumerate}
    \setcounter{enumi}{9}
    \item The initial $\mathrm{FFT}$ phase and amplitude are replaced by the corrected ones: 
    \newline
    $\mathrm{FFT}(k, E) = A_{0}(k, E) \, \mathrm{exp} \left(i \theta_{0}(k, E) \right) \longrightarrow \mathrm{FFT}_{c}(k, E) = A_{c}(k, E) \times \mathrm{exp} \left(i \theta_{c}(k, E) \right)$.
    \item We compute the 2D inverse Fourier transform (ifft2) and crop the resulting $\left|g_{c}^{(1)} \right|$ heat map to its original dimensions:
    $\left|g_{c}^{(1)}(\Delta r, \Delta t) \right|= \left|\mathrm{ifft2}\left( \mathrm{ifftshift} \left( \mathrm{FFT}_{c}(k, E) \right) \right)\right|$. The result is shown in Fig.~\ref{fig:sup_map_FFT_intensity}\textbf{b}.
\end{enumerate}

\end{algobox}

\clearpage

\subsection{Normalization of the coherence}
\label{sec:supmat_normalization}

\begin{figure}[b!]
    \centering
    \includegraphics[scale=0.55]{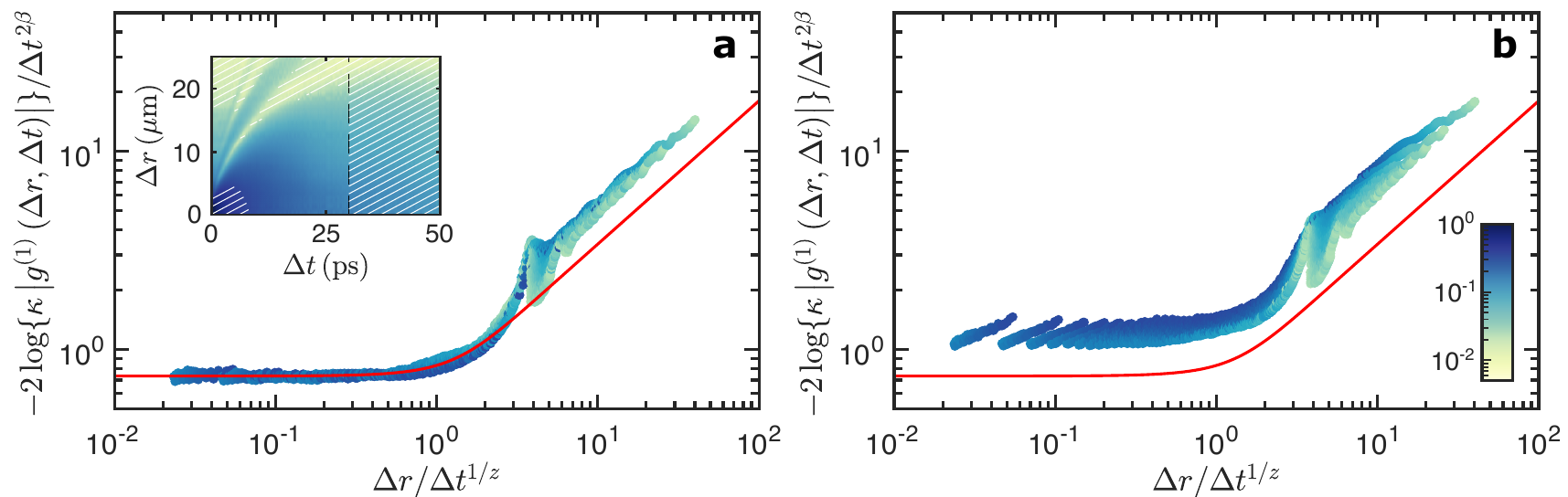}
    \caption{
    \textbf{Collapse of the experimental data onto the KPZ universal scaling function} (red solid line) \textbf{a} with and \textbf{b} without normalization $\kappa$. 
    The same region of the coherence heat map is used for the collapse of both datasets and is shown as an inset in Fig.~\textbf{a}.}
    \label{fig:sup_map_normalization}
\end{figure}

To identify KPZ scaling in the condensate spatio temporal coherence decay, we first examine the temporal decay of $\left|g^{(1)}(\Delta r,\Delta t)\right|$ for a fixed value of $\Delta r \simeq0$ and the spatial decay of $\left|g^{(1)}(\Delta r,\Delta t)\right|$ for $\Delta t =0$. We plot the spatial (temporal) experimental decay in logarithmic scales as a function of the expected KPZ scaling $\Delta r^{2\chi}$ ($\Delta t^{2\beta}$) (see Fig.~\ref{fig:scalingKPZ} of the main text). This first analysis is insensitive to the absolute normalization of the coherence, as it only offsets the data while preserving the stretched-exponential decay. The situation differs when testing the data collapse onto the universal KPZ scaling function. In this case, we plot on logarithmic axes versus the rescaled coordinate $\Delta r/\Delta t^{1/z}$:
\begin{equation*}
 -2\log \! \left(\kappa \left|g_{\rm exp}^{(1)}\right|\right)\!/\Delta t^{2\beta}= -2 \log \! \left( \left|g_{\rm norm}^{(1)}\right|\right)\!/\Delta t^{2\beta} -2\log \! (\kappa) /\Delta t^{2\beta}.
\end{equation*}
\noindent We see from this equation that the normalization factor $\kappa$ must be chosen carefully~\cite{bloch2026}. Indeed, this representation implicitly assumes that the KPZ temporal (resp. spatial) scaling extends down to $\Delta t = 0$ (resp. $\Delta r = 0$), with $\left|g_{\rm norm}^{(1)}(0,0)\right| = 1$. An incorrect normalization can therefore lead to misleading conclusions as the data collapse will be distorted by the term $-2\log \! (\kappa) /\Delta t^{2\beta}$. Experimentally, however, a short-time and short-distance transient precedes the onset of the KPZ regime. This transient originates from the short-range coherence of the photo-emission emitted by the  uncondensed polariton fraction. As a result, fits of the temporal coherence decay within the KPZ window systematically extrapolate to a value below unity at $\Delta t = 0$. To account for this effect, the entire coherence heat map is multiplied by a factor $\kappa > 1$, so that the extrapolated fit intercept is restored to 1. Figs.~\ref{fig:sup_map_normalization}\textbf{a} and \textbf{b} compare the resulting collapses for $\hbar\delta = -5.2~\mathrm{meV}$ and $P/P_{\mathrm{th}} = 1.25$, with and without this renormalization. Omitting the normalization significantly broadens both asymptotic branches of the collapse.


\clearpage

\subsection{\label{sec:supmat_phase_diagram}Computation of the phase diagram in Fig.~\ref{fig:Phase_Diagram}}

The phase diagram in Fig.~\ref{fig:Phase_Diagram} is obtained by comparing power-law and stretched-exponential fits to the temporal ($r \simeq 0$) coherence decay over the same fitting window using the Akaike information criterion (AIC), which balances goodness of fit against model complexity. The error bars are determined from the shot-to-shot reproducibility of the coherence radial profile at each time delay (see Sec.~S3.1.1). Assuming Gaussian-distributed errors, the likelihood measures how well the model fits the observed data:
\begin{equation}
    \mathcal{L} = \prod_{i =1}^{N} \frac{1}{\sqrt{2 \pi} \sigma_{i}} \, \mathrm{exp} \!\left[- \frac{\left(\left|g^{(1)}(0, \Delta t_{i})\right| - f(\Delta t_{i}) \right)^{2}}{2 \sigma_{i}^{2}} \right],
\end{equation}
\noindent where $N$ is the number of data points, $\sigma_{i}$ the standard deviation of the $i$th point and $f(\Delta t_{i})$ the fitted value at delay$ \Delta t_{i}$. The AIC is defined as: $\mathrm{AIC} = 2 n_{p} - 2 \mathrm{log}(\mathcal{L})$, where $n_{p}$ is  the number of model parameters. Substituting $\mathcal{L}$ gives: 
\begin{equation}
\boxed{
    \mathrm{AIC} = \chi^{2} + 2 p + \sum_{i = 1}^{N} \mathrm{ln} \! \left(2 \pi \sigma_{i}^{2} \right).
    }
    \label{eq:AIC}
\end{equation}
\noindent In Eq.~\eqref{eq:AIC}, $\chi^{2}$ is the weighted chi-square. Fixing the stretched-exponential exponent to its KPZ value ($\beta=0.24$) leaves both models with two fitting parameters. The difference in AIC between the stretched-exponential and power$\text{-}$law fits then reduces to the difference in their $\chi^{2}$ values, denoted $\Delta \chi^{2}$. We detail below the procedure used to evaluate $\Delta \chi^{2}$ and illustrate it using two datasets: one in the KPZ regime ($\hbar \delta \!=\! -6.8 \, \mathrm{meV}$) and the other in the EW regime ($\hbar \delta \!=\! -8.7 \, \mathrm{meV}$). Both datasets are included in the phase diagram of Fig.~\ref{fig:Phase_Diagram} of the main text.

\medskip

\begin{figure}[t!]
    \centering
    \includegraphics[scale=0.55]{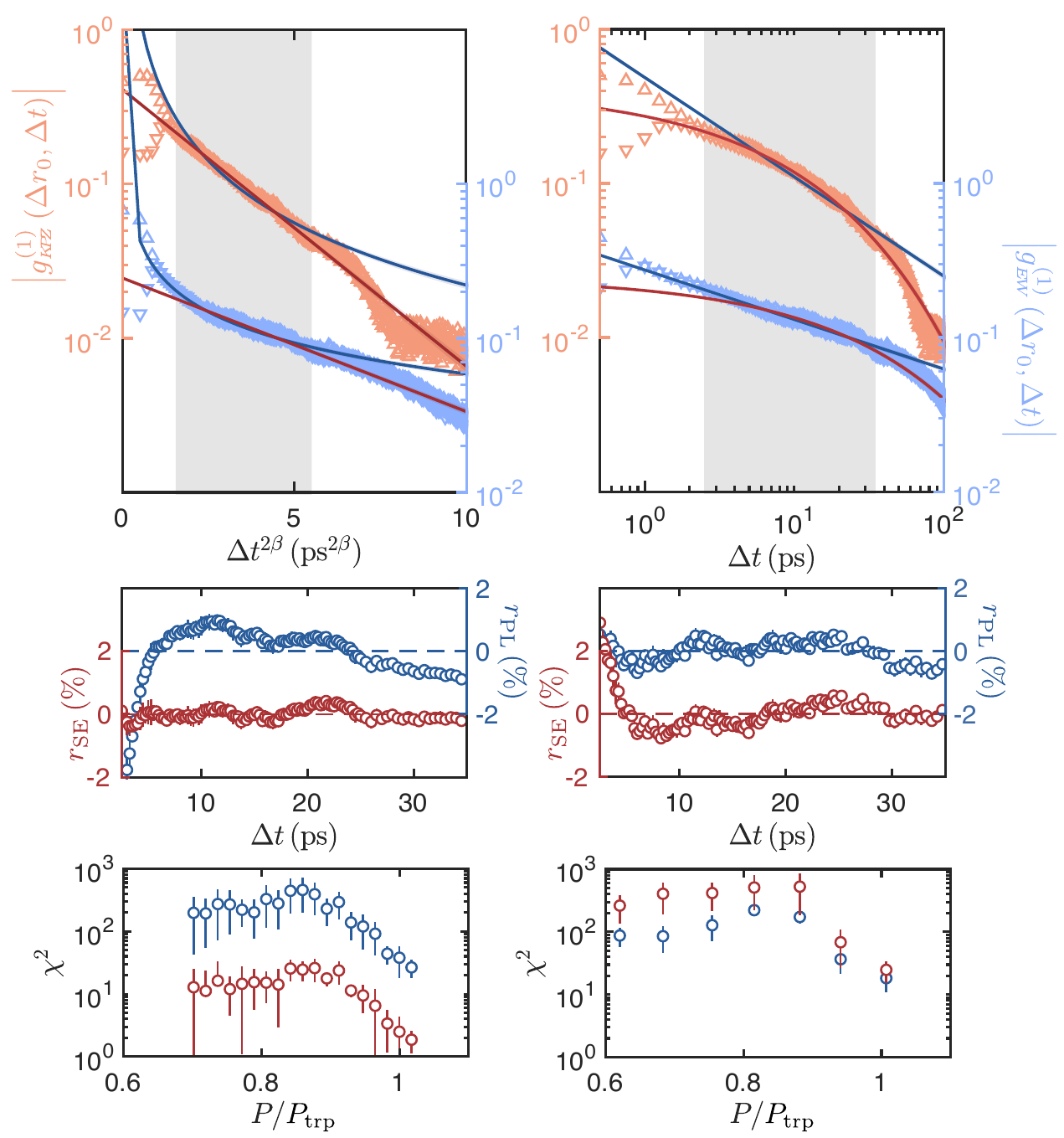}
    \caption{\textbf{Construction of the phase diagram: identifying the KPZ and EW regimes by comparing the goodness of stretched-exponential (SE) and power-law (PL) fits.}
    \textbf{a} Coherence decay versus $\Delta t^{2\beta}$ on a semi-logarithmic scale for representative KPZ ($\hbar \delta=-6.8 \, \mathrm{meV}$, orange) and EW ($\hbar \delta=-8.7 \, \mathrm{meV}$, blue) datasets.
    Upward- and downward-pointing triangles correspond to $\Delta r=1 \, \mu\mathrm{m}$ and $3 \, \mu\mathrm{m}$, respectively.
    \textbf{b} Same data plotted as a function of $\Delta t$ on a double-logarithmic scale.
    \textbf{c} Average residuals of the stretched-exponential ($r_{\mathrm{SE}}$, red circles) and power-law ($r_{\mathrm{PL}}$, blue circles) fits versus $\Delta t$, obtained by fitting the coherence decay in the KPZ regime for all spatial shifts in the range $1 \, \mu\mathrm{m} \leq \Delta r \leq 3 \, \mu\mathrm{m}$. 
    Errorbars indicate the $2\sigma$ variation of the residuals over this range. 
    \textbf{d} Same as panel \textbf{e} in the EW regime.
    \textbf{e-f} Chi-square values of the SE (red circles) and PL (blue circles) models as a function of $P/P_{\mathrm{trp}}$ for the KPZ and EW datasets, respectively. 
    Errorbars represent the combined $2\sigma$ uncertainty obtained by varying both the spatial shift ($1 \, \mu\mathrm{m} \leq \Delta r \leq 3 \, \mu\mathrm{m}$) and the temporal fitting window by $\pm1 \, \mathrm{ps}$.
    }
    \label{fig:sup_mat_phase_diagram}
\end{figure}

In Figs.~\ref{fig:sup_mat_phase_diagram}\textbf{a} and \textbf{b}, the orange (resp. blue) triangles show the temporal coherence decay in the KPZ (resp.~EW) regime for $P/P_{\mathrm{trp}} \approx 1.0$ (resp. $P/P_{\mathrm{trp}} \approx 0.9$) as a function of $\Delta t^{2 \beta}$ and $\Delta t$, in semi-logarithmic and logarithmic scales, respectively. The data within the gray shaded fitting window (extending from $2.5~\mathrm{ps}$ up to the onset of the Schawlow-–Townes exponential decay at $\Delta t_{ST} = 35 \pm 5~\mathrm{ps}$) are fitted with a stretched-exponential (SE) function (red line, $\beta \!=\! 0.24$) and a power-law (PL) function (blue line).  We repeat this fitting procedure for all spatial shifts ranging in $1~{\rm \mu m} \leq \Delta r \leq 3~{\rm \mu m}$. The coherence decays at the bounds of this range are shown as upward- and downward-pointing triangles in Figs.~\ref{fig:sup_mat_phase_diagram}\textbf{a}-\textbf{b}. This yields the $2 \sigma$ uncertainty on the residuals $r_{\mathrm{SE}}$ (red circles) and $r_{\mathrm{PL}}$ (blue circles) shown in Figs.~\ref{fig:sup_mat_phase_diagram}\textbf{c}-\textbf{d}. From the residuals, we finally compute $\chi^{2}$ as a function of the pump power, $P/P_{\mathrm{trp}}$, for the SE and PL models. The result is shown as red and blue circles in Figs.~\ref{fig:sup_mat_phase_diagram}\textbf{e}-\textbf{f}. The sensitivity of this analysis to the fitting window is evaluated by shifting the window by $\pm 1~\mathrm{ps}$. The resulting $\chi^{2}$ variations are included in the uncertainty estimate shown in Figs.~\ref{fig:sup_mat_phase_diagram}\textbf{e}-\textbf{f}. We immediately see on these plots that the SE model yields a lower $\chi^{2}$ than the PL model in the KPZ regime, while the opposite is observed in the EW regime. We can also easily evaluate at this stage the difference in $\chi^{2}$ between the two fitting models, $\Delta \chi^{2}$, and its associated uncertainty, $\delta \chi^2$. The colormap in the phase diagram of Fig.~\ref{fig:Phase_Diagram} then represents the ratio $\Delta \chi^{2}/\delta \chi^{2}$.


\clearpage

\subsection{\label{sec:additional_data}Additional data demonstrated 2D KPZ scaling in polariton condensates}

In this section, we provide additional results in the KPZ regime. We first demonstrate additional data collapses onto the universal KPZ scaling function for different pump powers, using the dataset presented in Fig.~2 of the main text. We then analyze an independent dataset acquired at $\hbar\delta=-6.8~\mathrm{meV}$, corresponding to an exciton fraction of $20 \%$, using the same analysis as in the main text.

\medskip

\begin{figure}[h!]
    \centering
    \includegraphics[scale=0.475]{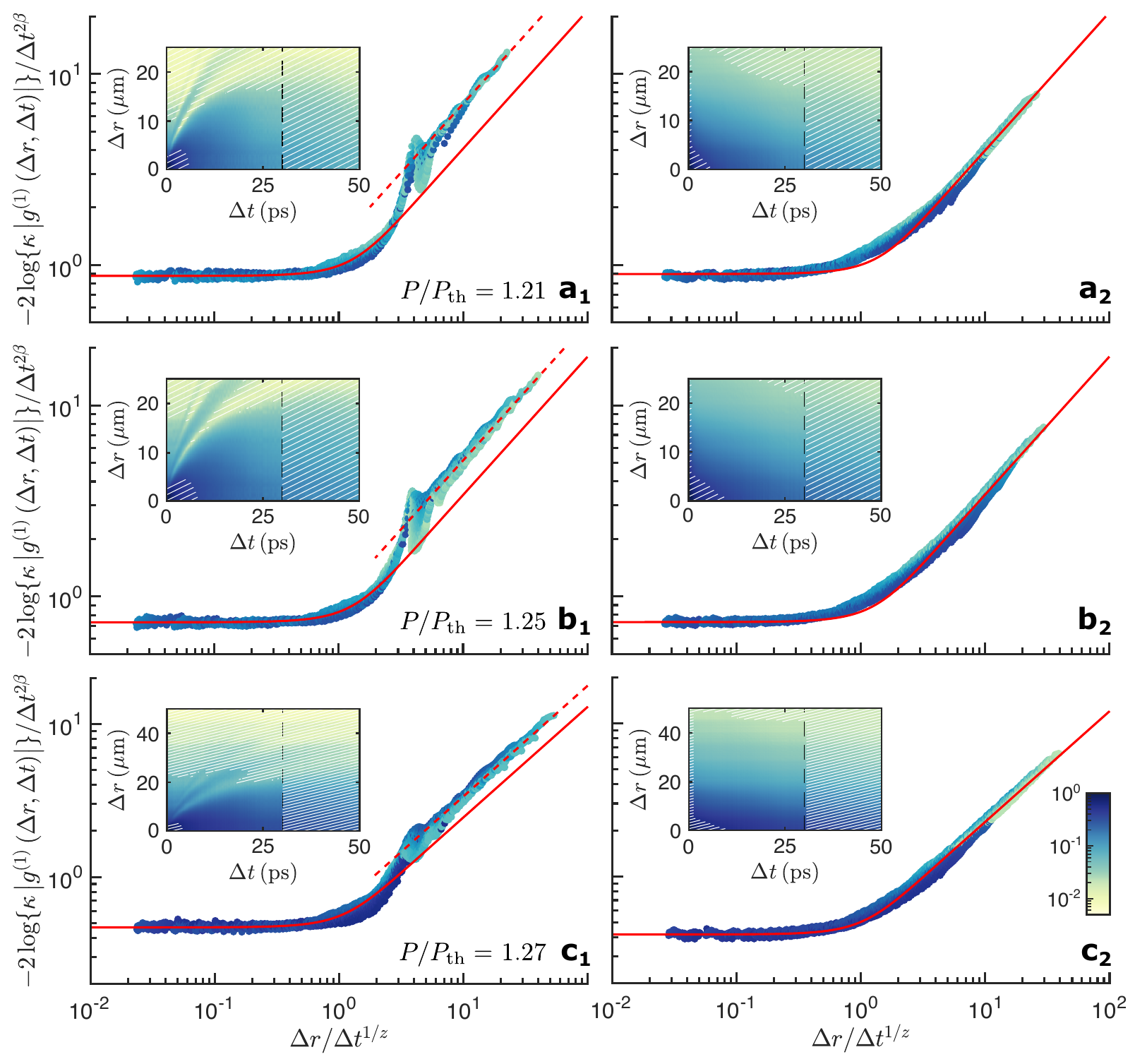}
    \caption{\textbf{Experimental data collapse onto the universal KPZ scaling function} at $\hbar \delta = -5.2 \, \mathrm{meV}$, for different pumping powers \textbf{a} $P/P_{\mathrm{th}} = 1.21$, \textbf{b} $P/P_{\mathrm{th}} = 1.25$ and \textbf{c} $P/P_{\mathrm{th}} = 1.27$.
    In all panels, we plot $-2 \mathrm{log}(\kappa|g^{(1)}(\Delta r, \Delta t)|)/\Delta t^{2 \beta}$ as a function of $\Delta r/\Delta t^{1/z}$, where $\kappa$ is a normalization constant (see, e.g., Sec.\ref{sec:supmat_normalization}). 
    The corresponding coherence heat maps are shown as insets; the non-hatched regions indicate the $(\Delta r,\Delta t)$ points selected for the collapse.
    The red curve is the universal KPZ scaling function.
    The first column displays the raw data, and the second the filtered data obtained following the procedure described in Sec.~\ref{sec:supmat_filtering_dispersive_branches}.
    From left to right, top to bottom: $\kappa = 2.45$ ($\boldsymbol{\mathrm{a_{1}}}$), 2.15 ($\boldsymbol{\mathrm{a_{2}}}$), 2.40 ($\boldsymbol{\mathrm{b_{1}}}$), 2.10 ($\boldsymbol{\mathrm{b_{2}}}$), 2.25 ($\boldsymbol{\mathrm{c_{1}}}$) and 2.00 ($\boldsymbol{\mathrm{c_{2}}}$).
    }
    \label{fig:sup_map_AD_collapse_KPZ}
\end{figure}

Figure~\ref{fig:sup_map_AD_collapse_KPZ} shows $-2\log \left(\kappa|g^{(1)}(\Delta r,\Delta t)|\right)/\Delta t^{2\beta}$ as a function of $\Delta r/\Delta t^{1/z}$, where $\kappa$ is the normalization constant given in the caption. From top to bottom, the pump power is $P/P_{\mathrm{th}}=1.21$ ($\boldsymbol{\mathrm{a_{1,2}}}$), $1.25$ ($\boldsymbol{\mathrm{b_{1,2}}}$), and $1.27$ ($\boldsymbol{\mathrm{c_{1,2}}}$). The first and second columns display the data collapse obtained from the raw and filtered data, respectively. Insets show the corresponding coherence heat maps, where the non-hatched regions indicate the $(\Delta r,\Delta t)$ points selected for the collapse. The selection is restricted to the range of coherence values over which KPZ scaling is observed, while all data acquired beyond the Schawlow--Townes onset time delay $\Delta t_{\mathrm{ST}}$ are discarded. 
The red curves correspond to the universal KPZ scaling function, $F=C_{0}F_{\mathrm{KPZ}}(y_{0} \Delta r/\Delta t^{1/z})$, with the scale factors $C_{0}$ and $y_{0}$ adjusted to the experimental data. The dispersive branches induce oscillations in the scaling collapse of the raw data and shift the tilted asymptote (red dashed line) away from the KPZ prediction. Both effects become less pronounced with increasing pump power as the visibility of the dispersive branches decreases. At high pump power, the condensate fraction dominates the emission, and the measured coherence is therefore primarily determined by the condensate itself. For all investigated powers, removing the dispersive branches yields data collapses in excellent agreement with the universal KPZ scaling function, with both the oscillations and the asymptotic offset eliminated.

\begin{figure}[h!]
    \centering
    \includegraphics[width=\linewidth]{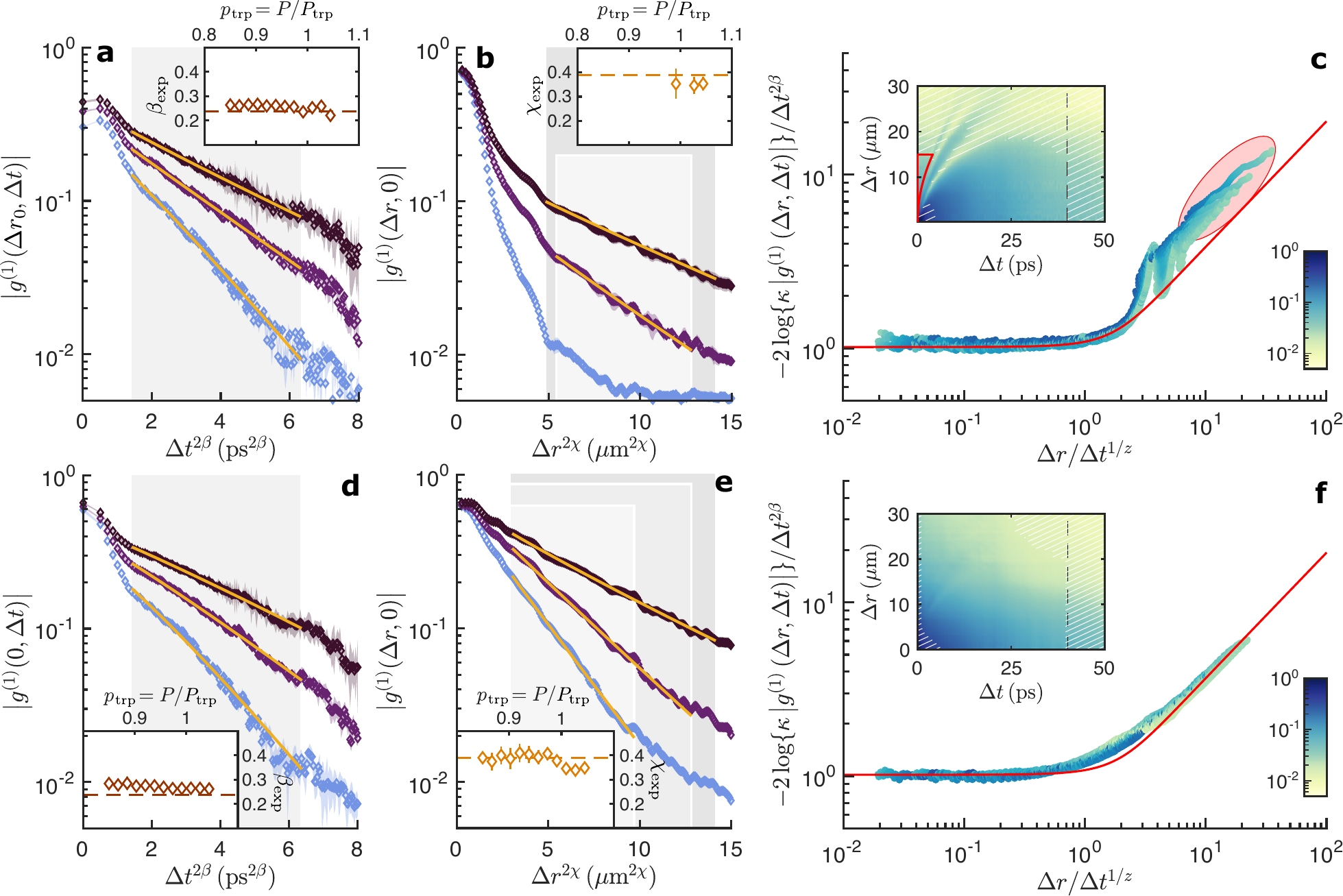}
    \caption{\textbf{KPZ universal scaling.}
    \textbf{a} (\textbf{b}): Measured values of $|g^{(1)}(\Delta r_{0}, \Delta t) |$ ($|g^{(1)}(\Delta r, 0)|$) as a function of $\Delta t^{2 \beta}$ ($\Delta r^{2 \chi}$)$\;$in semi-logarithmic scale  for $P/P_{\mathrm{trp}} = 0.87,\, 0.98, 1.04$ and $\Delta r_{0} = 1.3~\mathrm{\mu m}$. Straight orange lines: stretched exponential fits over the KPZ windows indicated by the gray-shaded areas. Errorbars are evaluated from multiple interferograms analysis.
    Insets in \textbf{a} (\textbf{b}): Experimental values of $\beta_{\mathrm{exp}}$ ($\chi_{\mathrm{exp}}$) shown in dark red (dark yellow) diamonds extracted from stretched exponential fits to the data shown in panel \textbf{a} (\textbf{b}), as a function of $P/P_{trp}$. Errorbars are obtained from the $\pm \sigma$ uncertainties of the stretched exponential fits.
    \textbf{c}: plot of $-2 \, \mathrm{log}(\kappa |g^{(1)}(\Delta r, \Delta t)|)/\Delta t^{2\beta}$ as a function of $\Delta r/\Delta t^{1/z}$  for measured values of $|g^{(1)}(\Delta r, \Delta t) |$  lying in the non-hatched region of the inset. 
    Inset of \textbf{c}: $|g^{(1)}(\Delta r, \Delta t) |$ in the $(\Delta r, \Delta t)$ plane.
    \textbf{d}-\textbf{e}: Same as panels \textbf{a}-\textbf{b} after filtering the dispersive branches.
    \textbf{f}: Same as \textbf{c} after filtering the dispersive branches. On both \textbf{c} and \textbf{f}, the red solid line corresponds to the KPZ scaling function $F = C_{0} F_{\mathrm{KPZ}} ( y_0 \Delta r/\Delta t^{1/z})$, adjusted to the experimental data by tuning the values of $C_{0}$ and $y_{0}$. In \textbf{c} and \textbf{f}, we set the renormalization $\kappa$ to $1.9$ and $1.8$, respectively, and $P/P_{\mathrm{trp}} = 0.98$. 
    All the data points are measured for $\hbar \delta = - 6.8~\mathrm{meV}$.
    }
    \label{fig:sup_map_AD_KPZ2}
\end{figure}

Figure~\ref{fig:sup_map_AD_KPZ2} presents the complete analysis of the spatiotemporal coherence decay for an independent dataset in the KPZ regime ($\hbar\delta=-6.8~\mathrm{meV}$, exciton fraction $20 \, \%$). As in the main text, we first analyze the temporal and spatial decays of $|g^{(1)}(0,\Delta t)|$ (\textbf{a}) and $|g^{(1)}(\Delta r,0)|$ (\textbf{b}), plotted as functions of $\Delta t^{2\beta}$ and $\Delta r^{2\chi}$, respectively, for which KPZ scaling is expected to appear as straight lines. The gray shaded regions identify the fitting windows used to extract the scaling exponents by stretched-exponential fits. The resulting values of $\beta_{\mathrm{exp}}$ and $\chi_{\mathrm{exp}}$ shown in the insets and are in good agreement with the KPZ predictions (dashed lines). At low pump power, the predominance of the photoluminescence emission at short $\Delta r$ prevents a reliable determination of the spatial exponent. We notice that the extracted growth exponent is systematically larger than the KPZ value, $\beta=0.24$, owing to weak oscillations of the temporal coherence over the fitting window, visible in the lowest-power data. These oscillations, which may originate from residual beating between two transverse polarization modes, are too weak to be filtered but nevertheless slightly distort the decay, leading to the observed bias in the fitted exponent. To probe the full spatio-temporal coherence dynamics, Fig.~\ref{fig:sup_map_AD_KPZ2}\textbf{c} tests the collapse of the raw data onto the universal KPZ scaling function (red solid curve) by plotting $-2\log \! \left(\kappa |g^{(1)}(\Delta r,\Delta t)|\right)/\Delta t^{2\beta}$ as a function of $\Delta r/\Delta t^{1/z}$ for all points within the non-hatched region of the $(\Delta r,\Delta t)$ plane (inset). As in the main text, the dispersive branches induce oscillations in the collapse together with an overall upward shift of the tilted asymptote relative to the KPZ prediction. Although the tilted branch appears more scattered than in Figs.~\ref{fig:sup_map_AD_collapse_KPZ}$\boldsymbol{\mathrm{a_{1}}}$-$\boldsymbol{\mathrm{b_{1}}}$, nearly all points lying above the red ellipse originate from the red-shaded region highlighted in the inset, where the contribution of the dispersive branches is strongest. In contrast, the collapse onto the horizontal asymptote remains excellent. 

\medskip

Figures~\ref{fig:sup_map_AD_KPZ2}\textbf{d}-\textbf{f} repeats the analysis presented in panels \textbf{a}-\textbf{c} after filtering out the dispersive branches. While the temporal coherence decay is only weakly affected, the spatial coherence changes substantially. The filtering suppresses the contribution of the emission from uncondensed polaritons, whose coherence otherwise masks that of the condensate. This filtering results in an upward shift of the spatial decay shown in Fig.~\ref{fig:sup_map_AD_KPZ2}\textbf{b}, and in an extension of the KPZ scaling window. It further allows fitting the spatial decay at lower powers, providing additional measurements of $\chi_{\mathrm{exp}}$ compared to Fig.~\ref{fig:sup_map_AD_KPZ2}\textbf{b}. Overall, the extracted exponents show good agreement with the KPZ predictions. Regarding the data collapse, filtering removes both the oscillatory deviations and the upward shift of the tilted asymptote observed in Fig.~\ref{fig:sup_map_AD_KPZ2}\textbf{c}. The crossover from the horizontal to the tilted asymptote is, however, smoother than that predicted by the KPZ scaling function. Nevertheless, all data points collapse onto a single, well-defined curve, demonstrating the emergence of spatio-temporal scaling governed by KPZ exponents.


\clearpage

\section{\label{sec:supmat_Numerics_method}Numerical Simulations - Method and Parameters}

\subsection{\label{sec:supmat_Numerics_method_parameters}Numerical scheme and parameters}

The generalized Gross-Pitaevskii equation (gGPE)~\eqref{eq:supmat_gGPE} is solved with periodic boundary conditions on a symmetric square lattice of parameter $dx = dy$ and surface $L_x \times L_y$. The numerical integration is performed relying on two numerical schemes, used to propagate in time its deterministic and stochastic contributions respectively. We denote $dt$  the numerical timestep. The deterministic part of Eq.~\eqref{eq:supmat_gGPE} is solved in Fourier space with a symmetrized split-step method. The stochastic contribution to Eq.~\eqref{eq:supmat_gGPE} is implemented with an Euler-Maruyama algorithm, adding to the deterministic solution, at every point in space and time, a complex random variable with Gaussian statistics of variance $2\sigma dt/dx^2$. The exciton reservoir dynamics~\ref{eq:supmat_xreservoir} is simply integrated  with a first-order Euler scheme.

\medskip

As detailed in Sec.~\ref{sec:supmat_model}, the polariton dispersion relation is approximated close to the bottom of the LPB by the parabola $\epsilon(\hat{\boldsymbol{k}}) = \hbar^2 \boldsymbol{\hat{k}}^2/(2m)$ with $m$ the polariton mass. Similarly, the $k$-dependent effective loss rate $\gamma(\hat{\boldsymbol{k}})$ at the bottom of the LPB is $\gamma(\hat{\boldsymbol{k}}) = \gamma_0 + \gamma_2 \boldsymbol{\hat{k}}^2$. We use the spatially dependent pump profile
\begin{equation}
    P(\boldsymbol{r}) = P \dfrac{\left( 1 + \tanh\left[ (R_0 + |\boldsymbol{r}|)/\sigma \right] \right)\left( 1 + \tanh\left[ (R_0 - |\boldsymbol{r}|)/\sigma \right] \right)}{\left(1 + \tanh\left[ R_0 \right] \right)^2} \, ,
\end{equation}
\noindent
where $R_0 = 25~{\rm\mu m}$ and $\sigma = 5~{\rm \mu m}$, which well approximates the experimental spatial profile of both the pump and the resulting condensate emission intensity. The pumping rate is fixed with respect to the mean-field transparency threshold of the spatially homogeneous system $P_{\rm trp} = \gamma_0\gamma_R/R$ through $P = p_{\rm trp} P_{\rm trp}$. Hence $p_{\rm trp}=1$ corresponds to the perfect compensation of gain and loss (see Sec.~\ref{sec:supmat_condensation_vs_transparency}). When the pump profile $P(\boldsymbol{r})$ is inhomogeneous in space, the mean-field threshold $P_{\rm trp}$ is increased with respect to $\gamma_0\gamma_R/R$ \cite{Bobrovska2014}. Its dependence on the microscopic parameters can be obtained from the steady state solution of Eqs.~\eqref{eq:supmat_gGPE}-\eqref{eq:supmat_xreservoir} with $\sigma=0$ \cite{Ostrovskaya2012}. In numerical simulations, the transparency threshold $P_{\rm trp}$ is further modified by fluctuations induced by the Gaussian noise $\xi(\boldsymbol{r},t)$ in Eq.~\eqref{eq:supmat_gGPE}. In the following, the overall shift is effectively absorbed into the control parameter $p_{\rm trp}$, so that gains compensate losses for $p_{\rm trp} > 1$ instead.

\medskip

The polariton-polariton interaction strength $g$ is approximated by $g\approx0$, its effect being negligible compared to the interaction between reservoir excitons and polaritons $g_R$ \cite{baboux2018}. The latter is fixed from the condensate blueshift through $g_R = \mu/(2 n_R)$. Above threshold and in the center of the pump profile, one finds $n_R\approx\gamma_0/R$. The parameters in both EW and KPZ regimes are summarized in Table~\ref{tab:supmat_params} below.

\begin{table}[h]
\centering
\small
\renewcommand{\arraystretch}{1.4}
\setlength{\tabcolsep}{4pt}

\resizebox{\linewidth}{!}{%
\begin{tabular}{c|c|c|c|c|c|c|c|}
\cline{2-8}
 & $m\,(m_e)$
 & $\hbar\gamma_0\,(\mu\mathrm{eV})$
 & $\hbar\gamma_2\,(\mu\mathrm{eV}\,\mu\mathrm{m}^2)$
 & $1/\gamma_R\,(\mathrm{ps})$ \cite{Deng_2003}
 & $R\,(\mu\mathrm{m}^2/\mathrm{ps})$
 & $\mu_{\mathrm{th}}\,(\mathrm{meV})$
 & $p_{\rm trp}$ \\
\hline

\multicolumn{1}{|c|}{EW}
& $4.7\times10^{-5}$
& $30$
& $150$
& $500$
& $1.3\times10^{-4}$
& $1.1$
& $1.3$ \\
\hline

\multicolumn{1}{|c|}{KPZ}
& $4\times10^{-5}$
& $20$
& $300$
& $500$
& $1.3\times10^{-3}$
& $4$
& $1.32$ \\
\hline

\end{tabular}%
}

\caption{\label{tab:supmat_params}
Parameters used in the EW and KPZ regimes, reproducing the experimental results of the main text. Numerical results are displayed in Fig.~\ref{fig:supmat_theory_EW_KPZ}.}
\end{table}

\noindent Additionally to the effective $k$-dependent loss rate $\gamma(\boldsymbol{\hat{k}})$ of Eq.~\eqref{eq:supmat_gGPE}, an $r$-dependent loss rate $\gamma_{e}(\boldsymbol{r})$ is added on the edges of the square lattice, outside of the pumping spot. Its value is set to $100\gamma_0$ on the outermost sites and decreases exponentially to $0$ over $5~{\rm \mu m}$ inward. This loss rate, defined outside of the condensed spot, mimics open boundary conditions, and dissipates plane waves emitted outside of the condensate \cite{Ohadi_PRX2016}. It allows to reduce the system size $L_x\times L_y$ around the pump profile $P(\boldsymbol{r})$ down to $L_x=L_y=40~{\rm \mu m}$ while preventing the condensate from self-interacting because of periodic boundary conditions. We observe that emitted plane waves are better damped using an exponential profile for $\gamma_{e}(\boldsymbol{r})$, while a hard-wall dissipation acts as a complex potential barrier, yielding reflection and interferences.

\medskip

The first-order correlation function is computed according to Eq.~\eqref{eq:supmat_g1_def} from approximately $2\times10^4$ independent realizations of the condensate dynamics, parallelized over typically 128 CPUs. Numerical simulations are performed on a spatial grid $dx=dy = 1~{\rm \mu m}$ with a time-step $dt\approx 10^{-2}~{\rm ps}$. Temporal correlations are sampled logarithmically, with up to $10^3$ points per decade. Density correlations $g^{(1)}_n$ and phase correlations $\langle e^{i\Delta \theta} \rangle$ are computed with the same conditions.

\newpage

\subsection{\label{sec:supmat_Numerics_method_vortex_tracking}Vortex tracking}

Vortices and antivortices are topological defects in the condensate phase $\theta(\boldsymbol{r},t)$ associated with a quantized $\pm2\pi$ charge. Their proliferation in two-dimensional polariton condensates have already been studied in numerical simulations, which have characterized their average number under various conditions \cite{Caputo2018_BKT_transition_incoherent_EPBEC, Comaron2021}, their nucleation under modulation instability \cite{Marzena_2021_first_order}, or the motion of spiral vortices in the absence of noise \cite{Gladilin_multivortex_state_2019}. 

\medskip

Vortices are also present in our simulations. In order to detect them and track their trajectories, we perform numerical simulations setting $dx=0.25~{\mu m}$, $dt\approx 10^{-2}~{\rm ps}$. The phase profile is extracted in space and time within a circle of radius $25~{\rm \mu m}$ corresponding to the size of the pumping spot $R_0$. This phase profile is smoothed out with a Gaussian filter of chosen width $\sigma_f$. We then compute the circulation $\mathfrak{C}$ for the stationary phase heat map at every time instant $t$. The circulation is defined according to $\mathfrak{C}(\boldsymbol{r},t) = \oint_{\mathcal{C}} \boldsymbol{\nabla}\theta(\boldsymbol{r},t)\cdot d\boldsymbol{\ell}$ where $\mathcal{C}$ is a small spatial contour defined around each pixel of the spatial grid. The location of vortices (respectively antivortices) is obtained identifying the $\pm 2\pi$ peaks in the circulation heat map $\mathfrak{C}(\boldsymbol{r},t)$ at each time $t$. Finally, vortex (respectively antivortex) trajectories are constructed by connecting together vortex (respectively antivortex) coordinates at one timestep to that of the vortex (respectively antivortex) lying the closest and in the immediate neighborhood at the next timestep.

\medskip

\begin{figure}[h!]
    \centering
    \includegraphics[scale=0.25]{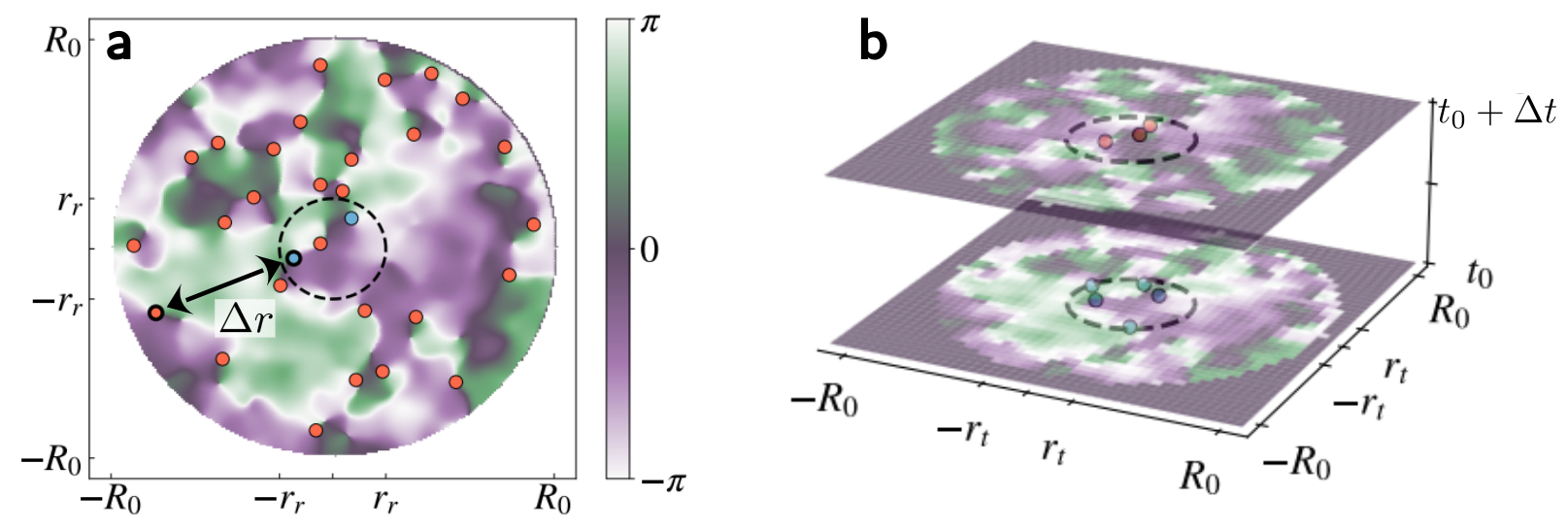}
    \caption{
    \textbf{Schematics of the vortex-antivortex pair correlation computation} \textbf{a} in space, \textbf{b} in time.}
    \label{fig:sup_map_schematics_pair_correlation}
\end{figure}

The correlation between vortices and antivortices can be probed in space and time through the pair correlation functions $g^{(2)}_{+-}(\Delta r)$, $g^{(2)}_{+-}(\Delta t)$. They are defined as in  Refs.~\cite{Carusotto_Castin_PRA2007_Semiclassical_field_method, Foster_Davis_PRA2010_Vortex_pairing_2dBose_gas} according to:
\begin{equation}
    g^{(2)}_{+-}(\Delta \boldsymbol{r}) = \dfrac{\big\langle \rho_+( \boldsymbol{r_{0}}, t_0)\rho_-(\Delta \boldsymbol{r} + \boldsymbol{r_{0}},t_0) \big\rangle_{r_r}}{\big\langle \rho_+(\boldsymbol{r_{0}}, t_0) \big\rangle_{r_r}\big\langle \rho_-(\Delta \boldsymbol{r} + \boldsymbol{r_{0}},t_0) \rangle_{r_r}},
\end{equation}
\begin{equation}
    g^{(2)}_{+-}(\Delta t) = \dfrac{\big\langle \rho_+( \boldsymbol{r_{0}},t_0)\rho_-(\boldsymbol{r_{0}},\Delta t + t_0) \big\rangle_{r_t}}{\big\langle \rho_+(\boldsymbol{r_{0}},t_0) \big\rangle_{r_t}\big\langle \rho_-(\boldsymbol{r_{0}}, \Delta t + t_0) \big\rangle_{r_t}},
\end{equation}
\noindent
where $\rho_{\pm}(\boldsymbol{r},t) = \overset{N_{\pm}}{\underset{i_{\pm}}{\sum}}\delta(\boldsymbol{r}-\boldsymbol{r}_{i_{\pm}})\delta(t-t_{i_{\pm}})$ is the vortex ($'+'$) or antivortex ($'-'$) density. The coordinates $(\boldsymbol{r}_{i_{\pm}}, t_{i_{\pm}})$ denote the position in space and time of the $i$-th vortex (antivortex). The averages $\langle \cdot \rangle_{r_{r,t}}$ are taken over independent noise realizations and over vortices whose positions $\boldsymbol{r_0}$ lie within a disk of radius $r_{r,t}$ centered on the condensate. We chose $r_r=2~{\rm \mu m}$ and $r_t=3~{\rm \mu m}$ so as to accumulate statistics over regions in which the condensate density remains approximately homogeneous. The vortex-antivortex pair correlation function $g^{(2)}_{+-}(\Delta r, \Delta t)$ represents the probability of finding, for each vortex within $r_{r,t}$, an anti-vortex at a relative distance $\Delta r$ (at the same absolute time) or time delay $\Delta t$ (within the circle of radius $r_t$). This is illustrated in Fig.~\ref{fig:sup_map_schematics_pair_correlation}. Similarly, we also define the vortex-vortex correlation functions $g_{++}(\Delta\boldsymbol{r})$ and $g_{++}(\Delta t)$. Pair correlation functions are expected to reach a plateau $g^{(2)}_{+\mp}\approx1$ as $\Delta t,\;\Delta r\to+\infty$. This illustrates that pairs are not correlated anymore at large temporal or spatial delays

\clearpage

\section{\label{sec:supmat_Numerics_discussion}Numerical simulations - Discussion}

\subsection{\label{sec:supmat_condensate_stability}Stability of the condensate}

Modulation instability fragments a homogeneous condensate into smaller mutually incoherent domains. Within Bogoliubov theory, it is signaled by a positive imaginary part of the excitation spectrum \cite{Wouters2007}. At short momenta, this is realized for a negative effective phase viscosity, $\nu<0$. From Eq.~\eqref{eq:supmat_effective_KPZ_params}, one gets:

\begin{equation}
    \nu = \frac{\gamma_2}{2} + \frac{g_{e}}{2mg_{i}} \, ,
\end{equation}

\noindent
so that its sign results from the competition between the dissipative contribution ($\gamma_2$) and the interaction term ($g_e/m$). Since the effective interaction strength is typically negative ($g_e<0$), owing to the coupling between the condensate and the exciton reservoir, negative-mass polaritons realized on a lattice of micropillars yield $\nu>0$ and therefore a stable condensate \cite{baboux2018}. By contrast, in planar microcavities ($m>0$), a condensate formed under a spatially homogeneous drive $P(\boldsymbol{r})=P$ can be unstable. This is the generic situation, when the $\gamma_2$ contribution is negligible compared to the interaction term. In our simulations, $\gamma_2/2 \approx 10^{-4} |g_{e}| / (2mg_{i})$, so that a spatially homogeneous drive yields $\nu<0$ and instability for any pumping rate above the transparency threshold $P_{\rm trp}$.

\medskip

Stable condensates can nevertheless be obtained with spatially inhomogeneous pump profiles. This has been reported from numerical simulations in both one and two-dimensions \cite{Bobrovska2014, Keeling2008}. In particular, modulation instability is recovered at fixed pumping rate as the pump profile is enlarged \cite{Bobrovska2014}. Similarly, increasing the pumping rate destabilizes the condensate once the Thomas–Fermi radius of the polariton cloud becomes smaller than the pump width \cite{Keeling2008}. Consequently, reducing the pump size broadens the range of pumping rates over which the condensate remains stable.

\medskip

The same behavior is observed in both our numerical simulations and experiments, as modulation instability is systematically found at sufficiently large pumping rates (see Sec.~\ref{sec:supmat_phase_diagram} and Sec.~\ref{sec:supmat_robustness_EW_KPZ}). Numerically, we also find that increasing the pump width $R_0$ enhances the condensate instability. Consequently, a wider pump profile reduces the parameter window to explore the emergence of EW/KPZ universal features in terms of pumping rate. Within the framework of the mapping in Eq.~\eqref{eq:supmap_KPZ_mapping}, the stabilization induced by a finite-size pump can be interpreted as a renormalization of the effective parameters Eq.~\eqref{eq:supmat_effective_KPZ_params}. In particular, stabilization can be viewed as a change in sign of the effective phase viscosity, from negative $\nu<0$ to positive $\nu_e>0$. The value of the effective viscosity $\nu_e$ is estimated from numerical results in Sec.~\ref{sec:supmat_accessing_EW_KPZ} below.

\newpage

\subsection{\label{sec:supmat_accessing_EW_KPZ}Accessing KPZ and EW regimes}

\begin{figure}[h!]
    \centering
    \includegraphics[width=\linewidth]{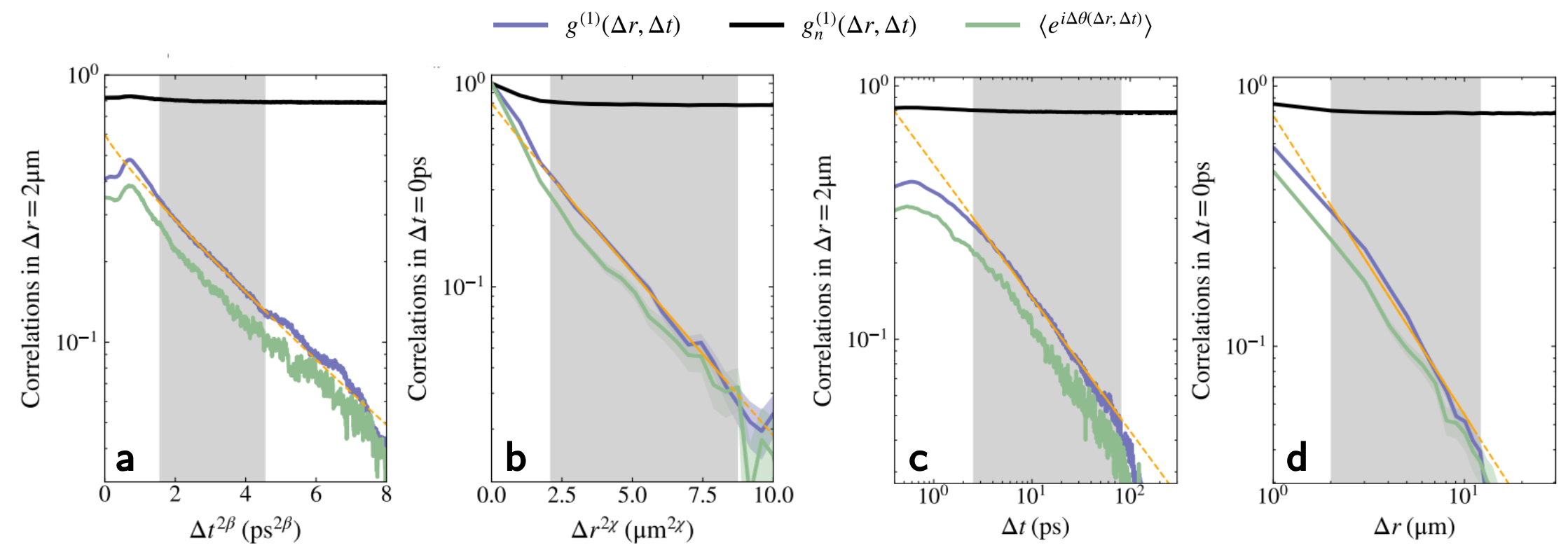}
    \caption{\textbf{Decay of the first-order coherence, phase, and density correlations in the KPZ (resp. EW) regime reproducing experimental results} as a function of \textbf{a} time and \textbf{b} space (resp. \textbf{c} and \textbf{d}). The orange lines indicate stretched exponential (resp. algebraic) fits performed in the universal scaling window (gray shades). In the KPZ regime, we obtain $\beta = 0.21\pm0.03$ and $\chi = 0.38\pm0.03$. In the EW regime, the fitted exponents are $a_s = 1.15\pm0.05$ and $a_t=0.5\pm3\times10^{-3}$, yielding the ratio $a_s/a_t=2.18\pm0.09$.
      }
    \label{fig:supmat_theory_EW_KPZ}
\end{figure}

In this section, we discuss the emergence of KPZ and EW regimes in numerical simulations and show that experimental results of Figs.~\ref{fig:scalingKPZ} and \ref{fig:scalingEW} of the main text are closely reproduced. The first-order correlation function $g^{(1)}(\Delta \textbf{r}, \Delta t)$ is computed according to Eq.~\eqref{eq:supmat_g1_def} using the parameters summarized in Tab.~\ref{tab:supmat_params} and close to experimental estimates. In the KPZ regime, we observe stretched exponential decays in both space and time (Figs.~\ref{fig:supmat_theory_EW_KPZ}.\textbf{a} and \textbf{b}). Fitting these decays over the same ranges as in Fig.~\ref{fig:scalingKPZ} yields universal exponents consistent with literature values, $\beta \approx 0.24$ and $\chi\approx 0.39$ \cite{Pagnani2015}. In the EW regime, the condensate coherence instead follows algebraic decays in both space and time, as shown in Figs.~\ref{fig:supmat_theory_EW_KPZ}.\textbf{c}-\textbf{d}. Fits performed over the ranges indicated by gray shades in Fig.~\ref{fig:scalingEW} confirm that $a_s>1/4$, while the ratio of exponents is close to $a_s/a_t=2$. Simulations also capture the long-time Schawlow--Townes exponential decay, which sets in beyond the universal temporal KPZ/EW scaling window \cite{Helluin2025} (see Sec.~\ref{sec:supmat_properties_EPBEC_coherence}). This long-time exponential decay is highlighted in the inset of Figs.~\ref{fig:sup_map_KPZ_EW_p}.\textbf{a}-\textbf{c} hereafter. We further compute phase and density correlations in both regimes, thereby providing a numerical benchmark for the approximations underlying the mapping of Eq.~\eqref{eq:supmat_g1_decoupling_density_phase}. Specifically, the results of Fig.~\ref{fig:supmat_theory_EW_KPZ} confirm that the coherence decay is governed by universal phase fluctuations over the identified scaling windows, in agreement with Eq.~\eqref{eq:supmat_g1_decoupling_density_phase}. In contrast, density correlations only contribute to the condensate coherence at short spatiotemporal scales, \textit{i.e.} before the onset of the EW/KPZ scaling regimes (see Sec.~\ref{sec:supmat_pair_correlation}). Hence, this analysis shows that all the assumptions underlying the mapping to the KPZ equation are satisfied, providing further support to the attribution of the coherence behavior to the KPZ and EW universality classes.

\begin{figure}[h!]
    \centering
    \includegraphics[width=\linewidth]{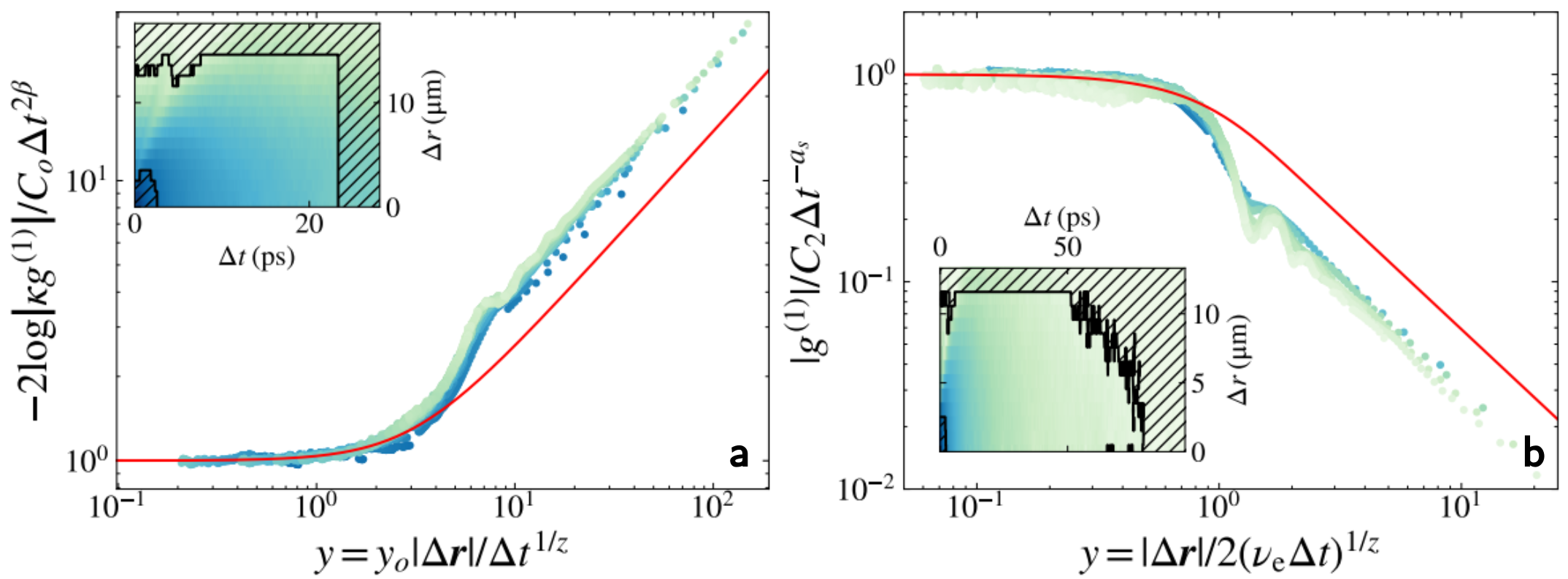}
    \caption{
    \textbf{Collapse of the first-order coherence} in the \textbf{a} KPZ with $\kappa\approx1.7$ ($6\times10^3$ data points) and \textbf{b} EW regimes ($8\times10^3$ data points). The space-time heat maps of the first-order correlation function are shown in insets. The red curves correspond to the KPZ and EW scaling function, $F_{\rm KPZ}(y)$ and $F_{\rm EW}(y;\;p_{\rm max})$ respectively.}
    \label{fig:sup_map_collapse_KPZ_EW}
\end{figure}

We now discuss the collapse of the first-order coherence onto a universal function in both the EW and the KPZ regime. In the latter, where phase correlations follow Eq.~\eqref{eq:supmat_KPZscaling}, the space-time heat map of the $g^{(1)}$ function can be made to collapse onto the KPZ universal function according to:
\begin{equation}
    -2\log|\kappa \;g^{(1)}(\Delta\boldsymbol{r},\Delta t)|/C_0 \Delta t^{2\beta} = F\left( y_0|\Delta \boldsymbol{r}|/\Delta t^{1/z} \right) \, . \label{eq:sup_mat_collapse_KPZ}
\end{equation}
\noindent As in Fig.~\ref{fig:sup_map_normalization}, the normalization factor $\kappa$ accounts for the fact that the KPZ scaling does not extend down to $\Delta t=0$, $\Delta r = 0$. Indeed, at short spatiotemporal scales, the decay of the first-order coherence is instead affected by density correlations, yielding a short-range photoluminescence peak (see Eq.\ref{eq:supmat_g1_decoupling_density_phase} and Sec.~\ref{sec:supmat_pair_correlation}). The collapse of Eq.~\eqref{eq:sup_mat_collapse_KPZ} is shown in Fig.~\ref{fig:sup_map_collapse_KPZ_EW}.\textbf{a}. The constant offset from the theoretical curve at large $y_0|\Delta \boldsymbol{r}|/\Delta t^{1/z}$ originates from the dispersive branches reported in Fig.~\ref{fig:condensation}.\textbf{c} of the main text, which are also reproduced in our numerical simulations (inset of Fig.~\ref{fig:sup_map_collapse_KPZ_EW}.\textbf{a}).

\medskip

Numerical data of the first-order coherence can be collapsed in the EW regime as well. The EW scaling function Eq.~\eqref{eq:supmat_EW_function} is computed numerically using a time-independent cutoff $p_{\rm max}=10^3$, such that 
\begin{equation}
    F_{\rm EW}\left( \frac{|\Delta \boldsymbol{r}|}{2\sqrt{\nu_e \Delta t}}; \, 2\Lambda\sqrt{\nu_e \Delta t}\right) = F_{\rm EW}\left(  \frac{|\Delta \boldsymbol{r}|}{2\sqrt{\nu_e \Delta t}} ; \, p_{\rm max}\right) + \log\left( \Lambda \sqrt{\nu_e \Delta t} \right) + C_1 \, ,
\end{equation}
\noindent with $C_1$ a non-universal constant chosen such that $F_{\rm EW}\left(  \frac{|\Delta \boldsymbol{r}|}{2\sqrt{\nu_e \Delta t}}=0 ; \, p_{\rm max}\right) = 0$, and $\nu_e$ the effective phase viscosity (Sec.~\ref{sec:supmat_condensate_stability}). Consequently, the first-order coherence is collapsed according to:
\begin{equation}
    g^{(1)}\left(\Delta \boldsymbol{r}, \Delta t \right)/C_2 \Delta t^{-a_t} = \exp\left[-a_s F_{\rm EW}\left( \frac{|\Delta \boldsymbol{r}|}{2\sqrt{\nu_e \Delta t}}; \, p_{\rm max}\right)\right] \, ,
\end{equation}
\noindent with $C_2$ a non-universal constant. The EW data collapse is displayed in Fig~\ref{fig:sup_map_collapse_KPZ_EW}.\textbf{b}, where dispersive branches yield again a constant offset from the power law decay at large $ {|\Delta \boldsymbol{r}|}/{2\sqrt{\nu_e \Delta t}}$. The horizontal adjustement of the data collapse allows one to extract the effective viscosity $\nu_e \approx 4~{\rm\mu m^2.ps^{-1}}$. The effective noise strength $D_e$ can also be estimated, using the fitted value for the spatial exponent $a_s$, giving $D_e \approx 29~{\rm\mu m^2.ps^{-1}}$.

\newpage

\subsection{\label{sec:supmat_robustness_EW_KPZ}Robustness of the KPZ and EW regimes}

\begin{figure}[h!]
    \centering
    \includegraphics[width=\linewidth]{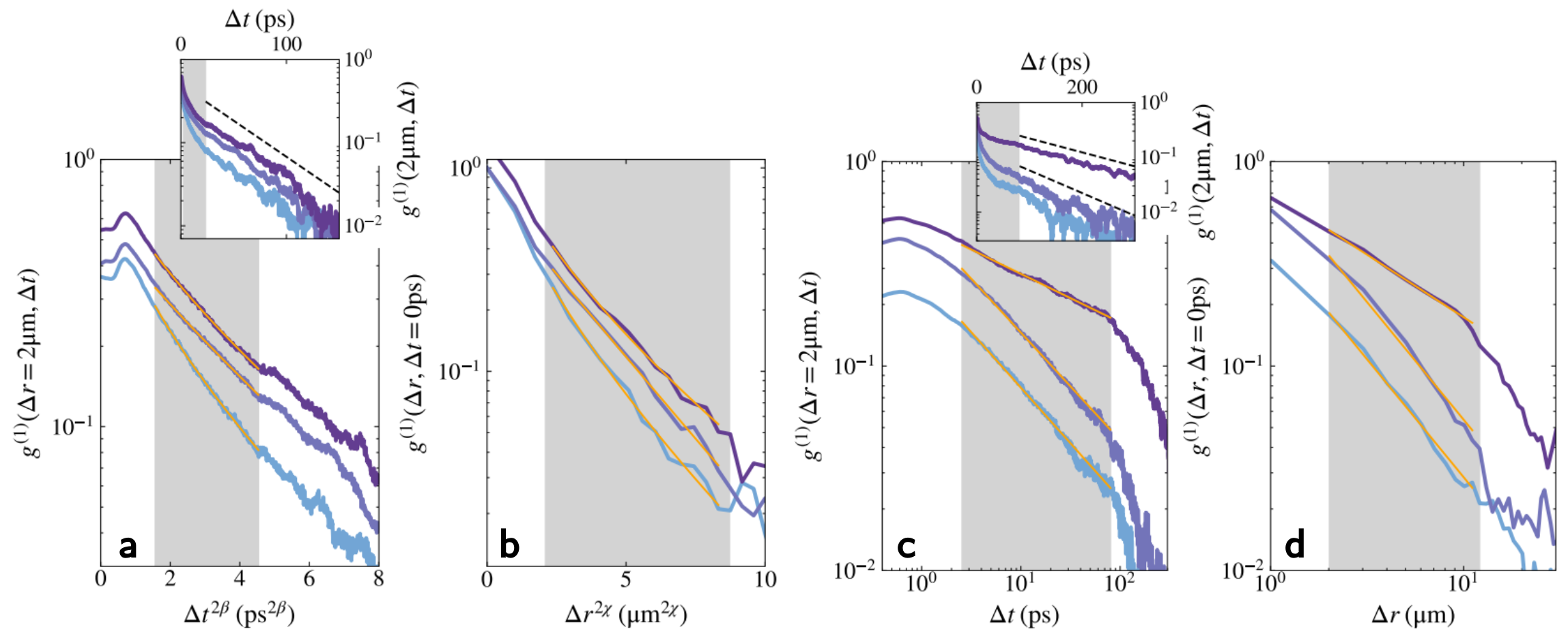}
    \caption{
    \textbf{Decay of the first-order coherence in the KPZ (resp. EW) regime for various pumping rates} as a function of \textbf{a} time and \textbf{b} space (resp. \textbf{c} and \textbf{d}). The different shades from light blue to purple correspond to increasing values of $p_{\rm trp}=\{1.05, 1.32, 1.37\}$ (resp. $p_{\rm trp}=\{1.1, 1.3, 1.5\}$). Insets: temporal decay of the first-order coherence on a semilog scale, higlighting the long-time exponential decay.}
    \label{fig:sup_map_KPZ_EW_p}
\end{figure}

Let us now comment on the robustness of our numerical observations against variations of the microscopic parameters, with respect to those of Table~\ref{tab:supmat_params}. We found that both KPZ and EW regimes are robust against variations of the pumping rate $p_{\rm trp}$, up to the onset of modulation instability. First-order correlation functions in the KPZ regime and for several pumping rates are displayed in Fig.~\ref{fig:sup_map_KPZ_EW_p}.\textbf{a}-\textbf{b}, and closely reproduce experimental observations. In this parameter regime, modulation instability is reached in the numerics for $p_{\rm trp}\gtrsim1.4$. At large times, beyond the temporal universal KPZ scaling window, the first-order coherence decays exponentially \cite{Helluin2025, Keeling2010, Amelio2024}. This Schawlow--Townes-like decay \cite{Schawlow1958} reproduces experimental observations (Fig.~\ref{fig:condensation}.\textbf{d} of the main text), and is evidenced in the inset of Fig.\ref{fig:sup_map_KPZ_EW_p}.\textbf{a}.

\medskip

Similar results, obtained in the EW regime, are shown in Fig.~\ref{fig:sup_map_KPZ_EW_p}.\textbf{c} and \textbf{d}, where the trend observed experimentally is again captured. There, modulation instability is observed in the numerics for $p_{\rm trp}\gtrsim1.6$. A Schawlow--Townes regime is also reported at late times, and evidenced in the inset of Fig.\ref{fig:sup_map_KPZ_EW_p}.\textbf{c}.
For every first-order correlation displayed in Fig.\ref{fig:sup_map_KPZ_EW_p}, we verified that the decay of the $g^{(1)}$ function is dominated by phase correlations, as in Fig.~\ref{fig:supmat_theory_EW_KPZ}. In addition, we checked that density correlations $g_n^{(1)}$ affect the condensate coherence over short space-time scales only, before KPZ or EW regimes set in (see also Fig.~\ref{fig:sup_map_defect_correlation}).

We note that, from Table~\ref{tab:supmat_params}, the KPZ regime is accessed starting from the EW regime by simultaneously increasing the interaction strength $g_R$ via $\mu$ \cite{Helluin2025}, and the loss rate $\gamma_2$. The EW regime is consistently recovered upon decreasing $\gamma_2$. As expected, the dispersive branches observed in the heat map of the $g^{(1)}$ (insets of Figs.~\ref{fig:sup_map_collapse_KPZ_EW}.\textbf{a} and \textbf{b}) are more contrasted for decreasing values of $\gamma_2$. Furthermore, our results are robust when a weak polariton–polariton interaction strength is introduced, which has been checked up to $g \approx 0.1 g_R$.

\begin{figure}[t!]
    \centering
    \includegraphics[scale=0.18]{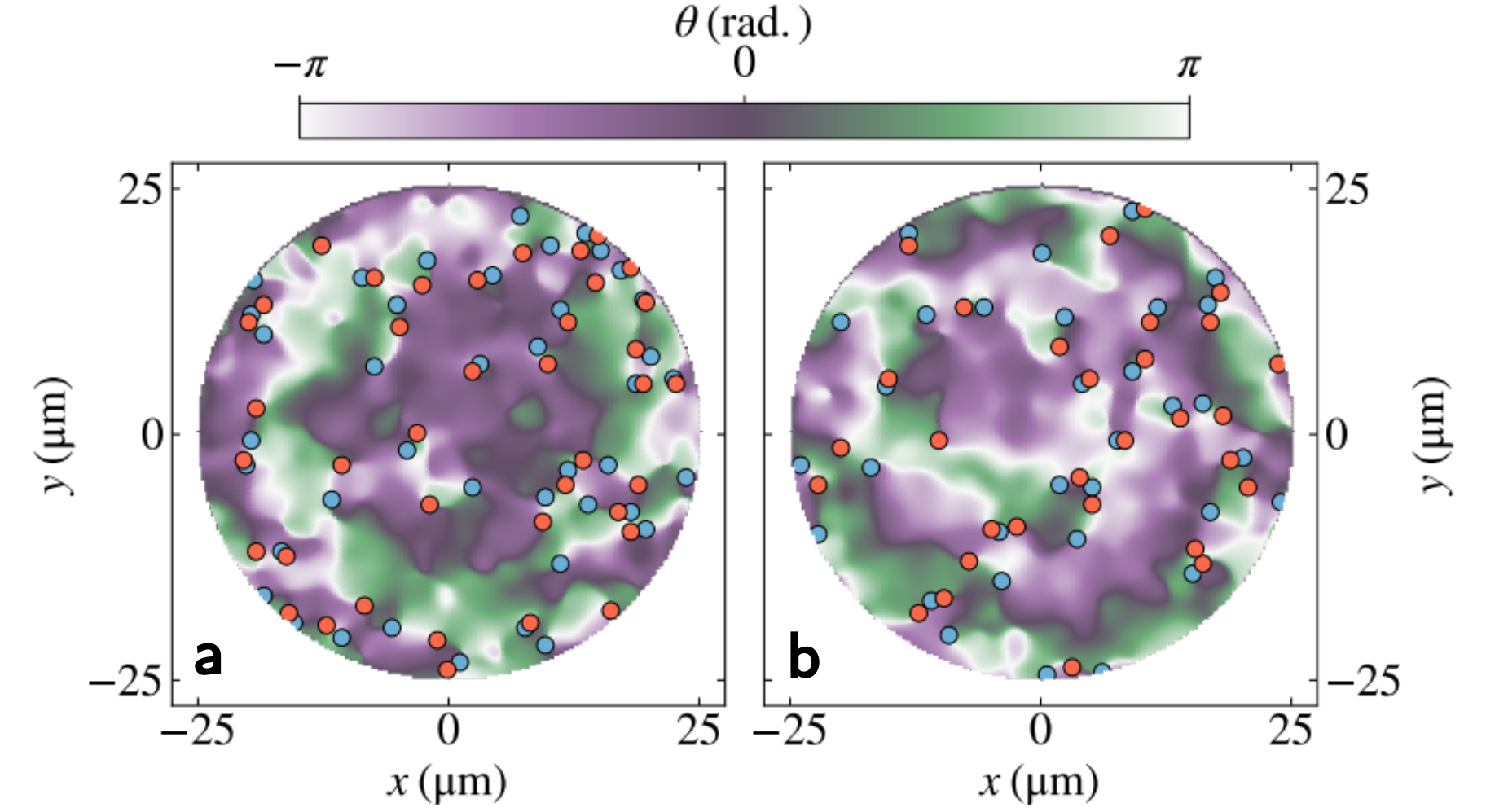}
    \caption{
    \textbf{Typical instantaneous phase map} in \textbf{a} the KPZ and \textbf{b} the EW regime. Vortices and antivortices are depicted by red and blue disks respectively.}
    \label{fig:sup_map_vortex_map}
\end{figure}

Finally, numerical simulations are also used to probe the presence of topological defects in the condensate phase. Typical phase heat maps in the KPZ and EW regimes are displayed in Figs.~\ref{fig:sup_map_vortex_map}.\textbf{a}-\textbf{b} respectively, where vortices and antivortices are represented by the blue and red dots. This observation is perfectly consistent with former numerical investigations, which reported a proliferation of topological defects close to threshold \cite{Caputo2018, Comaron2021}. In Sec.~\ref{sec:supmat_Nature_vortices}, it is shown that although vortices are abundant, they remain predominantly bound in pairs which do not affect the condensate macroscopic properties.
Specifically, correlation functions such as $g^{(1)}(\Delta \boldsymbol{r}, \Delta t) $ become progressively insensitive to short-scale structures that do not control their asymptotic behavior, as the separation increases. Consequently, the contribution of bound vortex pairs to the first-order correlation function is averaged out in the large-scale behavior of $ g^{(1)}(\Delta \boldsymbol{r}, \Delta t) \approx g^{(1)}_{n}(\Delta \boldsymbol{r}, \Delta t) \big\langle e^{i\Delta\theta(\Delta \boldsymbol{r}, \Delta t)} \big\rangle$. We note however that local observables such as the distribution of phase fluctuations remain sensitive to the presence vortices \cite{Fontaine2022, deligiannis2022}, preventing the identification of the full counting statistics from the unwrapped phase in current numerical datasets.

\clearpage

\subsection{\label{sec:supmat_Nature_vortices}Nature of the vortices}

In this section, we show that the EW and KPZ scaling behaviors are found in  a quasi-ordered regime, in which topological defects of opposite charges are bound in pairs. These defects therefore differ from the so-called spiral vortices of the two-dimensional complex Ginzburg Landau equation \cite{Aranson_RMP2002_world_CGLE, Chate_PhysicaA1996_phase_diagram_2DCGLE}, identified in former theoretical works as the free vortices of a non-equilibrium analog of a Berezinski--Kosterlitz--Thouless (BKT) disordered phase for polariton condensates \cite{Altman_2016_EM_duality, Sieberer_PRL2018_defects_anisotropic_driven-open_sys}. Spiral-vortices have a large core and long lifetimes \cite{Chate_PhysicaA1996_phase_diagram_2DCGLE}. Moreover, numerical simulations have shown that, in this disordered regime, density and phase fluctuations remain coupled over extended scales and destroy the condensate coherence \cite{Helluin2025}. By contrast, we argue that the topological defects observed in our simulations are nucleated by the noise $\xi$. Although they do affect the condensate coherence at short scales, they do not disrupt phase ordering at larger scales. Specifically, vortex dynamics plays no role at the scales where KPZ/EW regimes emerge. This is supported by the following observations:

\begin{itemize}

    \item Defects are predominantly paired in vortex-antivortex dipoles (Sec.~\ref{sec:supmat_vortex_clustering}). This microscopic arrangement of topological charges suggests that the phase of the condensate is quasi-ordered.
    
    \item Only a small fraction of vortices survive spatial coarse-graining (Sec.~\ref{sec:supmat_coarse_graining}). This indicates that, by analogy with the quasi-ordered BKT regime, vortex pairs can be averaged out at large spatial scales, where they do not affect the condensate coherence.
    
    \item Defects are very short-lived, with free defects being even more short-lived (Sec.~\ref{sec:supmat_lifetime}). This highlights the scale separation between the low energy (universal) phase fluctuations and the highly energetic vortex excitations, which do not affect the coherence at large temporal scales.
    
    \item Correlations between defect pairs are restricted to short space–time scales (Sec.~\ref{sec:supmat_pair_correlation}). Opposite charge defects exhibit a short scale attraction, while short scale repulsion is observed for same-sign defects. This confirms that defects are nucleated in pair of opposite charge by small scale fluctuations throughout the condensate dynamics. The spatiotemporal extent of vortex pair correlations correspond to the microscopic scales where density and phase fluctuations are coupled with each other, before a universal regime can set in.

\end{itemize}

\noindent Importantly, these results hold in both the EW and KPZ regimes, where the same behavior is observed. For this reason, we focus on the KPZ regime in the following, except in Sec.~\ref{sec:supmat_pair_correlation}, where the spacetime decay of the density correlation function $g^{(1)}_n$ is compared to that of the vortex pair correlation function $g^{(2)}_{+-}$ in both regimes.

\subsubsection{\label{sec:supmat_vortex_clustering}Vortex clustering}

In order to understand the microscopic arrangement of phase defects, we implement the vortex clustering algorithm described in Ref.~\cite{Valani2018}. This method classifies defects into three categories according to their spatial proximity and charge: free defect, dipole, or cluster. It works as follows :

\begin{algobox}

\begin{enumerate}

    \item For each defect, we locate the nearest opposite sign (NOS) defect. The corresponding distance is denoted $R_{\rm NOS}$.
    
    \item We then search for other same sign defects within a circle of radius $R_{\rm NOS}$.
    
    \item If no additional same sign defect is found within this circle, the vortex–antivortex pair is labeled as a dipole candidate.
    
    \item If other same sign defects are found within the circle, they are labeled as cluster candidates.
    
    \item Candidates are checked sequentially, to determine the subset of mutually agreeing dipoles and clusters candidates. For instance, if two opposite charge defects are mutually the nearest neighbors of each other, they are classified as forming a dipole. If not, they are left unclassified.
    
    \item Any defects that remain unclassified after this procedure are labeled as free defects.

\end{enumerate}
\end{algobox}

\medskip

\begin{figure}[h!]
    \centering
    \includegraphics[scale=0.22]{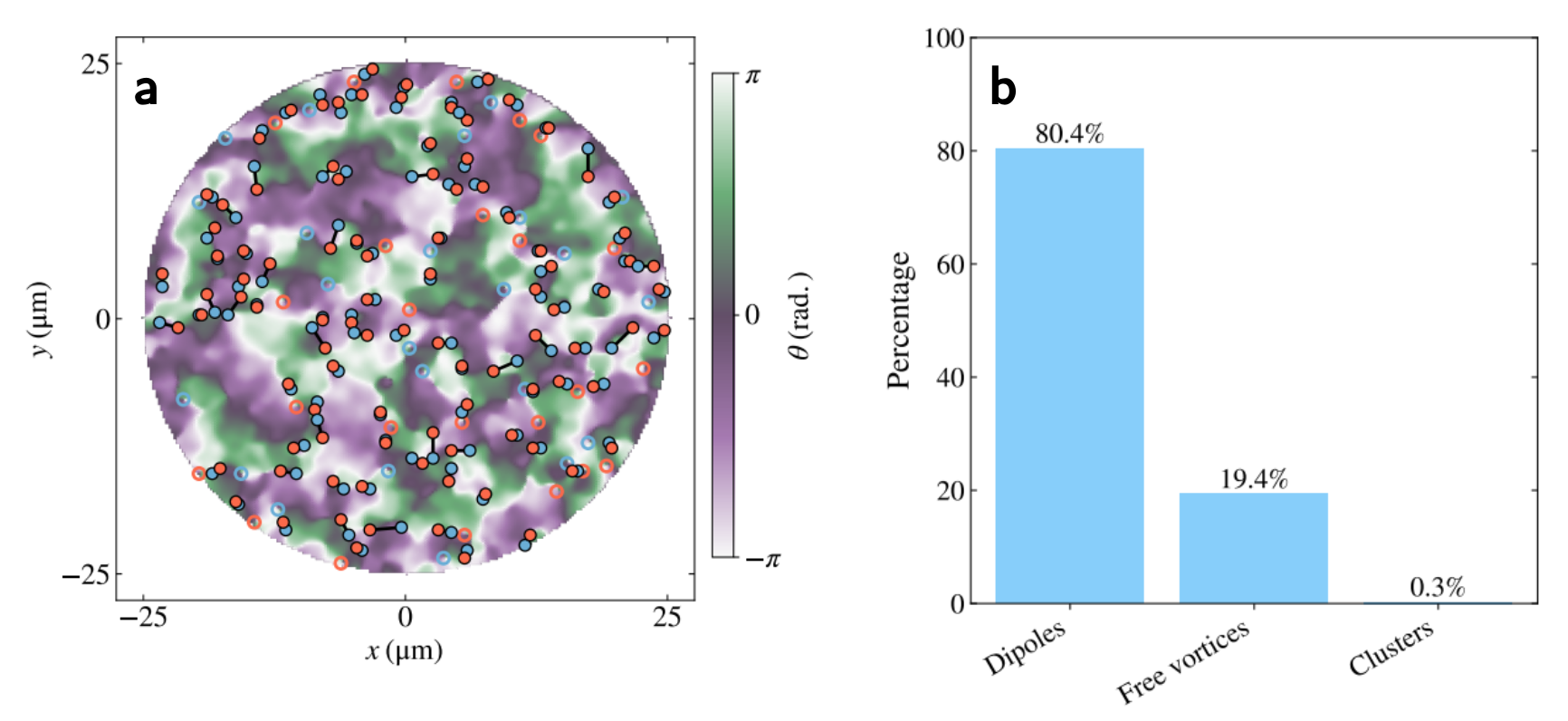}
    \caption{
    \textbf{Vortex clustering algorithm}. \textbf{a} Typical phase heat map in the steady state, where free defects are represented by an empty circle, while vortex-antivortex dipoles and clusters are highlighted by a black and white contour respectively. \textbf{b} Proportion of vortices classified as  free defects, paired in dipoles, or clustered.}
    \label{fig:sup_vortex_clustering}
\end{figure}

A typical instantaneous phase heat map after clustering is displayed in Fig.~\ref{fig:sup_vortex_clustering}.\textbf{a} for $\sigma_f = 0.25~{\rm \mu m}$. It immediately indicates that most defects are forming dipoles. Average over time and independent realizations yields the histogram in Fig.~\ref{fig:sup_vortex_clustering}\textbf{b}. It shows that about $80\%$ of defects are forming dipoles, while about $20\%$ of them are free. Less than $1\%$ of defects are identified as part of a cluster, which were always found composed of two defects. This suggests that defects of opposite charges are only rarely found in close proximity. They are instead typically brought together transiently during stochastic creation or annihilation events, throughout the dynamics. The statistics of Fig.~\ref{fig:sup_vortex_clustering}.\textbf{b} is found independent on the value of $\sigma_f$ (see also Secs.~\ref{sec:supmat_coarse_graining} and \ref{sec:supmat_lifetime}). 

\subsubsection{\label{sec:supmat_coarse_graining}Spatial coarse-graining}

\begin{figure}[t!]
    \centering
    \includegraphics[width=\linewidth]{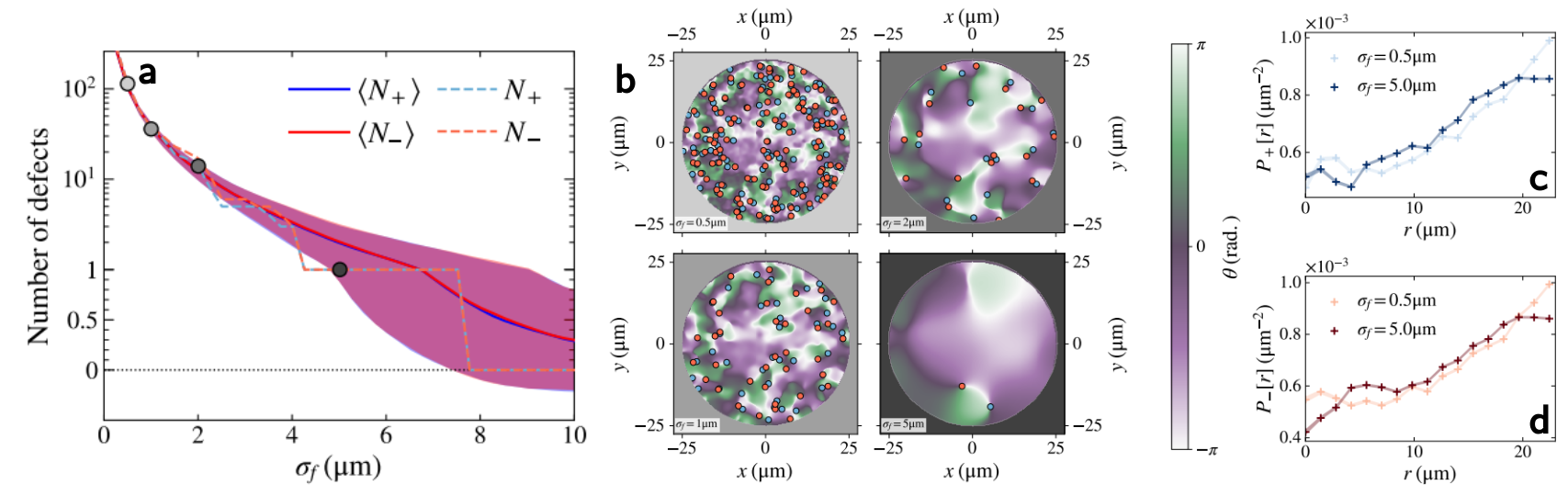}
    \caption{
    \textbf{Number of phase defects as a function of the Gaussian filtering $\sigma_f$}. \textbf{a} Number of vortices (blue) and antivortices (red) as a function of the width of the Gaussian filter $\sigma_f$. Plain lines represent the number of defects after averaging over 10 realizations and 120 ps of the dynamics. Dashed lines represent the number of defects for one instantaneous heat map of the phase. Phase heat maps corresponding to the grey dots of \textbf{a} are displayed in \textbf{b}. Radial probability density of \textbf{c} vortices and \textbf{d} antivortices for $\sigma_f = 0.5{\rm \mu m}$ and $\sigma_f = 5{\rm \mu m}$.
    }
    \label{fig:sup_map_Gaussian_smoothening}
\end{figure}

In this section, we build upon the work of Ref.~\cite{Foster_Davis_PRA2010_Vortex_pairing_2dBose_gas}, which investigates the BKT transition in semiclassical Monte Carlo simulations of 2D quasi-BECs at thermal equilibrium. Specifically, we numerically coarse-grain the condensate wavefunction using a Gaussian filter of width $\sigma_f$ in space, at every timestep of the dynamics independently. This acts as a low-pass filter in momentum space, gradually reducing the nucleation of dipoles (Fig.~\ref{fig:sup_vortex_clustering}). In doing so, the total number of vortices (respectively antivortices) for one configuration of the phase field decreases as $\sigma_f$ increases, gradually revealing the number of vortices (respectively antivortices) relevant at macroscopic scales. This procedure is illustrated in Fig.~\ref{fig:sup_map_Gaussian_smoothening}.\textbf{a}. A typical 2D phase heat map is shown in Fig.~\ref{fig:sup_map_Gaussian_smoothening}.\textbf{b} for increasing values of $\sigma_f$, showing that defects are indeed suppressed under spatial coarse-graining.

\medskip

For $\sigma_f\gtrsim2.5~{\rm \mu m}$, above which defects are not correlated (see Sec.~\ref{sec:supmat_pair_correlation}), the condensate contains on average less than ten defects of each charge. As detailed in Sec.~\ref{sec:supmat_lifetime} below, about $25\%$ of these vortices persist after a temporal coarse-graining, imposed by energy scales accessible in the experiment. This shows that, under the regimes considered, topological defects remaining at relevant spatiotemporal scales are virtually absent from the condensate.

\medskip

We now address the spatial location of topological defects within the condensate. The radial probability density of vortices (respectively antivortices), obtained from $\approx 7\times10^4$ trajectories, is displayed in Fig.~\ref{fig:sup_map_Gaussian_smoothening}.\textbf{c} (respectively \textbf{d}). It is found to weakly depend on $\sigma_f$, and shows that most defects remain on the outskirts of the pump profile, \textit{i.e.} where the condensate density is small. This indicates that, among the few vortices remaining after coarse-graining, most are in the peripheral low-density region of the condensate, thus not affecting the condensate coherence in the central region of the pump profile. These are presumably bound
to other defects of opposite charge in the tails of the condensate, at distances $r\gtrsim R_0$.

\medskip

Similar results were obtained alternatively using a real-space block decimation procedure, applied independently at each timestep. In this approach, the wavefunction defined on pixels of area $dx\times dx$ is coarse-grained onto meta-pixels of surface $\Delta x\times \Delta x$, by averaging over neighboring lattice sites. As a result, the minimum distance between paired defects is set by $\Delta x$. Finally, let us note that the average defect density obtained for $\sigma_f=1~{\rm \mu m}$ is compatible with that previously reported close to the mean-field transparency threshold \cite{Comaron2021}. We also checked that the average number of vortices decreases upon increasing the pumping rate $p_{\rm trp}$, though modulation instability is reached for smaller pumps than that required for defects to completely disappear from the condensate.

\subsubsection{\label{sec:supmat_lifetime}Defect lifetime}

\begin{figure}[h!]
    \centering
    \includegraphics[width=\linewidth]{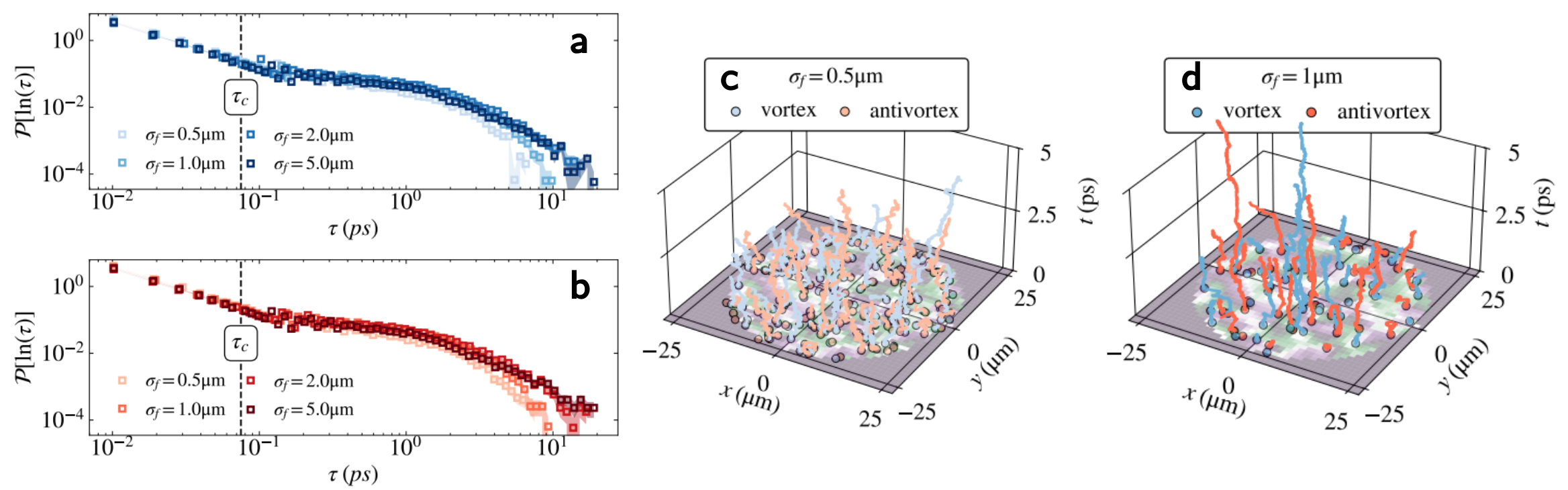}
    \caption{
    \textbf{Vortex and antivortex lifetime, temporal coarsening}. Probability distribution of the vortex \textbf{a} and antivortex \textbf{b} lifetime, for different Gaussian filters $\sigma_f$. Each distribution is obtained from approximately $1.7\times 10^5$ defect trajectories. The vertical dashed line corresponds to the cut-off time $\uptau_c$. Typical defect trajectories from one instantaneous phase heat map and $\sigma_f = 0.5~{\rm \mu m}$ \textbf{c}, $\sigma_f = 1~{\rm \mu m}$ \textbf{d}.}
    \label{fig:sup_map_vortex_lifetime}
\end{figure}

We define the lifetime $\uptau$ of tracked defects from the length of each individual trajectories. The probability distribution of vortex (respectively antivortex) lifetimes is displayed in Fig.~\ref{fig:sup_map_vortex_lifetime}.\textbf{a} (respectively \textbf{b}) for different $\sigma_f$. In each case, the distributions are obtained from approximately $1.7\times10^5$ trajectories. Note that increasing the width of the Gaussian filter $\sigma_f$ favors long-lived vortices with respect to short-lived defects, but that the probability distribution is not affected for short lifetimes. This is because the coarse-graining is applied in space only, and independently at every timestep of the dynamics. We checked that coarse-graining the condensate wavefunction using a Gaussian kernel simultaneously in space and in time gradually filters out short-lived defects. Consequently, the shape of the probability distribution at short lifetimes is modified upon filtering too, further increasing the relative probability of observing long-lived defects, but keeping their absolute number unaffected.

\medskip

The lifetime distributions are sharply peaked at the shortest timescales and decrease extremely rapidly : defects in the first time bin are about two orders of magnitude more probable to observe than those surviving for $\uptau \sim 1~{\rm ps}$. The shortest lifetimes are limited by the numerical timestep $dt$. Accordingly, it is observed that decreasing $dt$ increases the number of very short-lived defects. From Figs.~\ref{fig:sup_map_vortex_lifetime}.\textbf{a}-\textbf{b}, the mean lifetime of the full defect population is $\langle \uptau \rangle \approx 0.1~\mathrm{ps}$. 

\medskip

The shortest lifetimes resolved in the numerics, of order $dt$, correspond to excitations with energies $\sim 68~{\rm meV}$ above the condensate energy. In contrast, above the condensation threshold, experimental polariton spectra only exhibit excitations up to $\sim 4.5~{\rm meV}$, above which no signal persists. Accounting for the blueshift at threshold observed in the simulations, this corresponds to an energy cut-off of about $9~{\rm meV}$ in the numerical results. This mismatch defines a natural cut-off time $\uptau_c \approx 0.075~{\rm ps}$. Defects with lifetimes shorter than $\uptau_c$ should be regarded as artifacts of the simplified numerical model rather than physically meaningful excitations. This is a standard limitation of our semiclassical approach, which aims at reproducing the low-energy properties of polariton experiments. 
The cost of such a simplified approach is that unphysical ultraviolet divergences can appear, here a large proliferation of highly energetic defects, if no energy cutoff, or equivalently temporal filtering, is introduced \cite{Carusotto_Castin_PRA2007_Semiclassical_field_method}. In particular, we find that about $75\%$ of detected defects have a lifetime smaller than $\uptau_c$. This implies that, among the vortices remaining after spatial coarse-graining in Fig.~\ref{fig:sup_map_Gaussian_smoothening}.\textbf{a}, about $25\%$ correspond to excitations above the sharp time cutoff $\uptau_c$. Restricting the lifetime distribution to defects with $\uptau>\uptau_c$ yields an average lifetime $\langle \uptau \rangle \approx 0.37~{\rm ps}$. As pointed out above, temporal coarse-graining can alternatively be performed using a smooth Gaussian kernel instead, yielding comparable results. Typical vortex trajectories starting from one given phase configuration are shown in Fig.~\ref{fig:sup_map_vortex_lifetime}.\textbf{c}-\textbf{d}, for $\sigma_f = 0.5~{\rm \mu m}$ and $\sigma_f = 1~{\rm \mu m}$ respectively.

\medskip

Finally, focusing on free defects identified by clustering (see Sec.~\ref{sec:supmat_vortex_clustering}), we find that their average lifetime is smaller than that of the total defect population $\langle \uptau_{\mathrm{free}} \rangle \approx 0.035~{\rm ps}<\uptau_c$. This indicates that free defects identified microscopically are not only scarce, but predominantly correspond to high-energy excitations that are present in simulations but largely absent in this experiment. The large number of defects with lifetimes shorter than the cut-off timescale $\uptau_c$ observed in our simulations does not compromise comparison with experiments in the EW/KPZ regimes. In fact, defect dynamics primarily affects correlations at short spatiotemporal scales. As a consequence, it has a negligible impact on the  first-order correlation function at the scales where universal features emerge. This point is further discussed in the next section.

\subsubsection{\label{sec:supmat_pair_correlation}Defect pair correlation and discussion of the KPZ mapping}

\begin{figure}[h!]
    \centering
    \includegraphics[width=\linewidth]{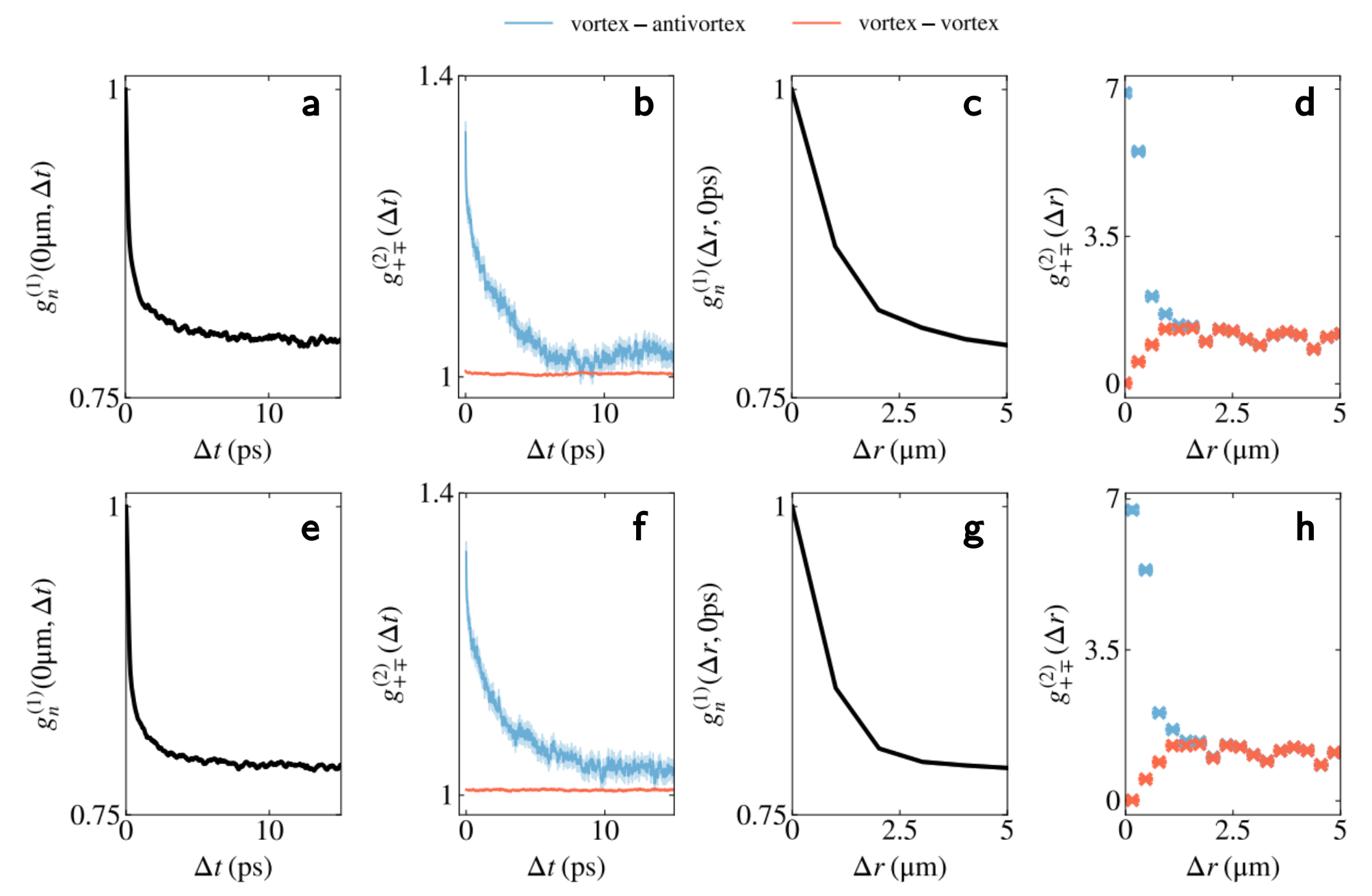}
    \caption{
    \textbf{Density correlation and defect pair correlation in space and time}. results in the KPZ regime are displayed on the upper panels, from \textbf{a} to \textbf{d}. Results in the EW regime are shown on the lower panels, from \textbf{e} to \textbf{h}.}
    \label{fig:sup_map_defect_correlation}
\end{figure}

In this section, we investigate the behavior of the defect pair correlation function in both space and time. Results in the KPZ regime are presented in Figs.~\ref{fig:sup_map_defect_correlation}.\textbf{b}-\textbf{d}, while those in the EW regime are shown in Figs.~\ref{fig:sup_map_defect_correlation}.\textbf{f}-\textbf{h}. In both regimes, we observe that vortices are preferentially paired with antivortices at short spatial and temporal separations. In contrast, vortex–vortex pairs are not correlated as a function of the time delay, and exhibit anti-correlations with space separation, indicating a short-range repulsive behavior. These findings are consistent with the vortex clustering analysis (Fig.~\ref{fig:sup_vortex_clustering}), \textit{i.e.} that defects are nucleated in dipoles by the noise. After a delay $\Delta t \sim 5~{\rm ps}$ and $\Delta r \sim 2.5~{\rm \mu m}$, the correlation functions approach a plateau close to unity, as expected for uncorrelated pairs.

\medskip

We also note the presence of a regular modulation of the plateau in $g^{(2)}_{+\mp}(\Delta r)$. These oscillations arise as an artifact of the finite circular region of radius $r_r$ used to compute pair correlations, within a locally homogeneous density region of the condensate (see Fig.~\ref{fig:sup_map_schematics_pair_correlation}). In a fully homogeneous system, these oscillations disappear, since $g^{(2)}_{+\mp}(\Delta r)$ can then be evaluated by averaging over all vortices in the system rather than over a small vortex population within a restricted spatial region \cite{Carusotto_Castin_PRA2007_Semiclassical_field_method}. 

\medskip

Taken together, pair correlations are short-ranged and decay rapidly, with negligible variation at the larger spatial and temporal scales where EW and KPZ scaling behaviors emerge. This observation is in agreement with the behavior of the first-order density correlation function $g^{(1)}_{n}(\Delta \boldsymbol{r}, \Delta t)$. Indeed, the vortex density fields $\rho{\pm}(\boldsymbol{r},t)$ can be viewed as a binarized negative image of the condensate density $|\psi(\boldsymbol{r},t)|^2$, taking the value 1 at vortex cores (where the density vanishes) and 0 elsewhere.

\medskip

Density correlations in space and time are shown in the KPZ regime in Fig.~\ref{fig:sup_map_defect_correlation}.\textbf{a} and \textbf{c}, and in the EW regime in Fig.~\ref{fig:sup_map_defect_correlation}.\textbf{e} and \textbf{g} (zoomed in compared to Fig.~\ref{fig:supmat_theory_EW_KPZ}). Similarly, they decay after very few picoseconds and micrometers. This behavior differs markedly from the slow decay of density correlations observed in numerical simulations of the spiral vortex phase \cite{Helluin2025}, highlighting the distinct nature of the defects reported here. In this context, the short-range nature of defect pair correlations supports one of the key assumptions underlying the observation of KPZ scaling. Namely, the decay of the first-order correlation function is governed by phase fluctuations, provided that density correlations remain negligible at large spatial and temporal scales.


\begin{thebibliography}{80}%
\makeatletter
\providecommand \@ifxundefined [1]{%
 \@ifx{#1\undefined}
}%
\providecommand \@ifnum [1]{%
 \ifnum #1\expandafter \@firstoftwo
 \else \expandafter \@secondoftwo
 \fi
}%
\providecommand \@ifx [1]{%
 \ifx #1\expandafter \@firstoftwo
 \else \expandafter \@secondoftwo
 \fi
}%
\providecommand \natexlab [1]{#1}%
\providecommand \enquote  [1]{``#1''}%
\providecommand \bibnamefont  [1]{#1}%
\providecommand \bibfnamefont [1]{#1}%
\providecommand \citenamefont [1]{#1}%
\providecommand \href@noop [0]{\@secondoftwo}%
\providecommand \href [0]{\begingroup \@sanitize@url \@href}%
\providecommand \@href[1]{\@@startlink{#1}\@@href}%
\providecommand \@@href[1]{\endgroup#1\@@endlink}%
\providecommand \@sanitize@url [0]{\catcode `\\12\catcode `\$12\catcode
  `\&12\catcode `\#12\catcode `\^12\catcode `\_12\catcode `\%12\relax}%
\providecommand \@@startlink[1]{}%
\providecommand \@@endlink[0]{}%
\providecommand \url  [0]{\begingroup\@sanitize@url \@url }%
\providecommand \@url [1]{\endgroup\@href {#1}{\urlprefix }}%
\providecommand \urlprefix  [0]{URL }%
\providecommand \Eprint [0]{\href }%
\providecommand \doibase [0]{https://doi.org/}%
\providecommand \selectlanguage [0]{\@gobble}%
\providecommand \bibinfo  [0]{\@secondoftwo}%
\providecommand \bibfield  [0]{\@secondoftwo}%
\providecommand \translation [1]{[#1]}%
\providecommand \BibitemOpen [0]{}%
\providecommand \bibitemStop [0]{}%
\providecommand \bibitemNoStop [0]{.\EOS\space}%
\providecommand \EOS [0]{\spacefactor3000\relax}%
\providecommand \BibitemShut  [1]{\csname bibitem#1\endcsname}%
\let\auto@bib@innerbib\@empty
\bibitem [{\citenamefont {Bose}(1924)}]{Bose1924}%
  \BibitemOpen
  \bibfield  {author} {\bibinfo {author} {\bibnamefont {Bose}},\ }\bibfield
  {title} {\bibinfo {title} {{Plancks Gesetz und Lichtquantenhypothese}},\
  }\href {https://doi.org/10.1007/BF01327326} {\bibfield  {journal} {\bibinfo
  {journal} {Zeitschrift f{\"{u}}r Physik}\ }\textbf {\bibinfo {volume} {26}},\
  \bibinfo {pages} {178} (\bibinfo {year} {1924})}\BibitemShut {NoStop}%
\bibitem [{\citenamefont {Anderson}\ \emph {et~al.}(1995)\citenamefont
  {Anderson}, \citenamefont {Ensher}, \citenamefont {Matthews}, \citenamefont
  {Wieman},\ and\ \citenamefont {Cornell}}]{Anderson1995}%
  \BibitemOpen
  \bibfield  {author} {\bibinfo {author} {\bibfnamefont {M.~H.}\ \bibnamefont
  {Anderson}}, \bibinfo {author} {\bibfnamefont {J.~R.}\ \bibnamefont
  {Ensher}}, \bibinfo {author} {\bibfnamefont {M.~R.}\ \bibnamefont
  {Matthews}}, \bibinfo {author} {\bibfnamefont {C.~E.}\ \bibnamefont
  {Wieman}},\ and\ \bibinfo {author} {\bibfnamefont {E.~A.}\ \bibnamefont
  {Cornell}},\ }\bibfield  {title} {\bibinfo {title} {Observation of
  bose-einstein condensation in a dilute atomic vapor},\ }\href
  {https://doi.org/10.1126/science.269.5221.198} {\bibfield  {journal}
  {\bibinfo  {journal} {Science}\ }\textbf {\bibinfo {volume} {269}},\ \bibinfo
  {pages} {198} (\bibinfo {year} {1995})}\BibitemShut {NoStop}%
\bibitem [{\citenamefont {Davis}\ \emph {et~al.}(1995)\citenamefont {Davis},
  \citenamefont {Mewes}, \citenamefont {Andrews}, \citenamefont {van Druten},
  \citenamefont {Durfee}, \citenamefont {Kurn},\ and\ \citenamefont
  {Ketterle}}]{Davis1995}%
  \BibitemOpen
  \bibfield  {author} {\bibinfo {author} {\bibfnamefont {K.~B.}\ \bibnamefont
  {Davis}}, \bibinfo {author} {\bibfnamefont {M.~O.}\ \bibnamefont {Mewes}},
  \bibinfo {author} {\bibfnamefont {M.~R.}\ \bibnamefont {Andrews}}, \bibinfo
  {author} {\bibfnamefont {N.~J.}\ \bibnamefont {van Druten}}, \bibinfo
  {author} {\bibfnamefont {D.~S.}\ \bibnamefont {Durfee}}, \bibinfo {author}
  {\bibfnamefont {D.~M.}\ \bibnamefont {Kurn}},\ and\ \bibinfo {author}
  {\bibfnamefont {W.}~\bibnamefont {Ketterle}},\ }\bibfield  {title} {\bibinfo
  {title} {Bose-einstein condensation in a gas of sodium atoms},\ }\href
  {https://doi.org/10.1103/PhysRevLett.75.3969} {\bibfield  {journal} {\bibinfo
   {journal} {Phys. Rev. Lett.}\ }\textbf {\bibinfo {volume} {75}},\ \bibinfo
  {pages} {3969} (\bibinfo {year} {1995})}\BibitemShut {NoStop}%
\bibitem [{\citenamefont {Berezinsky}(1971)}]{Berezinsky_1970}%
  \BibitemOpen
  \bibfield  {author} {\bibinfo {author} {\bibfnamefont {V.~L.}\ \bibnamefont
  {Berezinsky}},\ }\bibfield  {title} {\bibinfo {title} {{Destruction of long
  range order in one-dimensional and two-dimensional systems having a
  continuous symmetry group. I. Classical systems}},\ }\href
  {https://inspirehep.net/literature/61186} {\bibfield  {journal} {\bibinfo
  {journal} {Sov. Phys. JETP}\ }\textbf {\bibinfo {volume} {32}},\ \bibinfo
  {pages} {493} (\bibinfo {year} {1971})}\BibitemShut {NoStop}%
\bibitem [{\citenamefont {Butov}\ \emph {et~al.}(2002)\citenamefont {Butov},
  \citenamefont {Gossard},\ and\ \citenamefont {Chemla}}]{Butov2002}%
  \BibitemOpen
  \bibfield  {author} {\bibinfo {author} {\bibfnamefont {L.~V.}\ \bibnamefont
  {Butov}}, \bibinfo {author} {\bibfnamefont {A.~C.}\ \bibnamefont {Gossard}},\
  and\ \bibinfo {author} {\bibfnamefont {D.~S.}\ \bibnamefont {Chemla}},\
  }\bibfield  {title} {\bibinfo {title} {{Macroscopically ordered state in an
  exciton system}},\ }\href {https://doi.org/10.1038/nature00943} {\bibfield
  {journal} {\bibinfo  {journal} {Nature}\ }\textbf {\bibinfo {volume} {418}},\
  \bibinfo {pages} {751} (\bibinfo {year} {2002})}\BibitemShut {NoStop}%
\bibitem [{\citenamefont {Snoke}\ \emph {et~al.}(2002)\citenamefont {Snoke},
  \citenamefont {Denev}, \citenamefont {Liu}, \citenamefont {Pfeiffer},\ and\
  \citenamefont {West}}]{Snoke2002}%
  \BibitemOpen
  \bibfield  {author} {\bibinfo {author} {\bibfnamefont {D.}~\bibnamefont
  {Snoke}}, \bibinfo {author} {\bibfnamefont {S.}~\bibnamefont {Denev}},
  \bibinfo {author} {\bibfnamefont {Y.}~\bibnamefont {Liu}}, \bibinfo {author}
  {\bibfnamefont {L.}~\bibnamefont {Pfeiffer}},\ and\ \bibinfo {author}
  {\bibfnamefont {K.}~\bibnamefont {West}},\ }\bibfield  {title}
  {\bibinfo {title} {{Long-range transport in excitonic
  dark states in coupled quantum wells.}},\ }\href
  {https://doi.org/10.1038/nature00940} {\bibfield  {journal} {\bibinfo
  {journal} {Nature}\ }\textbf {\bibinfo {volume} {418}},\ \bibinfo {pages}
  {754} (\bibinfo {year} {2002})}\BibitemShut {NoStop}%
\bibitem [{\citenamefont {Dang}\ \emph {et~al.}(2020)\citenamefont {Dang},
  \citenamefont {Zamorano}, \citenamefont {Suffit}, \citenamefont {West},
  \citenamefont {Baldwin}, \citenamefont {Pfeiffer}, \citenamefont {Holzmann},\
  and\ \citenamefont {Dubin}}]{Dang2020}%
  \BibitemOpen
  \bibfield  {author} {\bibinfo {author} {\bibfnamefont {S.}~\bibnamefont
  {Dang}}, \bibinfo {author} {\bibfnamefont {M.}~\bibnamefont {Zamorano}},
  \bibinfo {author} {\bibfnamefont {S.}~\bibnamefont {Suffit}}, \bibinfo
  {author} {\bibfnamefont {K.}~\bibnamefont {West}}, \bibinfo {author}
  {\bibfnamefont {K.}~\bibnamefont {Baldwin}}, \bibinfo {author} {\bibfnamefont
  {L.}~\bibnamefont {Pfeiffer}}, \bibinfo {author} {\bibfnamefont
  {M.}~\bibnamefont {Holzmann}},\ and\ \bibinfo {author} {\bibfnamefont
  {F.~m.~c.}\ \bibnamefont {Dubin}},\ }\bibfield  {title} {\bibinfo {title}
  {Observation of algebraic time order for two-dimensional dipolar excitons},\
  }\href {https://doi.org/10.1103/PhysRevResearch.2.032013} {\bibfield
  {journal} {\bibinfo  {journal} {Phys. Rev. Res.}\ }\textbf {\bibinfo {volume}
  {2}},\ \bibinfo {pages} {032013} (\bibinfo {year} {2020})}\BibitemShut
  {NoStop}%
\bibitem [{\citenamefont {Kasprzak}\ \emph {et~al.}(2006)\citenamefont
  {Kasprzak}, \citenamefont {Richard}, \citenamefont {Kundermann},
  \citenamefont {Baas}, \citenamefont {Jeambrun}, \citenamefont {Keeling},
  \citenamefont {Marchetti}, \citenamefont {Szyma{\'{n}}ska}, \citenamefont
  {Andr{\'{e}}}, \citenamefont {Staehli}, \citenamefont {Savona}, \citenamefont
  {Littlewood}, \citenamefont {Deveaud},\ and\ \citenamefont
  {Dang}}]{Kasprzak2006}%
  \BibitemOpen
  \bibfield  {author} {\bibinfo {author} {\bibfnamefont {J.}~\bibnamefont
  {Kasprzak}}, \bibinfo {author} {\bibfnamefont {M.}~\bibnamefont {Richard}},
  \bibinfo {author} {\bibfnamefont {S.}~\bibnamefont {Kundermann}}, \bibinfo
  {author} {\bibfnamefont {A.}~\bibnamefont {Baas}}, \bibinfo {author}
  {\bibfnamefont {P.}~\bibnamefont {Jeambrun}}, \bibinfo {author}
  {\bibfnamefont {J.~M.~J.}\ \bibnamefont {Keeling}}, \bibinfo {author}
  {\bibfnamefont {F.~M.}\ \bibnamefont {Marchetti}}, \bibinfo {author}
  {\bibfnamefont {M.~H.}\ \bibnamefont {Szyma{\'{n}}ska}}, \bibinfo {author}
  {\bibfnamefont {R.}~\bibnamefont {Andr{\'{e}}}}, \bibinfo {author}
  {\bibfnamefont {J.~L.}\ \bibnamefont {Staehli}}, \bibinfo {author}
  {\bibfnamefont {V.}~\bibnamefont {Savona}}, \bibinfo {author} {\bibfnamefont
  {P.~B.}\ \bibnamefont {Littlewood}}, \bibinfo {author} {\bibfnamefont
  {B.}~\bibnamefont {Deveaud}},\ and\ \bibinfo {author} {\bibfnamefont {L.~S.}\
  \bibnamefont {Dang}},\ }\bibfield  {title} {\bibinfo {title}
  {{Bose–Einstein condensation of exciton polaritons}},\ }\href
  {https://doi.org/10.1038/nature05131} {\bibfield  {journal} {\bibinfo
  {journal} {Nature}\ }\textbf {\bibinfo {volume} {443}},\ \bibinfo {pages}
  {409} (\bibinfo {year} {2006})}\BibitemShut {NoStop}%
\bibitem [{\citenamefont {Hakala}\ \emph {et~al.}(2018)\citenamefont {Hakala},
  \citenamefont {Moilanen}, \citenamefont {V{\"{a}}kev{\"{a}}inen},
  \citenamefont {Guo}, \citenamefont {Martikainen}, \citenamefont {Daskalakis},
  \citenamefont {Rekola}, \citenamefont {Julku},\ and\ \citenamefont
  {T{\"{o}}rm{\"{a}}}}]{Hakala2018}%
  \BibitemOpen
  \bibfield  {author} {\bibinfo {author} {\bibfnamefont {T.~K.}\ \bibnamefont
  {Hakala}}, \bibinfo {author} {\bibfnamefont {A.~J.}\ \bibnamefont
  {Moilanen}}, \bibinfo {author} {\bibfnamefont {A.~I.}\ \bibnamefont
  {V{\"{a}}kev{\"{a}}inen}}, \bibinfo {author} {\bibfnamefont {R.}~\bibnamefont
  {Guo}}, \bibinfo {author} {\bibfnamefont {J.-P.}\ \bibnamefont
  {Martikainen}}, \bibinfo {author} {\bibfnamefont {K.~S.}\ \bibnamefont
  {Daskalakis}}, \bibinfo {author} {\bibfnamefont {H.~T.}\ \bibnamefont
  {Rekola}}, \bibinfo {author} {\bibfnamefont {A.}~\bibnamefont {Julku}},\ and\
  \bibinfo {author} {\bibfnamefont {P.}~\bibnamefont {T{\"{o}}rm{\"{a}}}},\
  }\bibfield  {title} {\bibinfo {title} {{Bose–Einstein condensation in a
  plasmonic lattice}},\ }\href {https://doi.org/10.1038/s41567-018-0109-9}
  {\bibfield  {journal} {\bibinfo  {journal} {Nature Physics}\ }\textbf
  {\bibinfo {volume} {14}},\ \bibinfo {pages} {739} (\bibinfo {year}
  {2018})}\BibitemShut {NoStop}%
\bibitem [{\citenamefont {Demokritov}\ \emph {et~al.}(2006)\citenamefont
  {Demokritov}, \citenamefont {Demidov}, \citenamefont {Dzyapko}, \citenamefont
  {Melkov}, \citenamefont {Serga}, \citenamefont {Hillebrands},\ and\
  \citenamefont {Slavin}}]{Demokritov2006}%
  \BibitemOpen
  \bibfield  {author} {\bibinfo {author} {\bibfnamefont {S.~O.}\ \bibnamefont
  {Demokritov}}, \bibinfo {author} {\bibfnamefont {V.~E.}\ \bibnamefont
  {Demidov}}, \bibinfo {author} {\bibfnamefont {O.}~\bibnamefont {Dzyapko}},
  \bibinfo {author} {\bibfnamefont {G.~A.}\ \bibnamefont {Melkov}}, \bibinfo
  {author} {\bibfnamefont {A.~A.}\ \bibnamefont {Serga}}, \bibinfo {author}
  {\bibfnamefont {B.}~\bibnamefont {Hillebrands}},\ and\ \bibinfo {author}
  {\bibfnamefont {A.~N.}\ \bibnamefont {Slavin}},\ }\bibfield  {title}
  {\bibinfo {title} {{Bose–Einstein condensation of quasi-equilibrium magnons
  at room temperature under pumping}},\ }\href
  {https://doi.org/10.1038/nature05117} {\bibfield  {journal} {\bibinfo
  {journal} {Nature}\ }\textbf {\bibinfo {volume} {443}},\ \bibinfo {pages}
  {430} (\bibinfo {year} {2006})}\BibitemShut {NoStop}%
\bibitem [{\citenamefont {Bloch}\ \emph {et~al.}(2022)\citenamefont {Bloch},
  \citenamefont {Carusotto},\ and\ \citenamefont {Wouters}}]{bloch2022}%
  \BibitemOpen
  \bibfield  {author} {\bibinfo {author} {\bibfnamefont {J.}~\bibnamefont
  {Bloch}}, \bibinfo {author} {\bibfnamefont {I.}~\bibnamefont {Carusotto}},\
  and\ \bibinfo {author} {\bibfnamefont {M.}~\bibnamefont {Wouters}},\
  }\bibfield  {title} {\bibinfo {title} {Non-equilibrium {B}ose--{E}instein
  condensation in photonic systems},\ }\href
  {https://doi.org/10.1038/s42254-022-00464-0} {\bibfield  {journal} {\bibinfo
  {journal} {Nature Reviews Physics}\ }\textbf {\bibinfo {volume} {4}},\
  \bibinfo {pages} {470} (\bibinfo {year} {2022})}\BibitemShut {NoStop}%
\bibitem [{\citenamefont {Altman}\ \emph
  {et~al.}(2015{\natexlab{a}})\citenamefont {Altman}, \citenamefont {Sieberer},
  \citenamefont {Chen}, \citenamefont {Diehl},\ and\ \citenamefont
  {Toner}}]{Altman2015}%
  \BibitemOpen
  \bibfield  {author} {\bibinfo {author} {\bibfnamefont {E.}~\bibnamefont
  {Altman}}, \bibinfo {author} {\bibfnamefont {L.~M.}\ \bibnamefont
  {Sieberer}}, \bibinfo {author} {\bibfnamefont {L.}~\bibnamefont {Chen}},
  \bibinfo {author} {\bibfnamefont {S.}~\bibnamefont {Diehl}},\ and\ \bibinfo
  {author} {\bibfnamefont {J.}~\bibnamefont {Toner}},\ }\bibfield  {title}
  {\bibinfo {title} {Two-dimensional superfluidity of exciton polaritons
  requires strong anisotropy},\ }\href
  {https://doi.org/10.1103/PhysRevX.5.011017} {\bibfield  {journal} {\bibinfo
  {journal} {Physical Review X}\ }\textbf {\bibinfo {volume} {5}},\ \bibinfo
  {pages} {011017} (\bibinfo {year} {2015}{\natexlab{a}})}\BibitemShut
  {NoStop}%
\bibitem [{\citenamefont {Zamora}\ \emph {et~al.}(2017)\citenamefont {Zamora},
  \citenamefont {Sieberer}, \citenamefont {Dunnett}, \citenamefont {Diehl},\
  and\ \citenamefont {Szyma\ifmmode~\acute{n}\else
  \'{n}\fi{}ska}}]{Zamora2017}%
  \BibitemOpen
  \bibfield  {author} {\bibinfo {author} {\bibfnamefont {A.}~\bibnamefont
  {Zamora}}, \bibinfo {author} {\bibfnamefont {L.~M.}\ \bibnamefont
  {Sieberer}}, \bibinfo {author} {\bibfnamefont {K.}~\bibnamefont {Dunnett}},
  \bibinfo {author} {\bibfnamefont {S.}~\bibnamefont {Diehl}},\ and\ \bibinfo
  {author} {\bibfnamefont {M.~H.}\ \bibnamefont {Szyma\ifmmode~\acute{n}\else
  \'{n}\fi{}ska}},\ }\bibfield  {title} {\bibinfo {title} {Tuning across
  universalities with a driven open condensate},\ }\href
  {https://doi.org/10.1103/PhysRevX.7.041006} {\bibfield  {journal} {\bibinfo
  {journal} {Physical Review X}\ }\textbf {\bibinfo {volume} {7}},\ \bibinfo
  {pages} {041006} (\bibinfo {year} {2017})}\BibitemShut {NoStop}%
\bibitem [{\citenamefont {Dagvadorj}\ \emph
  {et~al.}(2021{\natexlab{a}})\citenamefont {Dagvadorj}, \citenamefont
  {Kulczykowski}, \citenamefont {Szyma\ifmmode~\acute{n}\else \'{n}\fi{}ska},\
  and\ \citenamefont {Matuszewski}}]{Dagvadorj2021}%
  \BibitemOpen
  \bibfield  {author} {\bibinfo {author} {\bibfnamefont {G.}~\bibnamefont
  {Dagvadorj}}, \bibinfo {author} {\bibfnamefont {M.}~\bibnamefont
  {Kulczykowski}}, \bibinfo {author} {\bibfnamefont {M.}~\bibnamefont
  {Szyma\ifmmode~\acute{n}\else \'{n}\fi{}ska}},\ and\ \bibinfo {author}
  {\bibfnamefont {M.}~\bibnamefont {Matuszewski}},\ }\bibfield  {title}
  {\bibinfo {title} {First-order dissipative phase transition in an
  exciton-polariton condensate},\ }\href
  {https://doi.org/10.1103/PhysRevB.104.165301} {\bibfield  {journal} {\bibinfo
   {journal} {Physical Review B}\ }\textbf {\bibinfo {volume} {104}},\ \bibinfo
  {pages} {165301} (\bibinfo {year} {2021}{\natexlab{a}})}\BibitemShut
  {NoStop}%
\bibitem [{\citenamefont {Helluin}\ \emph {et~al.}(2025)\citenamefont
  {Helluin}, \citenamefont {Pinto-Dias}, \citenamefont {Fontaine},
  \citenamefont {Ravets}, \citenamefont {Bloch}, \citenamefont {Minguzzi},\
  and\ \citenamefont {Canet}}]{Helluin2025}%
  \BibitemOpen
  \bibfield  {author} {\bibinfo {author} {\bibfnamefont {F.}~\bibnamefont
  {Helluin}}, \bibinfo {author} {\bibfnamefont {D.}~\bibnamefont {Pinto-Dias}},
  \bibinfo {author} {\bibfnamefont {Q.}~\bibnamefont {Fontaine}}, \bibinfo
  {author} {\bibfnamefont {S.}~\bibnamefont {Ravets}}, \bibinfo {author}
  {\bibfnamefont {J.}~\bibnamefont {Bloch}}, \bibinfo {author} {\bibfnamefont
  {A.}~\bibnamefont {Minguzzi}},\ and\ \bibinfo {author} {\bibfnamefont
  {L.}~\bibnamefont {Canet}},\ }\bibfield  {title} {\bibinfo {title} {Phase
  diagram and universal scaling regimes of two-dimensional exciton--polariton
  {B}ose--{E}instein condensates},\ }\href {https://doi.org/10.1103/3gmk-xccn}
  {\bibfield  {journal} {\bibinfo  {journal} {Physical Review Res.}\ }\textbf
  {\bibinfo {volume} {7}},\ \bibinfo {pages} {033103} (\bibinfo {year}
  {2025})}\BibitemShut {NoStop}%
\bibitem [{\citenamefont {Szyma\ifmmode~\acute{n}\else \'{n}\fi{}ska}\ \emph
  {et~al.}(2006)\citenamefont {Szyma\ifmmode~\acute{n}\else \'{n}\fi{}ska},
  \citenamefont {Keeling},\ and\ \citenamefont {Littlewood}}]{Szymanska2006}%
  \BibitemOpen
  \bibfield  {author} {\bibinfo {author} {\bibfnamefont {M.~H.}\ \bibnamefont
  {Szyma\ifmmode~\acute{n}\else \'{n}\fi{}ska}}, \bibinfo {author}
  {\bibfnamefont {J.}~\bibnamefont {Keeling}},\ and\ \bibinfo {author}
  {\bibfnamefont {P.~B.}\ \bibnamefont {Littlewood}},\ }\bibfield  {title}
  {\bibinfo {title} {Nonequilibrium quantum condensation in an incoherently
  pumped dissipative system},\ }\href
  {https://doi.org/10.1103/PhysRevLett.96.230602} {\bibfield  {journal}
  {\bibinfo  {journal} {Phys. Rev. Lett.}\ }\textbf {\bibinfo {volume} {96}},\
  \bibinfo {pages} {230602} (\bibinfo {year} {2006})}\BibitemShut {NoStop}%
\bibitem [{\citenamefont {Dagvadorj}\ \emph {et~al.}(2015)\citenamefont
  {Dagvadorj}, \citenamefont {Fellows}, \citenamefont
  {Matyja\ifmmode~\acute{s}\else \'{s}\fi{}kiewicz}, \citenamefont {Marchetti},
  \citenamefont {Carusotto},\ and\ \citenamefont {Szyma\ifmmode~\acute{n}\else
  \'{n}\fi{}ska}}]{Dagvadorj2015}%
  \BibitemOpen
  \bibfield  {author} {\bibinfo {author} {\bibfnamefont {G.}~\bibnamefont
  {Dagvadorj}}, \bibinfo {author} {\bibfnamefont {J.~M.}\ \bibnamefont
  {Fellows}}, \bibinfo {author} {\bibfnamefont {S.}~\bibnamefont
  {Matyja\ifmmode~\acute{s}\else \'{s}\fi{}kiewicz}}, \bibinfo {author}
  {\bibfnamefont {F.~M.}\ \bibnamefont {Marchetti}}, \bibinfo {author}
  {\bibfnamefont {I.}~\bibnamefont {Carusotto}},\ and\ \bibinfo {author}
  {\bibfnamefont {M.~H.}\ \bibnamefont {Szyma\ifmmode~\acute{n}\else
  \'{n}\fi{}ska}},\ }\bibfield  {title} {\bibinfo {title} {Nonequilibrium phase
  transition in a two-dimensional driven open quantum system},\ }\href
  {https://doi.org/10.1103/PhysRevX.5.041028} {\bibfield  {journal} {\bibinfo
  {journal} {Phys. Rev. X}\ }\textbf {\bibinfo {volume} {5}},\ \bibinfo {pages}
  {041028} (\bibinfo {year} {2015})}\BibitemShut {NoStop}%
\bibitem [{\citenamefont {Comaron}\ \emph {et~al.}(2021)\citenamefont
  {Comaron}, \citenamefont {Carusotto}, \citenamefont {Szymańska},\ and\
  \citenamefont {Proukakis}}]{Comaron2021}%
  \BibitemOpen
  \bibfield  {author} {\bibinfo {author} {\bibfnamefont {P.}~\bibnamefont
  {Comaron}}, \bibinfo {author} {\bibfnamefont {I.}~\bibnamefont {Carusotto}},
  \bibinfo {author} {\bibfnamefont {M.~H.}\ \bibnamefont {Szymańska}},\ and\
  \bibinfo {author} {\bibfnamefont {N.~P.}\ \bibnamefont {Proukakis}},\
  }\bibfield  {title} {\bibinfo {title} {Non-equilibrium
  berezinskii-kosterlitz-thouless transition in driven-dissipative
  condensates(a)},\ }\href {https://doi.org/10.1209/0295-5075/133/17002}
  {\bibfield  {journal} {\bibinfo  {journal} {Europhysics Letters}\ }\textbf
  {\bibinfo {volume} {133}},\ \bibinfo {pages} {17002} (\bibinfo {year}
  {2021})}\BibitemShut {NoStop}%
\bibitem [{\citenamefont {Mei}\ \emph {et~al.}(2021)\citenamefont {Mei},
  \citenamefont {Ji},\ and\ \citenamefont {Wouters}}]{Mei2021}%
  \BibitemOpen
  \bibfield  {author} {\bibinfo {author} {\bibfnamefont {Q.}~\bibnamefont
  {Mei}}, \bibinfo {author} {\bibfnamefont {K.}~\bibnamefont {Ji}},\ and\
  \bibinfo {author} {\bibfnamefont {M.}~\bibnamefont {Wouters}},\ }\bibfield
  {title} {\bibinfo {title} {Spatiotemporal scaling of two-dimensional
  nonequilibrium exciton-polariton systems with weak interactions},\ }\href
  {https://doi.org/10.1103/PhysRevB.103.045302} {\bibfield  {journal} {\bibinfo
   {journal} {Physical Review B}\ }\textbf {\bibinfo {volume} {103}},\ \bibinfo
  {pages} {045302} (\bibinfo {year} {2021})}\BibitemShut {NoStop}%
\bibitem [{\citenamefont {Ferrier}\ \emph {et~al.}(2022)\citenamefont
  {Ferrier}, \citenamefont {Zamora}, \citenamefont {Dagvadorj},\ and\
  \citenamefont {Szyma\ifmmode~\acute{n}\else \'{n}\fi{}ska}}]{Ferrier2022}%
  \BibitemOpen
  \bibfield  {author} {\bibinfo {author} {\bibfnamefont {A.}~\bibnamefont
  {Ferrier}}, \bibinfo {author} {\bibfnamefont {A.}~\bibnamefont {Zamora}},
  \bibinfo {author} {\bibfnamefont {G.}~\bibnamefont {Dagvadorj}},\ and\
  \bibinfo {author} {\bibfnamefont {M.~H.}\ \bibnamefont
  {Szyma\ifmmode~\acute{n}\else \'{n}\fi{}ska}},\ }\bibfield  {title} {\bibinfo
  {title} {Searching for the {K}ardar--{P}arisi--{Z}hang phase in microcavity
  polaritons},\ }\href {https://doi.org/10.1103/PhysRevB.105.205301} {\bibfield
   {journal} {\bibinfo  {journal} {Physical Review B}\ }\textbf {\bibinfo
  {volume} {105}},\ \bibinfo {pages} {205301} (\bibinfo {year}
  {2022})}\BibitemShut {NoStop}%
\bibitem [{\citenamefont {Deligiannis}\ \emph {et~al.}(2022)\citenamefont
  {Deligiannis}, \citenamefont {Fontaine}, \citenamefont {Squizzato},
  \citenamefont {Richard}, \citenamefont {Ravets}, \citenamefont {Bloch},
  \citenamefont {Minguzzi},\ and\ \citenamefont {Canet}}]{deligiannis2022}%
  \BibitemOpen
  \bibfield  {author} {\bibinfo {author} {\bibfnamefont {K.}~\bibnamefont
  {Deligiannis}}, \bibinfo {author} {\bibfnamefont {Q.}~\bibnamefont
  {Fontaine}}, \bibinfo {author} {\bibfnamefont {D.}~\bibnamefont {Squizzato}},
  \bibinfo {author} {\bibfnamefont {M.}~\bibnamefont {Richard}}, \bibinfo
  {author} {\bibfnamefont {S.}~\bibnamefont {Ravets}}, \bibinfo {author}
  {\bibfnamefont {J.}~\bibnamefont {Bloch}}, \bibinfo {author} {\bibfnamefont
  {A.}~\bibnamefont {Minguzzi}},\ and\ \bibinfo {author} {\bibfnamefont
  {L.}~\bibnamefont {Canet}},\ }\bibfield  {title} {\bibinfo {title}
  {{K}ardar-{P}arisi-{Z}hang universality in discrete two-dimensional
  driven-dissipative exciton polariton condensates},\ }\href
  {https://doi.org/10.1103/PhysRevResearch.4.043207} {\bibfield  {journal}
  {\bibinfo  {journal} {Physical Review Res.}\ }\textbf {\bibinfo {volume}
  {4}},\ \bibinfo {pages} {043207} (\bibinfo {year} {2022})}\BibitemShut
  {NoStop}%
\bibitem [{\citenamefont {Kardar}\ \emph {et~al.}(1986)\citenamefont {Kardar},
  \citenamefont {Parisi},\ and\ \citenamefont {Zhang}}]{Kardar1986}%
  \BibitemOpen
  \bibfield  {author} {\bibinfo {author} {\bibfnamefont {M.}~\bibnamefont
  {Kardar}}, \bibinfo {author} {\bibfnamefont {G.}~\bibnamefont {Parisi}},\
  and\ \bibinfo {author} {\bibfnamefont {Y.}~\bibnamefont {Zhang}},\ }\bibfield
   {title} {\bibinfo {title} {Dynamic scaling of growing interfaces},\ }\href
  {https://doi.org/10.1103/PhysRevLett.56.889} {\bibfield  {journal} {\bibinfo
  {journal} {Physical Review Letters}\ }\textbf {\bibinfo {volume} {56}},\
  \bibinfo {pages} {889} (\bibinfo {year} {1986})}\BibitemShut {NoStop}%
\bibitem [{\citenamefont {He}\ \emph {et~al.}(2015{\natexlab{a}})\citenamefont
  {He}, \citenamefont {Sieberer}, \citenamefont {Altman},\ and\ \citenamefont
  {Diehl}}]{He2015}%
  \BibitemOpen
  \bibfield  {author} {\bibinfo {author} {\bibfnamefont {L.}~\bibnamefont
  {He}}, \bibinfo {author} {\bibfnamefont {L.~M.}\ \bibnamefont {Sieberer}},
  \bibinfo {author} {\bibfnamefont {E.}~\bibnamefont {Altman}},\ and\ \bibinfo
  {author} {\bibfnamefont {S.}~\bibnamefont {Diehl}},\ }\bibfield  {title}
  {\bibinfo {title} {Scaling properties of one-dimensional driven-dissipative
  condensates},\ }\href {https://doi.org/10.1103/PhysRevB.92.155307} {\bibfield
   {journal} {\bibinfo  {journal} {Phys. Rev. B}\ }\textbf {\bibinfo {volume}
  {92}},\ \bibinfo {pages} {155307} (\bibinfo {year}
  {2015}{\natexlab{a}})}\BibitemShut {NoStop}%
\bibitem [{\citenamefont {Ji}\ \emph {et~al.}(2015{\natexlab{a}})\citenamefont
  {Ji}, \citenamefont {Gladilin},\ and\ \citenamefont {Wouters}}]{Ji2015}%
  \BibitemOpen
  \bibfield  {author} {\bibinfo {author} {\bibfnamefont {K.}~\bibnamefont
  {Ji}}, \bibinfo {author} {\bibfnamefont {V.}~\bibnamefont {Gladilin}},\ and\
  \bibinfo {author} {\bibfnamefont {M.}~\bibnamefont {Wouters}},\ }\bibfield
  {title} {\bibinfo {title} {Temporal coherence of one-dimensional
  nonequilibrium quantum fluids},\ }\href
  {https://doi.org/10.1103/PhysRevB.91.045301} {\bibfield  {journal} {\bibinfo
  {journal} {Physical Review B}\ }\textbf {\bibinfo {volume} {91}},\ \bibinfo
  {pages} {045301} (\bibinfo {year} {2015}{\natexlab{a}})}\BibitemShut
  {NoStop}%
\bibitem [{\citenamefont {He}\ \emph {et~al.}(2017{\natexlab{a}})\citenamefont
  {He}, \citenamefont {Sieberer},\ and\ \citenamefont {Diehl}}]{He2017}%
  \BibitemOpen
  \bibfield  {author} {\bibinfo {author} {\bibfnamefont {L.}~\bibnamefont
  {He}}, \bibinfo {author} {\bibfnamefont {L.}~\bibnamefont {Sieberer}},\ and\
  \bibinfo {author} {\bibfnamefont {S.}~\bibnamefont {Diehl}},\ }\bibfield
  {title} {\bibinfo {title} {Space-time vortex driven crossover and vortex
  turbulence phase transition in one-dimensional driven open condensates},\
  }\href {https://doi.org/10.1103/PhysRevLett.118.085301} {\bibfield  {journal}
  {\bibinfo  {journal} {Physical Review Letters}\ }\textbf {\bibinfo {volume}
  {118}},\ \bibinfo {pages} {085301} (\bibinfo {year}
  {2017}{\natexlab{a}})}\BibitemShut {NoStop}%
\bibitem [{\citenamefont {Roumpos}\ \emph {et~al.}(2012)\citenamefont
  {Roumpos}, \citenamefont {Lohse}, \citenamefont {Nitsche}, \citenamefont
  {Keeling}, \citenamefont {Szymańska}, \citenamefont {Littlewood},
  \citenamefont {Löffler}, \citenamefont {Höfling}, \citenamefont
  {Worschech}, \citenamefont {Forchel},\ and\ \citenamefont
  {Yamamoto}}]{Roumpos2012}%
  \BibitemOpen
  \bibfield  {author} {\bibinfo {author} {\bibfnamefont {G.}~\bibnamefont
  {Roumpos}}, \bibinfo {author} {\bibfnamefont {M.}~\bibnamefont {Lohse}},
  \bibinfo {author} {\bibfnamefont {W.~H.}\ \bibnamefont {Nitsche}}, \bibinfo
  {author} {\bibfnamefont {J.}~\bibnamefont {Keeling}}, \bibinfo {author}
  {\bibfnamefont {M.~H.}\ \bibnamefont {Szymańska}}, \bibinfo {author}
  {\bibfnamefont {P.~B.}\ \bibnamefont {Littlewood}}, \bibinfo {author}
  {\bibfnamefont {A.}~\bibnamefont {Löffler}}, \bibinfo {author}
  {\bibfnamefont {S.}~\bibnamefont {Höfling}}, \bibinfo {author}
  {\bibfnamefont {L.}~\bibnamefont {Worschech}}, \bibinfo {author}
  {\bibfnamefont {A.}~\bibnamefont {Forchel}},\ and\ \bibinfo {author}
  {\bibfnamefont {Y.}~\bibnamefont {Yamamoto}},\ }\bibfield  {title} {\bibinfo
  {title} {Power-law decay of the spatial correlation function in
  exciton-polariton condensates},\ }\href
  {https://doi.org/10.1073/pnas.1107970109} {\bibfield  {journal} {\bibinfo
  {journal} {Proceedings of the National Academy of Sciences}\ }\textbf
  {\bibinfo {volume} {109}},\ \bibinfo {pages} {6467} (\bibinfo {year}
  {2012})}\BibitemShut {NoStop}%
\bibitem [{\citenamefont {Caputo}\ \emph
  {et~al.}(2018{\natexlab{a}})\citenamefont {Caputo}, \citenamefont
  {Ballarini}, \citenamefont {Dagvadorj}, \citenamefont {{S{\'{a}}nchez
  Mu{\~{n}}oz}}, \citenamefont {{De Giorgi}}, \citenamefont {Dominici},
  \citenamefont {West}, \citenamefont {Pfeiffer}, \citenamefont {Gigli},
  \citenamefont {Laussy}, \citenamefont {Szyma{\'{n}}ska},\ and\ \citenamefont
  {Sanvitto}}]{Caputo2018}%
  \BibitemOpen
  \bibfield  {author} {\bibinfo {author} {\bibfnamefont {D.}~\bibnamefont
  {Caputo}}, \bibinfo {author} {\bibfnamefont {D.}~\bibnamefont {Ballarini}},
  \bibinfo {author} {\bibfnamefont {G.}~\bibnamefont {Dagvadorj}}, \bibinfo
  {author} {\bibfnamefont {C.}~\bibnamefont {{S{\'{a}}nchez Mu{\~{n}}oz}}},
  \bibinfo {author} {\bibfnamefont {M.}~\bibnamefont {{De Giorgi}}}, \bibinfo
  {author} {\bibfnamefont {L.}~\bibnamefont {Dominici}}, \bibinfo {author}
  {\bibfnamefont {K.}~\bibnamefont {West}}, \bibinfo {author} {\bibfnamefont
  {L.~N.}\ \bibnamefont {Pfeiffer}}, \bibinfo {author} {\bibfnamefont
  {G.}~\bibnamefont {Gigli}}, \bibinfo {author} {\bibfnamefont {F.~P.}\
  \bibnamefont {Laussy}}, \bibinfo {author} {\bibfnamefont {M.~H.}\
  \bibnamefont {Szyma{\'{n}}ska}},\ and\ \bibinfo {author} {\bibfnamefont
  {D.}~\bibnamefont {Sanvitto}},\ }\bibfield  {title} {\bibinfo {title}
  {{Topological order and thermal equilibrium in polariton condensates}},\
  }\href {https://doi.org/10.1038/nmat5039} {\bibfield  {journal} {\bibinfo
  {journal} {Nature Materials}\ }\textbf {\bibinfo {volume} {17}},\ \bibinfo
  {pages} {145} (\bibinfo {year} {2018}{\natexlab{a}})}\BibitemShut {NoStop}%
\bibitem [{\citenamefont {Comaron}\ \emph {et~al.}(2025)\citenamefont
  {Comaron}, \citenamefont {Estrecho}, \citenamefont {Wurdack}, \citenamefont
  {Pieczarka}, \citenamefont {Steger}, \citenamefont {Snoke}, \citenamefont
  {West}, \citenamefont {Pfeiffer}, \citenamefont {Truscott}, \citenamefont
  {Matuszewski}, \citenamefont {Szyma{\'{n}}ska},\ and\ \citenamefont
  {Ostrovskaya}}]{Comaron2025}%
  \BibitemOpen
  \bibfield  {author} {\bibinfo {author} {\bibfnamefont {P.}~\bibnamefont
  {Comaron}}, \bibinfo {author} {\bibfnamefont {E.}~\bibnamefont {Estrecho}},
  \bibinfo {author} {\bibfnamefont {M.}~\bibnamefont {Wurdack}}, \bibinfo
  {author} {\bibfnamefont {M.}~\bibnamefont {Pieczarka}}, \bibinfo {author}
  {\bibfnamefont {M.}~\bibnamefont {Steger}}, \bibinfo {author} {\bibfnamefont
  {D.~W.}\ \bibnamefont {Snoke}}, \bibinfo {author} {\bibfnamefont
  {K.}~\bibnamefont {West}}, \bibinfo {author} {\bibfnamefont {L.~N.}\
  \bibnamefont {Pfeiffer}}, \bibinfo {author} {\bibfnamefont {A.~G.}\
  \bibnamefont {Truscott}}, \bibinfo {author} {\bibfnamefont {M.}~\bibnamefont
  {Matuszewski}}, \bibinfo {author} {\bibfnamefont {M.~H.}\ \bibnamefont
  {Szyma{\'{n}}ska}},\ and\ \bibinfo {author} {\bibfnamefont {E.~A.}\
  \bibnamefont {Ostrovskaya}},\ }\bibfield  {title} {\bibinfo {title}
  {{Coherence of a non-equilibrium polariton condensate across the
  interaction-mediated phase transition}},\ }\href
  {https://doi.org/10.1038/s42005-025-01977-7} {\bibfield  {journal} {\bibinfo
  {journal} {Communications Physics}\ }\textbf {\bibinfo {volume} {8}},\
  \bibinfo {pages} {94} (\bibinfo {year} {2025})}\BibitemShut {NoStop}%
\bibitem [{\citenamefont {Fontaine}\ \emph {et~al.}(2022)\citenamefont
  {Fontaine}, \citenamefont {Squizzato}, \citenamefont {Baboux}, \citenamefont
  {Amelio}, \citenamefont {Lema{\^i}tre}, \citenamefont {Morassi},
  \citenamefont {Sagnes}, \citenamefont {Le~Gratiet}, \citenamefont {Harouri},
  \citenamefont {Wouters}, \citenamefont {Carusotto}, \citenamefont {Amo},
  \citenamefont {Richard}, \citenamefont {Minguzzi}, \citenamefont {Canet},
  \citenamefont {Ravets},\ and\ \citenamefont {Bloch}}]{Fontaine2022}%
  \BibitemOpen
  \bibfield  {author} {\bibinfo {author} {\bibfnamefont {Q.}~\bibnamefont
  {Fontaine}}, \bibinfo {author} {\bibfnamefont {D.}~\bibnamefont {Squizzato}},
  \bibinfo {author} {\bibfnamefont {F.}~\bibnamefont {Baboux}}, \bibinfo
  {author} {\bibfnamefont {I.}~\bibnamefont {Amelio}}, \bibinfo {author}
  {\bibfnamefont {A.}~\bibnamefont {Lema{\^i}tre}}, \bibinfo {author}
  {\bibfnamefont {M.}~\bibnamefont {Morassi}}, \bibinfo {author} {\bibfnamefont
  {I.}~\bibnamefont {Sagnes}}, \bibinfo {author} {\bibfnamefont
  {L.}~\bibnamefont {Le~Gratiet}}, \bibinfo {author} {\bibfnamefont
  {A.}~\bibnamefont {Harouri}}, \bibinfo {author} {\bibfnamefont
  {M.}~\bibnamefont {Wouters}}, \bibinfo {author} {\bibfnamefont
  {I.}~\bibnamefont {Carusotto}}, \bibinfo {author} {\bibfnamefont
  {A.}~\bibnamefont {Amo}}, \bibinfo {author} {\bibfnamefont {M.}~\bibnamefont
  {Richard}}, \bibinfo {author} {\bibfnamefont {A.}~\bibnamefont {Minguzzi}},
  \bibinfo {author} {\bibfnamefont {L.}~\bibnamefont {Canet}}, \bibinfo
  {author} {\bibfnamefont {S.}~\bibnamefont {Ravets}},\ and\ \bibinfo {author}
  {\bibfnamefont {J.}~\bibnamefont {Bloch}},\ }\bibfield  {title} {\bibinfo
  {title} {{K}ardar--{P}arisi--{Z}hang universality in a one-dimensional
  polariton condensate},\ }\href {https://doi.org/10.1038/s41586-022-05001-8}
  {\bibfield  {journal} {\bibinfo  {journal} {Nature}\ }\textbf {\bibinfo
  {volume} {608}},\ \bibinfo {pages} {687} (\bibinfo {year}
  {2022})}\BibitemShut {NoStop}%
\bibitem [{\citenamefont {Widmann}\ \emph {et~al.}(2026)\citenamefont
  {Widmann}, \citenamefont {Dam}, \citenamefont {Düreth}, \citenamefont
  {Mayer}, \citenamefont {Daviet}, \citenamefont {Zelle}, \citenamefont
  {Laibacher}, \citenamefont {Emmerling}, \citenamefont {Kamp}, \citenamefont
  {Diehl}, \citenamefont {Betzold}, \citenamefont {Klembt},\ and\ \citenamefont
  {Höfling}}]{Widmann2026}%
  \BibitemOpen
  \bibfield  {author} {\bibinfo {author} {\bibfnamefont {S.}~\bibnamefont
  {Widmann}}, \bibinfo {author} {\bibfnamefont {S.}~\bibnamefont {Dam}},
  \bibinfo {author} {\bibfnamefont {J.}~\bibnamefont {Düreth}}, \bibinfo
  {author} {\bibfnamefont {C.~G.}\ \bibnamefont {Mayer}}, \bibinfo {author}
  {\bibfnamefont {R.}~\bibnamefont {Daviet}}, \bibinfo {author} {\bibfnamefont
  {C.~P.}\ \bibnamefont {Zelle}}, \bibinfo {author} {\bibfnamefont
  {D.}~\bibnamefont {Laibacher}}, \bibinfo {author} {\bibfnamefont
  {M.}~\bibnamefont {Emmerling}}, \bibinfo {author} {\bibfnamefont
  {M.}~\bibnamefont {Kamp}}, \bibinfo {author} {\bibfnamefont {S.}~\bibnamefont
  {Diehl}}, \bibinfo {author} {\bibfnamefont {S.}~\bibnamefont {Betzold}},
  \bibinfo {author} {\bibfnamefont {S.}~\bibnamefont {Klembt}},\ and\ \bibinfo
  {author} {\bibfnamefont {S.}~\bibnamefont {Höfling}},\ }\bibfield  {title}
  {\bibinfo {title} {Observation of kardar-parisi-zhang universal scaling in
  two dimensions},\ }\href {https://doi.org/10.1126/science.aeb4154} {\bibfield
   {journal} {\bibinfo  {journal} {Science}\ }\textbf {\bibinfo {volume}
  {392}},\ \bibinfo {pages} {221} (\bibinfo {year} {2026})}\BibitemShut
  {NoStop}%
\bibitem [{\citenamefont {Bloch}\ \emph {et~al.}(2026)\citenamefont {Bloch},
  \citenamefont {Escalera}, \citenamefont {Fontaine}, \citenamefont {Helluin},
  \citenamefont {Minguzzi}, \citenamefont {Canet},\ and\ \citenamefont
  {Ravets}}]{bloch2026}%
  \BibitemOpen
  \bibfield  {author} {\bibinfo {author} {\bibfnamefont {J.}~\bibnamefont
  {Bloch}}, \bibinfo {author} {\bibfnamefont {M.}~\bibnamefont {Escalera}},
  \bibinfo {author} {\bibfnamefont {Q.}~\bibnamefont {Fontaine}}, \bibinfo
  {author} {\bibfnamefont {F.}~\bibnamefont {Helluin}}, \bibinfo {author}
  {\bibfnamefont {A.}~\bibnamefont {Minguzzi}}, \bibinfo {author}
  {\bibfnamefont {L.}~\bibnamefont {Canet}},\ and\ \bibinfo {author}
  {\bibfnamefont {S.}~\bibnamefont {Ravets}},\ }\bibfield  {title} {\bibinfo
  {title} {Comment on ``observation of kardar--parisi--zhang universal scaling
  in two dimensions},\ }\href {https://arxiv.org/abs/2607.24152} {\bibfield
  {journal} {\bibinfo  {journal} {arXiv:2607.24152}\ } (\bibinfo {year}
  {2026})}\BibitemShut {NoStop}%
\bibitem [{\citenamefont {Carusotto}\ and\ \citenamefont
  {Ciuti}(2013)}]{carusotto2013}%
  \BibitemOpen
  \bibfield  {author} {\bibinfo {author} {\bibfnamefont {I.}~\bibnamefont
  {Carusotto}}\ and\ \bibinfo {author} {\bibfnamefont {C.}~\bibnamefont
  {Ciuti}},\ }\bibfield  {title} {\bibinfo {title} {Quantum fluids of light},\
  }\href {https://doi.org/10.1103/RevModPhys.85.299} {\bibfield  {journal}
  {\bibinfo  {journal} {Reviews of Modern Physics}\ }\textbf {\bibinfo {volume}
  {85}},\ \bibinfo {pages} {299} (\bibinfo {year} {2013})}\BibitemShut
  {NoStop}%
\bibitem [{\citenamefont {Wouters}\ and\ \citenamefont
  {Carusotto}(2007)}]{Wouters2007}%
  \BibitemOpen
  \bibfield  {author} {\bibinfo {author} {\bibfnamefont {M.}~\bibnamefont
  {Wouters}}\ and\ \bibinfo {author} {\bibfnamefont {I.}~\bibnamefont
  {Carusotto}},\ }\bibfield  {title} {\bibinfo {title} {Excitations in a
  nonequilibrium bose-einstein condensate of exciton polaritons},\ }\href
  {https://doi.org/10.1103/PhysRevLett.99.140402} {\bibfield  {journal}
  {\bibinfo  {journal} {Phys. Rev. Lett.}\ }\textbf {\bibinfo {volume} {99}},\
  \bibinfo {pages} {140402} (\bibinfo {year} {2007})}\BibitemShut {NoStop}%
\bibitem [{\citenamefont {Sieberer}\ \emph {et~al.}(2016)\citenamefont
  {Sieberer}, \citenamefont {Buchhold},\ and\ \citenamefont
  {Diehl}}]{Sieberer2016}%
  \BibitemOpen
  \bibfield  {author} {\bibinfo {author} {\bibfnamefont {L.~M.}\ \bibnamefont
  {Sieberer}}, \bibinfo {author} {\bibfnamefont {M.}~\bibnamefont {Buchhold}},\
  and\ \bibinfo {author} {\bibfnamefont {S.}~\bibnamefont {Diehl}},\ }\bibfield
   {title} {\bibinfo {title} {Keldysh field theory for driven open quantum
  systems},\ }\href {https://doi.org/10.1088/0034-4885/79/9/096001} {\bibfield
  {journal} {\bibinfo  {journal} {Reports on Progress in Physics}\ }\textbf
  {\bibinfo {volume} {79}},\ \bibinfo {pages} {096001} (\bibinfo {year}
  {2016})}\BibitemShut {NoStop}%
\bibitem [{\citenamefont {Gladilin}\ \emph {et~al.}(2014)\citenamefont
  {Gladilin}, \citenamefont {Ji},\ and\ \citenamefont
  {Wouters}}]{Gladilin_Wouters_PRA2014}%
  \BibitemOpen
  \bibfield  {author} {\bibinfo {author} {\bibfnamefont {V.~N.}\ \bibnamefont
  {Gladilin}}, \bibinfo {author} {\bibfnamefont {K.}~\bibnamefont {Ji}},\ and\
  \bibinfo {author} {\bibfnamefont {M.}~\bibnamefont {Wouters}},\ }\bibfield
  {title} {\bibinfo {title} {Spatial coherence of weakly interacting
  one-dimensional nonequilibrium bosonic quantum fluids},\ }\href
  {https://doi.org/10.1103/PhysRevA.90.023615} {\bibfield  {journal} {\bibinfo
  {journal} {Physical Review A}\ }\textbf {\bibinfo {volume} {90}},\ \bibinfo
  {pages} {023615} (\bibinfo {year} {2014})}\BibitemShut {NoStop}%
\bibitem [{\citenamefont {Ji}\ \emph {et~al.}(2015{\natexlab{b}})\citenamefont
  {Ji}, \citenamefont {Gladilin},\ and\ \citenamefont
  {Wouters}}]{Gladilin_2015_EW_to_KPZ}%
  \BibitemOpen
  \bibfield  {author} {\bibinfo {author} {\bibfnamefont {K.}~\bibnamefont
  {Ji}}, \bibinfo {author} {\bibfnamefont {V.}~\bibnamefont {Gladilin}},\ and\
  \bibinfo {author} {\bibfnamefont {M.}~\bibnamefont {Wouters}},\ }\bibfield
  {title} {\bibinfo {title} {Temporal coherence of one-dimensional
  nonequilibrium quantum fluids},\ }\href
  {https://doi.org/10.1103/PhysRevB.91.045301} {\bibfield  {journal} {\bibinfo
  {journal} {Physical Review B}\ }\textbf {\bibinfo {volume} {91}},\ \bibinfo
  {pages} {045301} (\bibinfo {year} {2015}{\natexlab{b}})}\BibitemShut
  {NoStop}%
\bibitem [{\citenamefont {He}\ \emph {et~al.}(2015{\natexlab{b}})\citenamefont
  {He}, \citenamefont {Sieberer}, \citenamefont {Altman},\ and\ \citenamefont
  {Diehl}}]{He_Diehl_roughness}%
  \BibitemOpen
  \bibfield  {author} {\bibinfo {author} {\bibfnamefont {L.}~\bibnamefont
  {He}}, \bibinfo {author} {\bibfnamefont {L.~M.}\ \bibnamefont {Sieberer}},
  \bibinfo {author} {\bibfnamefont {E.}~\bibnamefont {Altman}},\ and\ \bibinfo
  {author} {\bibfnamefont {S.}~\bibnamefont {Diehl}},\ }\bibfield  {title}
  {\bibinfo {title} {Scaling properties of one-dimensional driven-dissipative
  condensates},\ }\href {https://doi.org/10.1103/PhysRevB.92.155307} {\bibfield
   {journal} {\bibinfo  {journal} {Physical Review B}\ }\textbf {\bibinfo
  {volume} {92}},\ \bibinfo {pages} {155307} (\bibinfo {year}
  {2015}{\natexlab{b}})}\BibitemShut {NoStop}%
\bibitem [{\citenamefont {Squizzato}\ \emph {et~al.}(2018)\citenamefont
  {Squizzato}, \citenamefont {Canet},\ and\ \citenamefont
  {Minguzzi}}]{squizzato2018KPZsubclasses}%
  \BibitemOpen
  \bibfield  {author} {\bibinfo {author} {\bibfnamefont {D.}~\bibnamefont
  {Squizzato}}, \bibinfo {author} {\bibfnamefont {L.}~\bibnamefont {Canet}},\
  and\ \bibinfo {author} {\bibfnamefont {A.}~\bibnamefont {Minguzzi}},\
  }\bibfield  {title} {\bibinfo {title} {{K}ardar-{P}arisi-{Z}hang universality
  in the phase distributions of one-dimensional exciton-polaritons},\ }\href
  {https://doi.org/10.1103/PhysRevB.97.195453} {\bibfield  {journal} {\bibinfo
  {journal} {Physical Review B}\ }\textbf {\bibinfo {volume} {97}},\ \bibinfo
  {pages} {195453} (\bibinfo {year} {2018})}\BibitemShut {NoStop}%
\bibitem [{\citenamefont {Deligiannis}\ \emph {et~al.}(2021)\citenamefont
  {Deligiannis}, \citenamefont {Squizzato}, \citenamefont {Minguzzi},\ and\
  \citenamefont {Canet}}]{deligiannis2021KPZsubclasses}%
  \BibitemOpen
  \bibfield  {author} {\bibinfo {author} {\bibfnamefont {K.}~\bibnamefont
  {Deligiannis}}, \bibinfo {author} {\bibfnamefont {D.}~\bibnamefont
  {Squizzato}}, \bibinfo {author} {\bibfnamefont {A.}~\bibnamefont
  {Minguzzi}},\ and\ \bibinfo {author} {\bibfnamefont {L.}~\bibnamefont
  {Canet}},\ }\bibfield  {title} {\bibinfo {title} {Accessing
  {K}ardar-{P}arisi-{Z}hang universality sub-classes with exciton polaritons},\
  }\href {https://doi.org/10.1209/0295-5075/132/67004} {\bibfield  {journal}
  {\bibinfo  {journal} {Europhysics Letters}\ }\textbf {\bibinfo {volume}
  {132}},\ \bibinfo {pages} {67004} (\bibinfo {year} {2021})}\BibitemShut
  {NoStop}%
\bibitem [{\citenamefont {He}\ \emph {et~al.}(2017{\natexlab{b}})\citenamefont
  {He}, \citenamefont {Sieberer},\ and\ \citenamefont
  {Diehl}}]{Diehlspacetimevortex_PRL2017}%
  \BibitemOpen
  \bibfield  {author} {\bibinfo {author} {\bibfnamefont {L.}~\bibnamefont
  {He}}, \bibinfo {author} {\bibfnamefont {L.}~\bibnamefont {Sieberer}},\ and\
  \bibinfo {author} {\bibfnamefont {S.}~\bibnamefont {Diehl}},\ }\bibfield
  {title} {\bibinfo {title} {Space-time vortex driven crossover and vortex
  turbulence phase transition in one-dimensional driven open condensates},\
  }\href {https://doi.org/10.1103/PhysRevLett.118.085301} {\bibfield  {journal}
  {\bibinfo  {journal} {Physical Review Letters}\ }\textbf {\bibinfo {volume}
  {118}},\ \bibinfo {pages} {085301} (\bibinfo {year}
  {2017}{\natexlab{b}})}\BibitemShut {NoStop}%
\bibitem [{\citenamefont {Vercesi}\ \emph {et~al.}(2023)\citenamefont
  {Vercesi}, \citenamefont {Fontaine}, \citenamefont {Ravets}, \citenamefont
  {Bloch}, \citenamefont {Richard}, \citenamefont {Canet},\ and\ \citenamefont
  {Minguzzi}}]{Vercesi_PRR2023_1d_phase_diag}%
  \BibitemOpen
  \bibfield  {author} {\bibinfo {author} {\bibfnamefont {F.}~\bibnamefont
  {Vercesi}}, \bibinfo {author} {\bibfnamefont {Q.}~\bibnamefont {Fontaine}},
  \bibinfo {author} {\bibfnamefont {S.}~\bibnamefont {Ravets}}, \bibinfo
  {author} {\bibfnamefont {J.}~\bibnamefont {Bloch}}, \bibinfo {author}
  {\bibfnamefont {M.}~\bibnamefont {Richard}}, \bibinfo {author} {\bibfnamefont
  {L.}~\bibnamefont {Canet}},\ and\ \bibinfo {author} {\bibfnamefont
  {A.}~\bibnamefont {Minguzzi}},\ }\bibfield  {title} {\bibinfo {title} {Phase
  diagram of one-dimensional driven-dissipative exciton-polariton
  condensates},\ }\href {https://doi.org/10.1103/PhysRevResearch.5.043062}
  {\bibfield  {journal} {\bibinfo  {journal} {Physical Review Res.}\ }\textbf
  {\bibinfo {volume} {5}},\ \bibinfo {pages} {043062} (\bibinfo {year}
  {2023})}\BibitemShut {NoStop}%
\bibitem [{\citenamefont {Bobrovska}\ \emph {et~al.}(2014)\citenamefont
  {Bobrovska}, \citenamefont {Ostrovskaya},\ and\ \citenamefont
  {Matuszewski}}]{Bobrovska2014}%
  \BibitemOpen
  \bibfield  {author} {\bibinfo {author} {\bibfnamefont {N.}~\bibnamefont
  {Bobrovska}}, \bibinfo {author} {\bibfnamefont {E.~A.}\ \bibnamefont
  {Ostrovskaya}},\ and\ \bibinfo {author} {\bibfnamefont {M.}~\bibnamefont
  {Matuszewski}},\ }\bibfield  {title} {\bibinfo {title} {Stability and spatial
  coherence of nonresonantly pumped exciton-polariton condensates},\ }\href
  {https://doi.org/10.1103/PhysRevB.90.205304} {\bibfield  {journal} {\bibinfo
  {journal} {Phys. Rev. B}\ }\textbf {\bibinfo {volume} {90}},\ \bibinfo
  {pages} {205304} (\bibinfo {year} {2014})}\BibitemShut {NoStop}%
\bibitem [{\citenamefont {Baboux}\ \emph {et~al.}(2018)\citenamefont {Baboux},
  \citenamefont {Bernardis}, \citenamefont {Goblot}, \citenamefont {Gladilin},
  \citenamefont {Gomez}, \citenamefont {Galopin}, \citenamefont {Gratiet},
  \citenamefont {Lema\^{i}tre}, \citenamefont {Sagnes}, \citenamefont
  {Carusotto}, \citenamefont {Wouters}, \citenamefont {Amo},\ and\
  \citenamefont {Bloch}}]{baboux2018}%
  \BibitemOpen
  \bibfield  {author} {\bibinfo {author} {\bibfnamefont {F.}~\bibnamefont
  {Baboux}}, \bibinfo {author} {\bibfnamefont {D.~D.}\ \bibnamefont
  {Bernardis}}, \bibinfo {author} {\bibfnamefont {V.}~\bibnamefont {Goblot}},
  \bibinfo {author} {\bibfnamefont {V.~N.}\ \bibnamefont {Gladilin}}, \bibinfo
  {author} {\bibfnamefont {C.}~\bibnamefont {Gomez}}, \bibinfo {author}
  {\bibfnamefont {E.}~\bibnamefont {Galopin}}, \bibinfo {author} {\bibfnamefont
  {L.~L.}\ \bibnamefont {Gratiet}}, \bibinfo {author} {\bibfnamefont
  {A.}~\bibnamefont {Lema\^{i}tre}}, \bibinfo {author} {\bibfnamefont
  {I.}~\bibnamefont {Sagnes}}, \bibinfo {author} {\bibfnamefont
  {I.}~\bibnamefont {Carusotto}}, \bibinfo {author} {\bibfnamefont
  {M.}~\bibnamefont {Wouters}}, \bibinfo {author} {\bibfnamefont
  {A.}~\bibnamefont {Amo}},\ and\ \bibinfo {author} {\bibfnamefont
  {J.}~\bibnamefont {Bloch}},\ }\bibfield  {title} {\bibinfo {title} {Unstable
  and stable regimes of polariton condensation},\ }\href
  {https://doi.org/10.1364/OPTICA.5.001163} {\bibfield  {journal} {\bibinfo
  {journal} {Optica}\ }\textbf {\bibinfo {volume} {5}},\ \bibinfo {pages}
  {1163} (\bibinfo {year} {2018})}\BibitemShut {NoStop}%
\bibitem [{\citenamefont {Schawlow}\ and\ \citenamefont
  {Townes}(1958)}]{Schawlow1958}%
  \BibitemOpen
  \bibfield  {author} {\bibinfo {author} {\bibfnamefont {A.~L.}\ \bibnamefont
  {Schawlow}}\ and\ \bibinfo {author} {\bibfnamefont {C.~H.}\ \bibnamefont
  {Townes}},\ }\bibfield  {title} {\bibinfo {title} {Infrared and optical
  masers},\ }\href {https://doi.org/10.1103/PhysRev.112.1940} {\bibfield
  {journal} {\bibinfo  {journal} {Phys. Rev.}\ }\textbf {\bibinfo {volume}
  {112}},\ \bibinfo {pages} {1940} (\bibinfo {year} {1958})}\BibitemShut
  {NoStop}%
\bibitem [{\citenamefont {Keeling}\ \emph {et~al.}(2010)\citenamefont
  {Keeling}, \citenamefont {Szyma{\'{n}}ska},\ and\ \citenamefont
  {Littlewood}}]{Keeling2010}%
  \BibitemOpen
  \bibfield  {author} {\bibinfo {author} {\bibfnamefont {J.}~\bibnamefont
  {Keeling}}, \bibinfo {author} {\bibfnamefont {M.~H.}\ \bibnamefont
  {Szyma{\'{n}}ska}},\ and\ \bibinfo {author} {\bibfnamefont {P.~B.}\
  \bibnamefont {Littlewood}},\ }\bibinfo {title} {Keldysh {G}reen's function
  approach to coherence in a non-equilibrium steady state: connecting
  {B}ose-{E}instein condensation and lasing},\ in\ \href
  {https://doi.org/10.1007/978-3-642-12491-4_12} {\emph {\bibinfo {booktitle}
  {Optical Generation and Control of Quantum Coherence in Semiconductor
  Nanostructures}}},\ \bibinfo {editor} {edited by\ \bibinfo {editor}
  {\bibfnamefont {G.}~\bibnamefont {Slavcheva}}\ and\ \bibinfo {editor}
  {\bibfnamefont {P.}~\bibnamefont {Roussignol}}}\ (\bibinfo  {publisher}
  {Springer Berlin Heidelberg},\ \bibinfo {address} {Berlin, Heidelberg},\
  \bibinfo {year} {2010})\ pp.\ \bibinfo {pages} {293--329}\BibitemShut
  {NoStop}%
\bibitem [{\citenamefont {Pagnani}\ and\ \citenamefont
  {Parisi}(2015)}]{Pagnani2015}%
  \BibitemOpen
  \bibfield  {author} {\bibinfo {author} {\bibfnamefont {A.}~\bibnamefont
  {Pagnani}}\ and\ \bibinfo {author} {\bibfnamefont {G.}~\bibnamefont
  {Parisi}},\ }\bibfield  {title} {\bibinfo {title} {Numerical estimate of the
  {K}ardar-{P}arisi-{Z}hang universality class in (2+1) dimensions},\ }\href
  {https://doi.org/10.1103/PhysRevE.92.010101} {\bibfield  {journal} {\bibinfo
  {journal} {Physical Review E}\ }\textbf {\bibinfo {volume} {92}},\ \bibinfo
  {pages} {010101} (\bibinfo {year} {2015})}\BibitemShut {NoStop}%
\bibitem [{\citenamefont {Kloss}\ \emph {et~al.}(2012)\citenamefont {Kloss},
  \citenamefont {Canet},\ and\ \citenamefont
  {Wschebor}}]{Canet2012_Scaling_fct_amplitude_ratios_1d_2d_3d}%
  \BibitemOpen
  \bibfield  {author} {\bibinfo {author} {\bibfnamefont {T.}~\bibnamefont
  {Kloss}}, \bibinfo {author} {\bibfnamefont {L.}~\bibnamefont {Canet}},\ and\
  \bibinfo {author} {\bibfnamefont {N.}~\bibnamefont {Wschebor}},\ }\bibfield
  {title} {\bibinfo {title} {Nonperturbative renormalization group for the
  stationary {K}ardar-{P}arisi-{Z}hang equation: Scaling functions and
  amplitude ratios in 1+1, 2+1, and 3+1 dimensions},\ }\href
  {https://doi.org/10.1103/PhysRevE.86.051124} {\bibfield  {journal} {\bibinfo
  {journal} {Physical Review E}\ }\textbf {\bibinfo {volume} {86}},\ \bibinfo
  {pages} {051124} (\bibinfo {year} {2012})}\BibitemShut {NoStop}%
\bibitem [{\citenamefont {Nattermann}\ and\ \citenamefont
  {Tang}(1992)}]{Nattermann_PRA1992_solutionEW}%
  \BibitemOpen
  \bibfield  {author} {\bibinfo {author} {\bibfnamefont {T.}~\bibnamefont
  {Nattermann}}\ and\ \bibinfo {author} {\bibfnamefont {L.}~\bibnamefont
  {Tang}},\ }\bibfield  {title} {\bibinfo {title} {Kinetic surface roughening.
  {I}. the {K}ardar-{P}arisi-{Z}hang equation in the weak-coupling regime},\
  }\href {https://doi.org/10.1103/PhysRevA.45.7156} {\bibfield  {journal}
  {\bibinfo  {journal} {Physical Review A}\ }\textbf {\bibinfo {volume} {45}},\
  \bibinfo {pages} {7156} (\bibinfo {year} {1992})}\BibitemShut {NoStop}%
\bibitem [{\citenamefont {Foster}\ \emph {et~al.}(2010)\citenamefont {Foster},
  \citenamefont {Blakie},\ and\ \citenamefont
  {Davis}}]{Foster_Davis_PRA2010_Vortex_pairing_2dBose_gas}%
  \BibitemOpen
  \bibfield  {author} {\bibinfo {author} {\bibfnamefont {C.~J.}\ \bibnamefont
  {Foster}}, \bibinfo {author} {\bibfnamefont {P.~B.}\ \bibnamefont {Blakie}},\
  and\ \bibinfo {author} {\bibfnamefont {M.~J.}\ \bibnamefont {Davis}},\
  }\bibfield  {title} {\bibinfo {title} {Vortex pairing in two-dimensional
  {B}ose gases},\ }\href {https://doi.org/10.1103/PhysRevA.81.023623}
  {\bibfield  {journal} {\bibinfo  {journal} {Phys. Rev. A}\ }\textbf {\bibinfo
  {volume} {81}},\ \bibinfo {pages} {023623} (\bibinfo {year}
  {2010})}\BibitemShut {NoStop}%
\bibitem [{\citenamefont {Chat\'e}\ and\ \citenamefont
  {Manneville}(1996)}]{Chate_PhysicaA1996_phase_diagram_2DCGLE}%
  \BibitemOpen
  \bibfield  {author} {\bibinfo {author} {\bibfnamefont {H.}~\bibnamefont
  {Chat\'e}}\ and\ \bibinfo {author} {\bibfnamefont {P.}~\bibnamefont
  {Manneville}},\ }\bibfield  {title} {\bibinfo {title} {Phase diagram of the
  two-dimensional complex {G}inzburg-{L}andau equation},\ }\href
  {https://doi.org/https://doi.org/10.1016/0378-4371(95)00361-4} {\bibfield
  {journal} {\bibinfo  {journal} {Physica A: Statistical Mechanics and its
  Applications}\ }\textbf {\bibinfo {volume} {224}},\ \bibinfo {pages} {348}
  (\bibinfo {year} {1996})},\ \bibinfo {note} {dynamics of Complex
  Systems}\BibitemShut {NoStop}%
\bibitem [{\citenamefont {Almeida}\ \emph {et~al.}(2014)\citenamefont
  {Almeida}, \citenamefont {Ferreira}, \citenamefont {Oliveira},\ and\
  \citenamefont {Reis}}]{almeida2014}%
  \BibitemOpen
  \bibfield  {author} {\bibinfo {author} {\bibfnamefont {R.~A.~L.}\
  \bibnamefont {Almeida}}, \bibinfo {author} {\bibfnamefont {S.~O.}\
  \bibnamefont {Ferreira}}, \bibinfo {author} {\bibfnamefont {T.~J.}\
  \bibnamefont {Oliveira}},\ and\ \bibinfo {author} {\bibfnamefont {F.~D. A.
  A.~a.}\ \bibnamefont {Reis}},\ }\bibfield  {title} {\bibinfo {title}
  {Universal fluctuations in the growth of semiconductor thin films},\ }\href
  {https://doi.org/10.1103/PhysRevB.89.045309} {\bibfield  {journal} {\bibinfo
  {journal} {Physical Review B}\ }\textbf {\bibinfo {volume} {89}},\ \bibinfo
  {pages} {045309} (\bibinfo {year} {2014})}\BibitemShut {NoStop}%
\bibitem [{\citenamefont {Almeida}\ \emph {et~al.}(2017)\citenamefont
  {Almeida}, \citenamefont {Ferreira}, \citenamefont {Ferraz},\ and\
  \citenamefont {Oliveira}}]{Almeida2017}%
  \BibitemOpen
  \bibfield  {author} {\bibinfo {author} {\bibfnamefont {R.~A.~L.}\
  \bibnamefont {Almeida}}, \bibinfo {author} {\bibfnamefont {S.~O.}\
  \bibnamefont {Ferreira}}, \bibinfo {author} {\bibfnamefont {I.}~\bibnamefont
  {Ferraz}},\ and\ \bibinfo {author} {\bibfnamefont {T.~J.}\ \bibnamefont
  {Oliveira}},\ }\bibfield  {title} {\bibinfo {title} {Initial pseudo-steady
  state {\&} asymptotic {KPZ} universality in semiconductor on polymer
  deposition},\ }\href {https://doi.org/10.1038/s41598-017-03843-1} {\bibfield
  {journal} {\bibinfo  {journal} {Scientific Reports}\ }\textbf {\bibinfo
  {volume} {7}},\ \bibinfo {pages} {3773} (\bibinfo {year} {2017})}\BibitemShut
  {NoStop}%
\bibitem [{\citenamefont {Stazzu}\ \emph {et~al.}(2026)\citenamefont {Stazzu},
  \citenamefont {Sacchetto},\ and\ \citenamefont {Carusotto}}]{Stazzu2026}%
  \BibitemOpen
  \bibfield  {author} {\bibinfo {author} {\bibfnamefont {E.}~\bibnamefont
  {Stazzu}}, \bibinfo {author} {\bibfnamefont {G.~A.~P.}\ \bibnamefont
  {Sacchetto}},\ and\ \bibinfo {author} {\bibfnamefont {I.}~\bibnamefont
  {Carusotto}},\ }\bibfield  {title} {\bibinfo {title} {Opening a gap in the
  dispersion of the collective excitations of a driven-dissipative condensate
  subject to an external coherent drive},\ }\href
  {https://doi.org/10.1103/5kr9-ry4f} {\bibfield  {journal} {\bibinfo
  {journal} {Phys. Rev. A}\ }\textbf {\bibinfo {volume} {113}},\ \bibinfo
  {pages} {053303} (\bibinfo {year} {2026})}\BibitemShut {NoStop}%
\bibitem [{\citenamefont {Squizzato}\ and\ \citenamefont
  {Canet}(2019)}]{Squizzato2019}%
  \BibitemOpen
  \bibfield  {author} {\bibinfo {author} {\bibfnamefont {D.}~\bibnamefont
  {Squizzato}}\ and\ \bibinfo {author} {\bibfnamefont {L.}~\bibnamefont
  {Canet}},\ }\bibfield  {title} {\bibinfo {title} {{K}ardar-{P}arisi-{Z}hang
  equation with temporally correlated noise: A nonperturbative renormalization
  group approach},\ }\href {https://doi.org/10.1103/PhysRevE.100.062143}
  {\bibfield  {journal} {\bibinfo  {journal} {Phys. Rev. E}\ }\textbf {\bibinfo
  {volume} {100}},\ \bibinfo {pages} {062143} (\bibinfo {year}
  {2019})}\BibitemShut {NoStop}%
\bibitem [{\citenamefont {Gladilin}\ and\ \citenamefont
  {Wouters}(2020)}]{Wouters2020}%
  \BibitemOpen
  \bibfield  {author} {\bibinfo {author} {\bibfnamefont {V.~N.}\ \bibnamefont
  {Gladilin}}\ and\ \bibinfo {author} {\bibfnamefont {M.}~\bibnamefont
  {Wouters}},\ }\bibfield  {title} {\bibinfo {title} {Classical field model for
  arrays of photon condensates},\ }\href@noop {} {\bibfield  {journal}
  {\bibinfo  {journal} {Phys. Rev. A}\ }\textbf {\bibinfo {volume} {101}},\
  \bibinfo {pages} {043814} (\bibinfo {year} {2020})}\BibitemShut {NoStop}%
\bibitem [{\citenamefont {Porras}\ \emph {et~al.}(2002)\citenamefont {Porras},
  \citenamefont {Ciuti}, \citenamefont {Baumberg},\ and\ \citenamefont
  {Tejedor}}]{Porras2002}%
  \BibitemOpen
  \bibfield  {author} {\bibinfo {author} {\bibfnamefont {D.}~\bibnamefont
  {Porras}}, \bibinfo {author} {\bibfnamefont {C.}~\bibnamefont {Ciuti}},
  \bibinfo {author} {\bibfnamefont {J.~J.}\ \bibnamefont {Baumberg}},\ and\
  \bibinfo {author} {\bibfnamefont {C.}~\bibnamefont {Tejedor}},\ }\bibfield
  {title} {\bibinfo {title} {Polariton dynamics and {B}ose-{E}instein
  condensation in semiconductor microcavities},\ }\href
  {https://doi.org/10.1103/PhysRevB.66.085304} {\bibfield  {journal} {\bibinfo
  {journal} {Phys. Rev. B}\ }\textbf {\bibinfo {volume} {66}},\ \bibinfo
  {pages} {085304} (\bibinfo {year} {2002})}\BibitemShut {NoStop}%
\bibitem [{\citenamefont {Wouters}\ and\ \citenamefont
  {Savona}(2009)}]{Wouters2009}%
  \BibitemOpen
  \bibfield  {author} {\bibinfo {author} {\bibfnamefont {M.}~\bibnamefont
  {Wouters}}\ and\ \bibinfo {author} {\bibfnamefont {V.}~\bibnamefont
  {Savona}},\ }\bibfield  {title} {\bibinfo {title} {Stochastic classical field
  model for polariton condensates},\ }\href
  {https://doi.org/10.1103/PhysRevB.79.165302} {\bibfield  {journal} {\bibinfo
  {journal} {Phys. Rev. B}\ }\textbf {\bibinfo {volume} {79}},\ \bibinfo
  {pages} {165302} (\bibinfo {year} {2009})}\BibitemShut {NoStop}%
\bibitem [{\citenamefont {Gardiner}\ and\ \citenamefont
  {Davis}(2003)}]{Gardiner2003}%
  \BibitemOpen
  \bibfield  {author} {\bibinfo {author} {\bibfnamefont {C.~W.}\ \bibnamefont
  {Gardiner}}\ and\ \bibinfo {author} {\bibfnamefont {M.~J.}\ \bibnamefont
  {Davis}},\ }\bibfield  {title} {\bibinfo {title} {The stochastic
  {G}ross–{P}itaevskii equation: {II}},\ }\href
  {https://doi.org/10.1088/0953-4075/36/23/010} {\bibfield  {journal} {\bibinfo
   {journal} {Journal of Physics B: Atomic, Molecular and Optical Physics}\
  }\textbf {\bibinfo {volume} {36}},\ \bibinfo {pages} {4731} (\bibinfo {year}
  {2003})}\BibitemShut {NoStop}%
\bibitem [{\citenamefont {Altman}\ \emph
  {et~al.}(2015{\natexlab{b}})\citenamefont {Altman}, \citenamefont {Sieberer},
  \citenamefont {Chen}, \citenamefont {Diehl},\ and\ \citenamefont
  {Toner}}]{Altman_PRX2015_2D_superfluidity_anisotropy}%
  \BibitemOpen
  \bibfield  {author} {\bibinfo {author} {\bibfnamefont {E.}~\bibnamefont
  {Altman}}, \bibinfo {author} {\bibfnamefont {L.~M.}\ \bibnamefont
  {Sieberer}}, \bibinfo {author} {\bibfnamefont {L.}~\bibnamefont {Chen}},
  \bibinfo {author} {\bibfnamefont {S.}~\bibnamefont {Diehl}},\ and\ \bibinfo
  {author} {\bibfnamefont {J.}~\bibnamefont {Toner}},\ }\bibfield  {title}
  {\bibinfo {title} {Two-dimensional superfluidity of exciton polaritons
  requires strong anisotropy},\ }\href
  {https://doi.org/10.1103/PhysRevX.5.011017} {\bibfield  {journal} {\bibinfo
  {journal} {Physical Review X}\ }\textbf {\bibinfo {volume} {5}},\ \bibinfo
  {pages} {011017} (\bibinfo {year} {2015}{\natexlab{b}})}\BibitemShut
  {NoStop}%
\bibitem [{\citenamefont {Edwards}\ and\ \citenamefont
  {Wilkinson}(1982)}]{edwards1982surface}%
  \BibitemOpen
  \bibfield  {author} {\bibinfo {author} {\bibfnamefont {S.}~\bibnamefont
  {Edwards}}\ and\ \bibinfo {author} {\bibfnamefont {D.}~\bibnamefont
  {Wilkinson}},\ }\bibfield  {title} {\bibinfo {title} {The surface statistics
  of a granular aggregate},\ }\href {https://doi.org/10.1098/rspa.1982.0056}
  {\bibfield  {journal} {\bibinfo  {journal} {Proceedings of the Royal Society
  of London. A. Mathematical and Physical Sciences}\ }\textbf {\bibinfo
  {volume} {381}},\ \bibinfo {pages} {17} (\bibinfo {year} {1982})}\BibitemShut
  {NoStop}%
\bibitem [{\citenamefont {Tang}\ \emph {et~al.}(1992)\citenamefont {Tang},
  \citenamefont {Forrest},\ and\ \citenamefont
  {Wolf}}]{Tang_PRA1992_hybercube_stacking_EW-KPZ_crossover}%
  \BibitemOpen
  \bibfield  {author} {\bibinfo {author} {\bibfnamefont {L.}~\bibnamefont
  {Tang}}, \bibinfo {author} {\bibfnamefont {B.}~\bibnamefont {Forrest}},\ and\
  \bibinfo {author} {\bibfnamefont {D.}~\bibnamefont {Wolf}},\ }\bibfield
  {title} {\bibinfo {title} {Kinetic surface roughening. {II}.
  {H}ypercube-stacking models},\ }\href
  {https://doi.org/10.1103/PhysRevA.45.7162} {\bibfield  {journal} {\bibinfo
  {journal} {Physical Review A}\ }\textbf {\bibinfo {volume} {45}},\ \bibinfo
  {pages} {7162} (\bibinfo {year} {1992})}\BibitemShut {NoStop}%
\bibitem [{\citenamefont {Nicoli}\ \emph {et~al.}(2010)\citenamefont {Nicoli},
  \citenamefont {Vivo},\ and\ \citenamefont
  {Cuerno}}]{Cuerno_KPZasympotics_2DKS}%
  \BibitemOpen
  \bibfield  {author} {\bibinfo {author} {\bibfnamefont {M.}~\bibnamefont
  {Nicoli}}, \bibinfo {author} {\bibfnamefont {E.}~\bibnamefont {Vivo}},\ and\
  \bibinfo {author} {\bibfnamefont {R.}~\bibnamefont {Cuerno}},\ }\bibfield
  {title} {\bibinfo {title} {{K}ardar-{P}arisi-{Z}hang asymptotics for the
  two-dimensional noisy {K}uramoto-{S}ivashinsky equation},\ }\href
  {https://doi.org/10.1103/PhysRevE.82.045202} {\bibfield  {journal} {\bibinfo
  {journal} {Physical Review E}\ }\textbf {\bibinfo {volume} {82}},\ \bibinfo
  {pages} {045202} (\bibinfo {year} {2010})}\BibitemShut {NoStop}%
\bibitem [{\citenamefont {Savenko}\ \emph {et~al.}(2013)\citenamefont
  {Savenko}, \citenamefont {Liew},\ and\ \citenamefont
  {Shelykh}}]{Shelykh_PRL2013_stochastic_GPE_dynamical_thermalization}%
  \BibitemOpen
  \bibfield  {author} {\bibinfo {author} {\bibfnamefont {I.~G.}\ \bibnamefont
  {Savenko}}, \bibinfo {author} {\bibfnamefont {T.~C.~H.}\ \bibnamefont
  {Liew}},\ and\ \bibinfo {author} {\bibfnamefont {I.~A.}\ \bibnamefont
  {Shelykh}},\ }\bibfield  {title} {\bibinfo {title} {Stochastic
  {G}ross-{P}itaevskii equation for the dynamical thermalization of
  {B}ose-{E}instein condensates},\ }\href
  {https://doi.org/10.1103/PhysRevLett.110.127402} {\bibfield  {journal}
  {\bibinfo  {journal} {Physical Review Letters}\ }\textbf {\bibinfo {volume}
  {110}},\ \bibinfo {pages} {127402} (\bibinfo {year} {2013})}\BibitemShut
  {NoStop}%
\bibitem [{\citenamefont {Fr\'erot}\ \emph {et~al.}(2023)\citenamefont
  {Fr\'erot}, \citenamefont {Vashisht}, \citenamefont {Morassi}, \citenamefont
  {Lema\^{\i}tre}, \citenamefont {Ravets}, \citenamefont {Bloch}, \citenamefont
  {Minguzzi},\ and\ \citenamefont
  {Richard}}]{Frerot_Richard_PRX2023_Bogo_excitations_thermal_phonons}%
  \BibitemOpen
  \bibfield  {author} {\bibinfo {author} {\bibfnamefont {I.}~\bibnamefont
  {Fr\'erot}}, \bibinfo {author} {\bibfnamefont {A.}~\bibnamefont {Vashisht}},
  \bibinfo {author} {\bibfnamefont {M.}~\bibnamefont {Morassi}}, \bibinfo
  {author} {\bibfnamefont {A.}~\bibnamefont {Lema\^{\i}tre}}, \bibinfo {author}
  {\bibfnamefont {S.}~\bibnamefont {Ravets}}, \bibinfo {author} {\bibfnamefont
  {J.}~\bibnamefont {Bloch}}, \bibinfo {author} {\bibfnamefont
  {A.}~\bibnamefont {Minguzzi}},\ and\ \bibinfo {author} {\bibfnamefont
  {M.}~\bibnamefont {Richard}},\ }\bibfield  {title} {\bibinfo {title}
  {Bogoliubov excitations driven by thermal lattice phonons in a quantum fluid
  of light},\ }\href {https://doi.org/10.1103/PhysRevX.13.041058} {\bibfield
  {journal} {\bibinfo  {journal} {Physical Review X}\ }\textbf {\bibinfo
  {volume} {13}},\ \bibinfo {pages} {041058} (\bibinfo {year}
  {2023})}\BibitemShut {NoStop}%
\bibitem [{\citenamefont {Chiocchetta}\ and\ \citenamefont
  {Carusotto}(2013)}]{chiocchetta2013}%
  \BibitemOpen
  \bibfield  {author} {\bibinfo {author} {\bibfnamefont {A.}~\bibnamefont
  {Chiocchetta}}\ and\ \bibinfo {author} {\bibfnamefont {I.}~\bibnamefont
  {Carusotto}},\ }\bibfield  {title} {\bibinfo {title} {Non-equilibrium
  quasi-condensates in reduced dimensions},\ }\href
  {https://doi.org/10.1209/0295-5075/102/67007} {\bibfield  {journal} {\bibinfo
   {journal} {Europhysics Letters}\ }\textbf {\bibinfo {volume} {102}},\
  \bibinfo {pages} {67007} (\bibinfo {year} {2013})}\BibitemShut {NoStop}%
\bibitem [{\citenamefont {Amelio}\ \emph {et~al.}(2024)\citenamefont {Amelio},
  \citenamefont {Chiocchetta},\ and\ \citenamefont {Carusotto}}]{Amelio2024}%
  \BibitemOpen
  \bibfield  {author} {\bibinfo {author} {\bibfnamefont {I.}~\bibnamefont
  {Amelio}}, \bibinfo {author} {\bibfnamefont {A.}~\bibnamefont
  {Chiocchetta}},\ and\ \bibinfo {author} {\bibfnamefont {I.}~\bibnamefont
  {Carusotto}},\ }\bibfield  {title} {\bibinfo {title}
  {{K}ardar-{P}arisi-{Z}hang universality in the coherence time of
  nonequilibrium one-dimensional quasicondensates},\ }\href
  {https://doi.org/10.1103/PhysRevE.109.014104} {\bibfield  {journal} {\bibinfo
   {journal} {Phys. Rev. E}\ }\textbf {\bibinfo {volume} {109}},\ \bibinfo
  {pages} {014104} (\bibinfo {year} {2024})}\BibitemShut {NoStop}%
\bibitem [{\citenamefont {Grynberg}\ \emph {et~al.}(2010)\citenamefont
  {Grynberg}, \citenamefont {Aspect},\ and\ \citenamefont {Fabre}}]{Fabre2010}%
  \BibitemOpen
  \bibfield  {author} {\bibinfo {author} {\bibfnamefont {G.}~\bibnamefont
  {Grynberg}}, \bibinfo {author} {\bibfnamefont {A.}~\bibnamefont {Aspect}},\
  and\ \bibinfo {author} {\bibfnamefont {C.}~\bibnamefont {Fabre}},\ }\bibinfo
  {title} {Complement 3d: The spectral width of a laser: the
  {S}chawlow–{T}ownes limit},\ in\ \href@noop {} {\emph {\bibinfo {booktitle}
  {Introduction to Quantum Optics: From the Semi-classical Approach to
  Quantized Light}}}\ (\bibinfo  {publisher} {Cambridge University Press},\
  \bibinfo {year} {2010})\ p.\ \bibinfo {pages} {257–260}\BibitemShut
  {NoStop}%
\bibitem [{\citenamefont {Deng}\ \emph {et~al.}(2010)\citenamefont {Deng},
  \citenamefont {Haug},\ and\ \citenamefont {Yamamoto}}]{Deng2010}%
  \BibitemOpen
  \bibfield  {author} {\bibinfo {author} {\bibfnamefont {H.}~\bibnamefont
  {Deng}}, \bibinfo {author} {\bibfnamefont {H.}~\bibnamefont {Haug}},\ and\
  \bibinfo {author} {\bibfnamefont {Y.}~\bibnamefont {Yamamoto}},\ }\bibfield
  {title} {\bibinfo {title} {Exciton-polariton {B}ose-{E}instein
  condensation},\ }\href {https://doi.org/10.1103/RevModPhys.82.1489}
  {\bibfield  {journal} {\bibinfo  {journal} {Rev. Mod. Phys.}\ }\textbf
  {\bibinfo {volume} {82}},\ \bibinfo {pages} {1489} (\bibinfo {year}
  {2010})}\BibitemShut {NoStop}%
\bibitem [{\citenamefont {Ostrovskaya}\ \emph {et~al.}(2012)\citenamefont
  {Ostrovskaya}, \citenamefont {Abdullaev}, \citenamefont {Desyatnikov},
  \citenamefont {Fraser},\ and\ \citenamefont {Kivshar}}]{Ostrovskaya2012}%
  \BibitemOpen
  \bibfield  {author} {\bibinfo {author} {\bibfnamefont {E.~A.}\ \bibnamefont
  {Ostrovskaya}}, \bibinfo {author} {\bibfnamefont {J.}~\bibnamefont
  {Abdullaev}}, \bibinfo {author} {\bibfnamefont {A.~S.}\ \bibnamefont
  {Desyatnikov}}, \bibinfo {author} {\bibfnamefont {M.~D.}\ \bibnamefont
  {Fraser}},\ and\ \bibinfo {author} {\bibfnamefont {Y.~S.}\ \bibnamefont
  {Kivshar}},\ }\bibfield  {title} {\bibinfo {title} {Dissipative solitons and
  vortices in polariton bose-einstein condensates},\ }\href
  {https://doi.org/10.1103/PhysRevA.86.013636} {\bibfield  {journal} {\bibinfo
  {journal} {Phys. Rev. A}\ }\textbf {\bibinfo {volume} {86}},\ \bibinfo
  {pages} {013636} (\bibinfo {year} {2012})}\BibitemShut {NoStop}%
\bibitem [{\citenamefont {Deng}\ \emph {et~al.}(2003)\citenamefont {Deng},
  \citenamefont {Weihs}, \citenamefont {Snoke}, \citenamefont {Bloch},\ and\
  \citenamefont {Yamamoto}}]{Deng_2003}%
  \BibitemOpen
  \bibfield  {author} {\bibinfo {author} {\bibfnamefont {H.}~\bibnamefont
  {Deng}}, \bibinfo {author} {\bibfnamefont {G.}~\bibnamefont {Weihs}},
  \bibinfo {author} {\bibfnamefont {D.}~\bibnamefont {Snoke}}, \bibinfo
  {author} {\bibfnamefont {J.}~\bibnamefont {Bloch}},\ and\ \bibinfo {author}
  {\bibfnamefont {Y.}~\bibnamefont {Yamamoto}},\ }\bibfield  {title} {\bibinfo
  {title} {Polariton lasing vs. photon lasing in a semiconductor microcavity},\
  }\href {https://doi.org/10.1073/pnas.2634328100} {\bibfield  {journal}
  {\bibinfo  {journal} {Proceedings of the National Academy of Sciences}\
  }\textbf {\bibinfo {volume} {100}},\ \bibinfo {pages} {15318} (\bibinfo
  {year} {2003})}\BibitemShut {NoStop}%
\bibitem [{\citenamefont {Ohadi}\ \emph {et~al.}(2016)\citenamefont {Ohadi},
  \citenamefont {Gregory}, \citenamefont {Freegarde}, \citenamefont {Rubo},
  \citenamefont {Kavokin}, \citenamefont {Berloff},\ and\ \citenamefont
  {Lagoudakis}}]{Ohadi_PRX2016}%
  \BibitemOpen
  \bibfield  {author} {\bibinfo {author} {\bibfnamefont {H.}~\bibnamefont
  {Ohadi}}, \bibinfo {author} {\bibfnamefont {R.~L.}\ \bibnamefont {Gregory}},
  \bibinfo {author} {\bibfnamefont {T.}~\bibnamefont {Freegarde}}, \bibinfo
  {author} {\bibfnamefont {Y.~G.}\ \bibnamefont {Rubo}}, \bibinfo {author}
  {\bibfnamefont {A.~V.}\ \bibnamefont {Kavokin}}, \bibinfo {author}
  {\bibfnamefont {N.~G.}\ \bibnamefont {Berloff}},\ and\ \bibinfo {author}
  {\bibfnamefont {P.~G.}\ \bibnamefont {Lagoudakis}},\ }\bibfield  {title}
  {\bibinfo {title} {Nontrivial phase coupling in polariton multiplets},\
  }\href {https://doi.org/10.1103/PhysRevX.6.031032} {\bibfield  {journal}
  {\bibinfo  {journal} {Phys. Rev. X}\ }\textbf {\bibinfo {volume} {6}},\
  \bibinfo {pages} {031032} (\bibinfo {year} {2016})}\BibitemShut {NoStop}%
\bibitem [{\citenamefont {Caputo}\ \emph
  {et~al.}(2018{\natexlab{b}})\citenamefont {Caputo}, \citenamefont
  {Ballarini}, \citenamefont {Dagvadorj}, \citenamefont
  {S{\'a}nchez~Mu{\~{n}}oz}, \citenamefont {De~Giorgi}, \citenamefont
  {Dominici}, \citenamefont {West}, \citenamefont {Pfeiffer}, \citenamefont
  {Gigli}, \citenamefont {Laussy}, \citenamefont {Szyma{\'{n}}ska},\ and\
  \citenamefont {Sanvitto}}]{Caputo2018_BKT_transition_incoherent_EPBEC}%
  \BibitemOpen
  \bibfield  {author} {\bibinfo {author} {\bibfnamefont {D.}~\bibnamefont
  {Caputo}}, \bibinfo {author} {\bibfnamefont {D.}~\bibnamefont {Ballarini}},
  \bibinfo {author} {\bibfnamefont {G.}~\bibnamefont {Dagvadorj}}, \bibinfo
  {author} {\bibfnamefont {C.}~\bibnamefont {S{\'a}nchez~Mu{\~{n}}oz}},
  \bibinfo {author} {\bibfnamefont {M.}~\bibnamefont {De~Giorgi}}, \bibinfo
  {author} {\bibfnamefont {L.}~\bibnamefont {Dominici}}, \bibinfo {author}
  {\bibfnamefont {K.}~\bibnamefont {West}}, \bibinfo {author} {\bibfnamefont
  {L.}~\bibnamefont {Pfeiffer}}, \bibinfo {author} {\bibfnamefont
  {G.}~\bibnamefont {Gigli}}, \bibinfo {author} {\bibfnamefont
  {F.}~\bibnamefont {Laussy}}, \bibinfo {author} {\bibfnamefont
  {M.}~\bibnamefont {Szyma{\'{n}}ska}},\ and\ \bibinfo {author} {\bibfnamefont
  {D.}~\bibnamefont {Sanvitto}},\ }\bibfield  {title} {\bibinfo {title}
  {Topological order and thermal equilibrium in polariton condensates},\ }\href
  {https://doi.org/10.1038/nmat5039} {\bibfield  {journal} {\bibinfo  {journal}
  {Nature Materials}\ }\textbf {\bibinfo {volume} {17}},\ \bibinfo {pages}
  {145} (\bibinfo {year} {2018}{\natexlab{b}})}\BibitemShut {NoStop}%
\bibitem [{\citenamefont {Dagvadorj}\ \emph
  {et~al.}(2021{\natexlab{b}})\citenamefont {Dagvadorj}, \citenamefont
  {Kulczykowski}, \citenamefont {Szyma\ifmmode~\acute{n}\else \'{n}\fi{}ska},\
  and\ \citenamefont {Matuszewski}}]{Marzena_2021_first_order}%
  \BibitemOpen
  \bibfield  {author} {\bibinfo {author} {\bibfnamefont {G.}~\bibnamefont
  {Dagvadorj}}, \bibinfo {author} {\bibfnamefont {M.}~\bibnamefont
  {Kulczykowski}}, \bibinfo {author} {\bibfnamefont {M.}~\bibnamefont
  {Szyma\ifmmode~\acute{n}\else \'{n}\fi{}ska}},\ and\ \bibinfo {author}
  {\bibfnamefont {M.}~\bibnamefont {Matuszewski}},\ }\bibfield  {title}
  {\bibinfo {title} {First-order dissipative phase transition in an
  exciton-polariton condensate},\ }\href
  {https://doi.org/10.1103/PhysRevB.104.165301} {\bibfield  {journal} {\bibinfo
   {journal} {Physical Review B}\ }\textbf {\bibinfo {volume} {104}},\ \bibinfo
  {pages} {165301} (\bibinfo {year} {2021}{\natexlab{b}})}\BibitemShut
  {NoStop}%
\bibitem [{\citenamefont {Gladilin}\ and\ \citenamefont
  {Wouters}(2019)}]{Gladilin_multivortex_state_2019}%
  \BibitemOpen
  \bibfield  {author} {\bibinfo {author} {\bibfnamefont {V.}~\bibnamefont
  {Gladilin}}\ and\ \bibinfo {author} {\bibfnamefont {M.}~\bibnamefont
  {Wouters}},\ }\bibfield  {title} {\bibinfo {title} {Multivortex states and
  dynamics in nonequilibrium polariton condensates},\ }\href
  {https://doi.org/10.1088/1751-8121/ab3abc} {\bibfield  {journal} {\bibinfo
  {journal} {Journal of Physics A: Mathematical and Theoretical}\ }\textbf
  {\bibinfo {volume} {52}},\ \bibinfo {pages} {395303} (\bibinfo {year}
  {2019})}\BibitemShut {NoStop}%
\bibitem [{\citenamefont {Giorgetti}\ \emph {et~al.}(2007)\citenamefont
  {Giorgetti}, \citenamefont {Carusotto},\ and\ \citenamefont
  {Castin}}]{Carusotto_Castin_PRA2007_Semiclassical_field_method}%
  \BibitemOpen
  \bibfield  {author} {\bibinfo {author} {\bibfnamefont {L.}~\bibnamefont
  {Giorgetti}}, \bibinfo {author} {\bibfnamefont {I.}~\bibnamefont
  {Carusotto}},\ and\ \bibinfo {author} {\bibfnamefont {Y.}~\bibnamefont
  {Castin}},\ }\bibfield  {title} {\bibinfo {title} {Semiclassical field method
  for the equilibrium {B}ose gas and application to thermal vortices in two
  dimensions},\ }\href {https://doi.org/10.1103/PhysRevA.76.013613} {\bibfield
  {journal} {\bibinfo  {journal} {Phys. Rev. A}\ }\textbf {\bibinfo {volume}
  {76}},\ \bibinfo {pages} {013613} (\bibinfo {year} {2007})}\BibitemShut
  {NoStop}%
\bibitem [{\citenamefont {Keeling}\ and\ \citenamefont
  {Berloff}(2008)}]{Keeling2008}%
  \BibitemOpen
  \bibfield  {author} {\bibinfo {author} {\bibfnamefont {J.}~\bibnamefont
  {Keeling}}\ and\ \bibinfo {author} {\bibfnamefont {N.~G.}\ \bibnamefont
  {Berloff}},\ }\bibfield  {title} {\bibinfo {title} {Spontaneous rotating
  vortex lattices in a pumped decaying condensate},\ }\href
  {https://doi.org/10.1103/PhysRevLett.100.250401} {\bibfield  {journal}
  {\bibinfo  {journal} {Phys. Rev. Lett.}\ }\textbf {\bibinfo {volume} {100}},\
  \bibinfo {pages} {250401} (\bibinfo {year} {2008})}\BibitemShut {NoStop}%
\bibitem [{\citenamefont {Aranson}\ and\ \citenamefont
  {Kramer}(2002)}]{Aranson_RMP2002_world_CGLE}%
  \BibitemOpen
  \bibfield  {author} {\bibinfo {author} {\bibfnamefont {I.}~\bibnamefont
  {Aranson}}\ and\ \bibinfo {author} {\bibfnamefont {L.}~\bibnamefont
  {Kramer}},\ }\bibfield  {title} {\bibinfo {title} {The world of the complex
  {G}inzburg-{L}andau equation},\ }\href
  {https://doi.org/10.1103/RevModPhys.74.99} {\bibfield  {journal} {\bibinfo
  {journal} {Reviews of Modern Physics}\ }\textbf {\bibinfo {volume} {74}},\
  \bibinfo {pages} {99} (\bibinfo {year} {2002})}\BibitemShut {NoStop}%
\bibitem [{\citenamefont {Wachtel}\ \emph {et~al.}(2016)\citenamefont
  {Wachtel}, \citenamefont {Sieberer}, \citenamefont {Diehl},\ and\
  \citenamefont {Altman}}]{Altman_2016_EM_duality}%
  \BibitemOpen
  \bibfield  {author} {\bibinfo {author} {\bibfnamefont {G.}~\bibnamefont
  {Wachtel}}, \bibinfo {author} {\bibfnamefont {L.~M.}\ \bibnamefont
  {Sieberer}}, \bibinfo {author} {\bibfnamefont {S.}~\bibnamefont {Diehl}},\
  and\ \bibinfo {author} {\bibfnamefont {E.}~\bibnamefont {Altman}},\
  }\bibfield  {title} {\bibinfo {title} {Electrodynamic duality and vortex
  unbinding in driven-dissipative condensates},\ }\href
  {https://doi.org/10.1103/PhysRevB.94.104520} {\bibfield  {journal} {\bibinfo
  {journal} {Physical Review B}\ }\textbf {\bibinfo {volume} {94}},\ \bibinfo
  {pages} {104520} (\bibinfo {year} {2016})}\BibitemShut {NoStop}%
\bibitem [{\citenamefont {Sieberer}\ and\ \citenamefont
  {Altman}(2018)}]{Sieberer_PRL2018_defects_anisotropic_driven-open_sys}%
  \BibitemOpen
  \bibfield  {author} {\bibinfo {author} {\bibfnamefont {L.~M.}\ \bibnamefont
  {Sieberer}}\ and\ \bibinfo {author} {\bibfnamefont {E.}~\bibnamefont
  {Altman}},\ }\bibfield  {title} {\bibinfo {title} {Topological defects in
  anisotropic driven open systems},\ }\href
  {https://doi.org/10.1103/PhysRevLett.121.085704} {\bibfield  {journal}
  {\bibinfo  {journal} {Physical Review Letters}\ }\textbf {\bibinfo {volume}
  {121}},\ \bibinfo {pages} {085704} (\bibinfo {year} {2018})}\BibitemShut
  {NoStop}%
\bibitem [{\citenamefont {Valani}\ \emph {et~al.}(2018)\citenamefont {Valani},
  \citenamefont {Groszek},\ and\ \citenamefont {Simula}}]{Valani2018}%
  \BibitemOpen
  \bibfield  {author} {\bibinfo {author} {\bibfnamefont {R.~N.}\ \bibnamefont
  {Valani}}, \bibinfo {author} {\bibfnamefont {A.~J.}\ \bibnamefont
  {Groszek}},\ and\ \bibinfo {author} {\bibfnamefont {T.~P.}\ \bibnamefont
  {Simula}},\ }\bibfield  {title} {\bibinfo {title} {{E}instein–{B}ose
  condensation of {O}nsager vortices},\ }\href
  {https://doi.org/10.1088/1367-2630/aac0bb} {\bibfield  {journal} {\bibinfo
  {journal} {New Journal of Physics}\ }\textbf {\bibinfo {volume} {20}},\
  \bibinfo {pages} {053038} (\bibinfo {year} {2018})}\BibitemShut {NoStop}%
\end{thebibliography}
\end{document}